\documentclass[12pt, a4paper]{article}

\usepackage[height=25cm,width=16cm]{geometry}
\usepackage{feynmp}
\usepackage{amsmath}
\usepackage{ascmac}
\usepackage{dcolumn}
\usepackage{bm,here}
\usepackage{subfig}
\usepackage{comment}
\usepackage{ifpdf}
\usepackage{slashed}
\usepackage{colortbl}
\usepackage{color}
\usepackage{ulem}
\usepackage{comment}
\usepackage{braket}
\usepackage[mathscr]{eucal}
\usepackage[sort&compress, numbers, merge]{natbib}
\usepackage{cancel}

\ifpdf
\usepackage{graphicx}     
\usepackage[bookmarksopen,colorlinks=true,linkcolor=bblue,citecolor=ppink,urlcolor=ppink]{hyperref}
\else     
\usepackage[dvipdfmx]{graphicx}     
\usepackage[dvipdfmx,bookmarksopen,colorlinks=true,linkcolor=bblue,citecolor=ppink,urlcolor=ppink]{hyperref}
\fi

\usepackage{multicol}
\definecolor{red}{rgb}{1,0,0}
\definecolor{ppink}{rgb}{0.921545,0.440586,0.687243}
\definecolor{bblue}{rgb}{0.400000,0.400000,1.000000}
\usepackage[charter]{mathdesign}
\usepackage{soul}
\usepackage{wrapfig}
\usepackage{arydshln}

\newcommand{\abs}[1]{\left\vert {#1} \right\vert}

\newcommand\blfootnote[1]{%
	\begingroup
	\renewcommand\thefootnote{}\footnote{#1}%
	\addtocounter{footnote}{-1}%
	\endgroup
}

\begin{document}

\begin{titlepage}

\begin{flushright}
\end{flushright}

\begin{center}

\vskip 1.5cm
{\large \bf Light WIMPs and MeV Gamma-ray Detection with COSI}

\vskip 2.0cm
{\large
Yu Watanabe$^ {1,*}$\blfootnote{$^*$yu.watanabe@ipmu.jp},
Shigeki Matsumoto$^1$, Christopher M. Karwin$^2$, Tom Melia$^1$, Michela Negro$^3$, Thomas Siegert$^4$, Yuki Watanabe$^1$, Hiroki Yoneda$^4$ and Tadayuki Takahashi$^1$}

\vskip 1.5cm
$^1${\sl Kavli IPMU (WPI), UTIAS, University of Tokyo, Kashiwa, 277-8583, Japan} \\ [.3em]
$^2${\sl NASA Goddard Space Flight Center, 8800 Greenbelt Road, Greenbelt, MD 20771, USA} \\ [.3em]
$^3${\sl Louisiana State University, Baton Rouge, LA 70803, USA} \\ [.3em]
$^4${\sl Julius-Maximilians-Universit\"{a}t W\"{u}rzburg, Fakult\"{a}t f\"{u}r Physik und Astronomie, Institut f\"{u}r Theoretische Physik und Astrophysik, Lehrstuhl f\"{u}r Astronomie, Emil-Fischer-Str. 31, D-97074 W\"{u}rzburg, Germany} \\ [.3em]

\vskip 2.5cm
\begin{abstract}
\noindent
Light weakly interacting massive particles (WIMPs), whose masses are in the sub-GeV scale, have been attracting more attention due to the negative results searching for traditional WIMPs. The light WIMPs are expected to produce gamma rays from annihilation in the MeV energy region. Advancements in technology have opened up possibilities to precisely detect MeV gamma rays, leading to the upcoming space-based mission of the Compton Spectrometer and Imager (COSI). We comprehensively and quantitatively study the phenomenology of light WIMPs to determine if the COSI observations will probe their viable model parameter regions. We first construct models to describe light WIMPs based on the minimality and renormalizability of quantum field theory. Next, we impose various constraints on the models obtained from cosmological observations (CMB, BBN) and dark matter searches (accelerator, underground, astrophysical experiments, etc.). Finally, we identify viable parameter regions in each model and discuss whether or not COSI will be sensitive to the parameter regions. We find that a velocity-dependent annihilation cross-section is predicted in some regions, enabling COSI to detect the dark matter signal while avoiding severe constraints from cosmological observations.
\end{abstract}

\end{center}
		
\end{titlepage}

\tableofcontents
\newpage
\setcounter{page}{1}

\section{Introduction}
\label{sec: intro}

Revealing the nature of cosmic dark matter in our universe is one of the most important problems in astrophysics, cosmology, and particle physics. Numerous candidates have been proposed, one of the most attractive candidates being the thermal dark matter. Thermal dark matter is a candidate that experienced equilibrium with the standard model (SM) particles in the early universe. Hence, it is free from the initial condition problem concerning the dark matter density observed today. Thermal dark matter created by the freeze-out mechanism in the early universe\,\cite{Srednicki:1988ce, Bernstein:1985th} is generally referred to as a weakly interacting massive particle (WIMP), and its mass is predicted to be between $\mathcal{O}(1)$\,MeV and $\mathcal{O}(100)$\,TeV\,\cite{Giovanetti:2021izc, Sabti:2021reh, Chu:2022xuh, Griest:1989wd}. Historically, WIMPs with electroweak-scale masses and interactions have attracted much attention, as they may have a possible connection to new physics related to the origin of electroweak (EW) symmetry breaking. Additionally, they offer a natural explanation for the observed dark matter density, a phenomenon known as the ‘WIMP Miracle.' So, such a WIMP has been intensively searched for with many experiments and observations using various probes\,\cite{Feng:2010gw, Roszkowski:2017nbc}.

However, despite substantial efforts, no robust signals of the EW-scale WIMP have been detected, severely reducing its viable parameter space\,\cite{Arcadi:2017kky, Giusti:1998gz}. Consequently, there is growing interest in a lighter WIMP with a mass less than ${\cal O}(1)$\,GeV, and a heavier WIMP with a mass exceeding ${\cal O}(1)$\,TeV. Given the increasing attention and new experiments planned to search for these candidates, we focus on the light WIMP in this article. In direct dark matter detection at underground laboratories, conventional approaches utilizing a noble liquid such as xenon as a target material quickly lose sensitivity for the light WIMP due to its small recoil energy falling below the detector threshold. Many significant efforts to overcome this difficulty are underway, which include using detectors with lower energy thresholds\,\cite{CDEX:2019hzn, CRESST:2019jnq, EDELWEISS:2019vjv, SuperCDMS:2018gro, SuperCDMS:2014cds}, considering the Migdal effect\,\cite{Ibe:2017yqa, Dolan:2017xbu}, and exploring the WIMP-electron scattering\,\cite{Essig:2011nj, Essig:2015cda}. The high-energy community uses accelerator experiments with increased luminosity to search for feebly interacting light particles such as the light WIMP\,\cite{Lanfranchi:2020crw, Antel:2023hkf}. Indirect dark matter detection utilizing astrophysical observations will also develop in the near future. Light WIMPs typically produces gamma rays with an energy of $\sim \mathcal{O}(1)$ MeV, which is known to be challenging to detect efficiently. Currently, data from COMPTEL\,\cite{schonfelder1993instrument} and SPI\,\cite{Winkler:2003nn} are available, with sensitivities worse than those of different energy ranges, a phenomenon known as the `MeV gap.' Nevertheless, advancements in technology and theoretical studies have led to numerous proposed experiments to fill this gap, including NASA's next gamma-ray instrument currently being developed,  the Compton Spectrometer and Imager (COSI)\,\cite{Tomsick:2021wed}. COSI, a wide-FOV ($\geq 25\%$ of the sky) telescope designed to survey the gamma-ray sky at 0.2–-5\,MeV, is planned to be launched in 2027 as a NASA Small Explorer (SMEX) satellite mission. Therefore, we study the phenomenology of the light WIMP comprehensively and quantitatively to determine if the COSI observations can probe (at least in part) its viable parameter region.

We consider the light WIMP from the perspective of renormalizability and minimality, imposing the $Z_2$ symmetry to stabilize the WIMP, wherein the WIMP is odd while others are even. For the WIMP to be light, it should be singlet under the SM gauge symmetry. In such a case, the most minimal model, which involves only the addition of the WIMP to the SM, known as the Higgs portal scenario, has already been invalidated by stringent collider constraints when the WIMP mass is less than half of the Higgs mass\,\cite{Guo:2010hq, GAMBIT:2017gge}. Therefore, we need a further extension, and the next minimal model is the extension of the SM, adding the WIMP and one more new particle called a mediator, which connects the WIMP with the SM particles. Moreover, the mediator particle must be as light as or lighter than the WIMP to satisfy the relic abundance condition, resulting in being singlet to avoid severe constraints from collider experiments. Furthermore, the mediator particle should be bosonic to have a renormalizable WIMP-WIMP-Mediator interaction vertex. Hence, as concrete models of the light WIMP, we consider two cases of the singlet WIMP with a spin-0 (a real/complex scalar) and a spin-1/2 (a Majorana/Dirac fermion) as representatives of bosonic and fermionic WIMP, along with the two cases of singlet mediator particle with, a spin-0 and 1.

For each model, we construct the Lagrangian and investigate whether there is a parameter region that explains the observed dark matter density via the freeze-out mechanism and survives the experimental constraints mentioned above, along with various cosmological conditions and constraints. The light WIMP faces numerous cosmological constraints, which are very different from the traditional WIMP, as it remains in equilibrium at late times and may impact the Big Bang Nucleosynthesis\,(BBN) and Cosmic Microwave Background\,(CMB). The most striking constraint is from the CMB anisotropy observation, limiting the injection of electromagnetic particles at the recombination epoch. This results in the WIMP annihilation cross-section at that time being $\lesssim 4.1\times 10^{-27} (m_{\rm WIMP}/{\rm GeV})\,{\rm cm^3/s}$\,\cite{Slatyer:2015jla}. In comparison to the canonical value of the cross-section at the freeze-out era for the dark matter density, $10^{-26}\,{\rm cm^3/s}$, a naive WIMP scenario predicting a constant annihilation cross-section over the entire region of a non-relativistic WIMP does not work. However, in each model above, thanks to the existence of the mediator particle, we can find viable parameter regions with a non-trivial velocity-dependence satisfying both requirements by enhancing (suppressing) the annihilation cross-section at the freeze-out (recombination) era. The nature of the WIMP and the mediator particle characterize the regions: "p-wave"-suppressed annihilation, the annihilation via the $s$-channel resonance of the mediator particle with $m_{\rm MED} \sim 2m_{\rm WIMP}$, and the annihilation via the forbidden channel into a pair of mediator particles with $m_{\rm MED} \gtrsim m_{\rm WIMP}$. We investigate the COSI detectability for each of the regions in each model above and find that COSI has the potential to probe a diverse range of the regions: those annihilating via the forbidden channel and via the $s$-channel resonance can produce a detectable continuum gamma-ray through the final state radiation from the $e^- e^+$ final state. Moreover, in some models annihilating via the s-channel resonance, the COSI can detect a line gamma-ray signal generated by the WIMP annihilation.

Our analysis is organized as follows. In section\,\ref{sec: candidates}, we provide the setup of our models, i.e., the extension of the SM with a singlet WIMP having a spin of 0 or 1/2 and a singlet mediator with a spin of 0 or 1. For each model, we detail interactions predicted by the models and present formulae for various phenomenologically important observables. In section\,\ref{sec: conditions and constraints}, we discuss the cosmological and experimental constraints on the models and present how we implement them in our analysis. Section\,\ref{sec: Analysis} presents the result of the comprehensive analysis considering all constraints mentioned in the previous section and discusses the detectability of the light WIMP in the COSI observation. Section\,\ref{sec: summary} summarizes our discussion.

\section{Light WIMP candidates}
\label{sec: candidates}

The most minimal model of the light thermal WIMP is the extension of the SM, adding only the WIMP particle. For the WIMP to be light, it should be singlet under the SM gauge symmetry. In such a case, the WIMP must be a scalar to make the theory renormalizable if we impose the $Z_2$ symmetry to stabilize the WIMP, where the WIMP is odd, and the SM particles are even under this symmetry. Then, the WIMP interacts with the SM particles only through the Higgs boson, i.e., the WIMP has only the so-called Higgs portal interaction, described by the operator $X(x)^2 |H(x)|^2$ (or $|X(x)|^2 |H(x)|^2$) with $X(x)$ and $H(x)$ being the WIMP and the Higgs doublet fields, respectively\,\cite{McDonald:1993ex}. The coupling of the interaction must be strong enough to obtain the correct relic abundance of the WIMP. In contrast, such a strong coupling has already been ruled out by the invisible decay search of the Higgs boson at the LHC experiment when the WIMP mass is smaller than half of the Higgs mass\,\cite{Guo:2010hq, GAMBIT:2017gge}.

Therefore, we need a further extension, and the next minimal model is the extension of the SM, adding the WIMP and one more new particle called a mediator, which connects the WIMP with the SM particles. It is the minimal model for a fermionic WIMP because no renormalizable interaction between the WIMP and the SM particles can be written due to the SM gauge symmetry and the $Z_2$ symmetry. The mediator particle must be as light as or lither than the WIMP. Otherwise, satisfying the relic abundance condition is difficult, so it must also be singlet under the SM gauge symmetry to avoid severe constraints from collider experiments. As a result, to have a WIMP-WIMP-Mediator interaction vertex, the mediator particle must be bosonic and even under the $Z_2$ symmetry. In this article, as concrete light WIMP models to discuss, we consider two cases for the spin of the WIMP, a spin-0 (real/complex scalar) and a spin-1/2 (Majorana/Dirac fermion) as representatives of bosonic and fermionic WIMPs, with two cases for the mediator particle, a spin-0 and 1. To summarize, to make the discussion in this article concrete, quantitative, and systematic, we consider models of the light WIMP based on the following guidelines:
\begin{itemize}
    \item The models involve the standard model (SM) particles, WIMP, and a mediator particle.
    \item All interactions involved in the models are renormalizable unless otherwise stated.
    \item Both the WIMP and the mediator particle are singlets under the SM gauge interactions.
    \item A $Z_2$ symmetry is imposed. The WIMP is odd, and others are even under the symmetry.
    \item The spin of the WIMP is either zero (scalar WIMP) or one-half (fermion WIMP).
    \item The spin of the mediator particle is either zero (scalar) or one (vector).
\end{itemize}

\begin{figure}[t]
    \centering
    \includegraphics[keepaspectratio, scale=0.288]{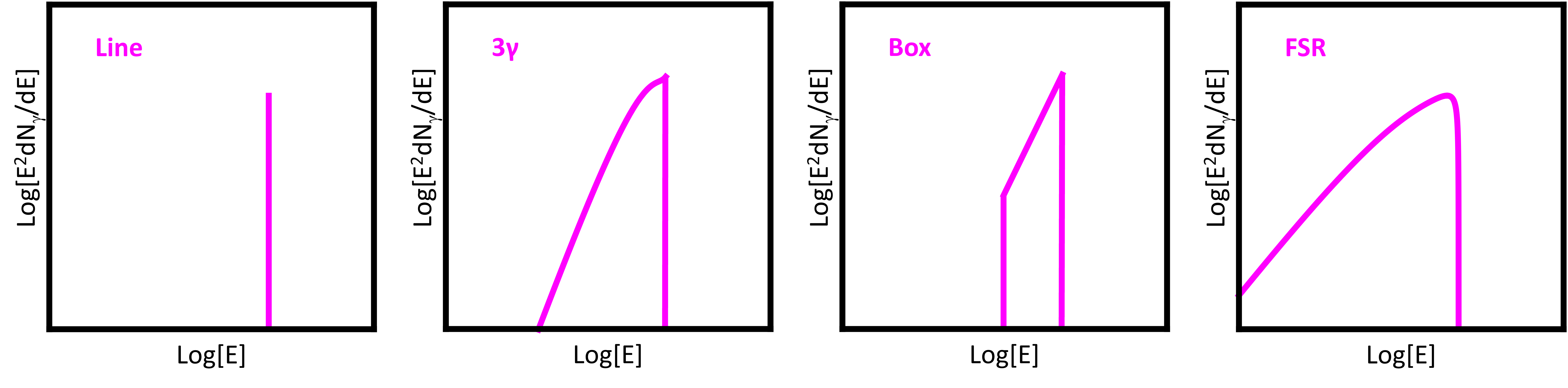}
    \caption{\small \sl Energy spectra giving a harder MeV gamma-ray signal than backgrounds' are shown. The spectra with a line, 3$\gamma$, box, and FSR shapes are shown from the left to the right panels.}
    \label{fig: dN/dE}
\end{figure}

We consider the WIMP mass much lighter than the electroweak scale; we set an upper limit on the mass to be 100\,MeV in this article to make a discussion simple, i.e., to avoid WIMPs annihilating into multiple hadrons. On the other hand, we do not a priori limit the mediator particle mass; other conditions and constraints mentioned below will restrict it. On top of that, we impose constraints on the models obtained by cosmological observations (CMB, BBN) and new particle searches, which are obtained by astrophysical, collider, and underground experiments. We also impose the condition that the dark matter density observed today is from the freeze-out of the WIMP annihilation in the early universe. Then, it turns out that the WIMP annihilation in the present universe predicts a distinct MeV gamma-ray signal, namely its energy spectrum ($dN_\gamma/dE$) is harder than backgrounds', such as the line, $3\gamma$, box, the final state radiation (FSR) spectra, as seen in Fig.\,\ref{fig: dN/dE}.\footnote{
    The differential flux of the WIMP signal is proportional to the energy spectrum, i.e., $d{\cal F}/dE \propto dN_\gamma/dE$.}
To summarize, we focus on the parameter region relevant to dark matter detection at MeV gamma-ray (Compton telescope) observation in the above models, which is characterized as follows:
\begin{itemize}
    \item The parameter region with the WIMP mass smaller than 100\,MeV. The mediator particle mass is not a priori limited but limited by the constraints and conditions below.
    \item The parameter region consistent with cosmological observations (CMB, BBN).
    \item The parameter region consistent with constraints obtained by new particle searches.
    \item The parameter region that all the dark matter density observed today, $\Omega_{\rm DM} h^2 \simeq 0.1$, is from the freeze-out phenomenon of the WIMP annihilation in the early universe.
\end{itemize}

Below, we will introduce the Lagrangians of all the models addressed above and present several observables crucial to discussing the phenomenology of the light WIMP.

\subsection{Models with a scalar mediator}
\label{subsec: scalar mediator}

The most general and renormalizable Lagrangian describing interactions between the mediator particle (i.e., a real singlet scalar particle) and the SM particles is given as follows:
\begin{align}
    & {\cal L}_S = {\cal L}_{\rm SM} + \frac{1}{2} (\partial_\mu S)^2
    - \mu_{S H} S |H|^2
    - \frac{\lambda_{S H}}{2} S^2 |H|^2
    - \mu_1^3 S - \frac{\mu_S^2}{2} S^2
    - \frac{\mu_3}{3!} S^3
    - \frac{\lambda_S}{4!} S^4,
    \label{eq: S}
\end{align}
where ${\cal L}_{\rm SM}$ is the SM Lagrangian. The field describing the mediator particle is denoted by $S$, while $H$ is the Higgs doublet. After the electroweak symmetry breaking and taking the unitary gauge, i.e., $H = (0, v_H + h')^T/\sqrt{2}$ with $v_H \simeq 246$\,GeV being the vacuum expectation value of $H$, while $S = v_S + \varsigma'$ with $v_S = 0$ being that of $S$,\footnote{
    The vacuum expectation value $v_S$ can be zero without loss of generality, leading to $2\mu_1^3 + \mu_{S H} v_H^2 = 0$.}
the mediator mixes with the Higgs boson. The quadratic terms of $h'$ and $\varsigma'$ in the above Lagrangian are given by
\begin{align}
    {\cal L}_S \, \supset \,
	\frac{1}{2}
	\begin{pmatrix} h' & \varsigma' \end{pmatrix}
	\begin{pmatrix} m^2_{h' h'} & m^2_{h' \varsigma'} \\ m^2_{h' \varsigma'} & m^2_{\varsigma' \varsigma'}         \end{pmatrix}
	\begin{pmatrix} h' \\ \varsigma' \end{pmatrix}
	\, = \,
	\frac{1}{2}
	\begin{pmatrix} h & \varsigma \end{pmatrix}
	\begin{pmatrix} m_h^2 & 0 \\ 0 & m_\varsigma^2 \end{pmatrix}
	\begin{pmatrix} h \\ \varsigma \end{pmatrix},
\end{align}
where $m_{h' h'}^2 = \lambda_H v_H^2$, $m_{h' \varsigma'}^2 = \mu_{S H} v_H$ and $m_{\varsigma' \varsigma'}^2 = \mu_S^2 + \lambda_{S H} v_H^2/2$, respectively. The coupling constant $\lambda_H$ is from the Higgs potential in the SM; ${\cal L}_{\rm SM} \supset -(\lambda_H/2)(|H|^2 - v_H^2/2)^2$. The mixing matrix diagonalizing the mass matrix and relating the states $(h', \varsigma')$ to $(h, \varsigma)$ is written as
\begin{align}
	\begin{pmatrix} h \\ \varsigma \end{pmatrix} =
	\begin{pmatrix} \cos\theta & -\sin\theta \\ \sin\theta & \cos\theta \end{pmatrix}
	\begin{pmatrix} h' \\ \varsigma' \end{pmatrix}.
\end{align}
Mass eigenstates and the mixing angle are $m_h^2,\,m_\varsigma^2 = [(m_{h' h'}^2 + m_{\varsigma' \varsigma'}^2) \pm \{(m_{h' h'}^2 - m_{\varsigma' \varsigma'}^2)^2 + 4m_{h' \varsigma'}^4\}^{1/2}]/2$ and $\tan\, (2\theta) = 2m_{h' \varsigma'}^2/(m_{\varsigma' \varsigma'}^2 - m_{h' h'}^2)$ in terms of Lagrangian parameters. Since the mixing angle $\theta$ is constrained to be small, as we will see in the following sections, $h$ is almost the SM Higgs boson, and its mass is given by $m_h \simeq 125$\,GeV as confirmed by the collider (LHC) experiment. All the interactions in the above Lagrangian are written as
\begin{align}
    {\cal L}_S \, \supset \,
	=
	&
	-\frac{s_\theta \varsigma + c_\theta h}{v_H} \sum_f m_f \bar{f}f
	+\left[\frac{s_\theta \varsigma + c_\theta h}{v_H} + \frac{(s_\theta \varsigma + c_\theta h)^2}{2v_H^2}\right] \left(2m_W^2 W_\mu^\dagger W^\mu + m_Z^2 Z_\mu Z^\mu\right)
    \nonumber \\
    &
	-\frac{C_{h h h}}{3!} h^3
	-\frac{C_{\varsigma h h}}{2} \varsigma h^2
	-\frac{C_{\varsigma \varsigma h}}{2} \varsigma^2 h
	-\frac{C_{\varsigma \varsigma \varsigma }}{3!} \varsigma^3
	\nonumber\\
	&
	-\frac{C_{h h h h}}{4!} h^4
	-\frac{C_{\varsigma h h h}}{3!} \varsigma h^3
        -\frac{C_{\varsigma \varsigma h h}}{4} \varsigma^2 h^2
        -\frac{C_{\varsigma \varsigma \varsigma h}}{3!} \varsigma^3 h
        -\frac{C_{\varsigma \varsigma \varsigma \varsigma}}{4!} \varsigma^4
	+ \cdots,
	\label{eq: S interactions}
\end{align}
where $c_\theta\equiv \cos\theta$, $s_\theta\equiv \sin\theta$ and, ``$\dots$'' represent other interactions in ${\cal L}_{\rm SM}$ which are not affected by the mixing between $H$ and $S$, e.g., the gauge interactions of the SM fermions. The SM fermions (quarks and leptons) are denoted by $f$, while $W_\mu$ and $Z_\mu$ are the SM weak gauge bosons ($W$ and $Z$ bosons) with $m_W$ and $m_Z$ being their masses. The coefficients of the scalar interactions, i.e., the interactions among $h$ and $\varsigma$, are given in appendix\,\ref{app: scalar interactions}.

The mediator particle behaves like a light SM Higgs boson, interacting with the SM particles via mixing between $H$ and $S$. Its decay width is generally given by $\Gamma\,(\varsigma \to {\rm SMs}) = \sin^2 \theta\,\Gamma(h_{\rm SM} \to {\rm SMs})$, with $m_{h_{\rm SM}}$ being replaced by $m_\varsigma$. Here, $h_{\rm SM}$ is the SM Higgs boson. First, the mediator particle decays mainly into a pair of photons when $m_\varsigma \leq 2m_e$, with $m_e$ being the electron mass: We use the formula developed in Ref.\,\cite{Leutwyler:1989tn} to calculate its partial decay width $\Gamma(\varsigma \to \gamma \gamma)$. In the region of $2 m_e \leq m_\varsigma \leq 2 m_\mu$ with $m_\mu$ being the muon mass, it decays mainly into a pair of electrons. Its partial decay width at the leading order is
\begin{equation}
	\Gamma(\varsigma \to e^- e^+) = \sin^2 \theta \times \frac{m_e^2 m_\varsigma}{8 \pi v_H^2} 
	\left( 1 - \frac{4 m_e^2}{m_\varsigma^2} \right)^{3/2}.
	\label{eq: electrons}
\end{equation}
In addition, the mediator particle also decays into a pair of electrons associated with a photon, $\varsigma \to e^- e^+ \gamma$. The following formula gives its differential partial decay width\,\cite{Coogan:2019qpu}:
{\small
\begin{align}
    &\frac{d\Gamma(\varsigma \to e^- e^+ \gamma)}{dE}
    =\frac{2 \alpha}{\pi m_\varsigma} \Gamma(\varsigma \to e^- e^+)
    \times {\rm FSRS}(2E/m_\varsigma, m_e/m_\varsigma),
    \label{eq: s to eegamma}
    \\
    &{\rm FSRS}(x, \mu)
    =\frac{2(1-x-6\mu^2)+(x+4\mu^2)^2}{x\,(1 - 4\mu^2)^{3/2}}
    \log\left[\frac{1 + \sqrt{1 - \frac{4\mu^2}{1 - x}}}{1 - \sqrt{1 - \frac{4\mu^2}{1 - x}}}\right]
    -\frac{2(1 - 4\mu^2)(1 - x)}{x\,(1 - 4\mu^2)^{3/2}}\sqrt{1-\frac{4\mu^2}{1-x}},
    \nonumber
\end{align}
}\noindent
where $E$ is the energy of the photon emitted. Though this is a next-to-leading order process, it contributes to the MeV gamma-ray signal in some light WIMP scenarios. Meanwhile, the mass region of $m_\varsigma \geq 2 m_\mu$ is irrelevant to our discussion, so we do not consider it.

\begin{figure}[t]
    \centering
    \includegraphics[keepaspectratio, scale=0.8]{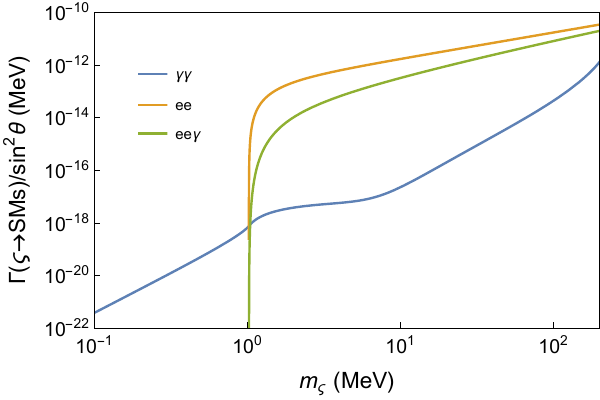}
    \caption{\small \sl The partial decay widths of the scalar mediator particle $\varsigma$ into the SM particles.}
    \label{fig: s to SMs}
\end{figure}

All the above partial decay widths ($\varsigma \to \gamma\gamma$, $e^- e^+$, and $e^- e^+ \gamma$) are summarized in Fig.\,\ref{fig: s to SMs}.\footnote{
    Because of the infrared divergence (i.e., the divergence at $E \to 0$) in eq.\,(\ref{eq: s to eegamma}), which is regularized by an interference term between the leading and the next-leading order diagrams of the $\varsigma \to e^- e^+$ process, the partial decay width of $\Gamma(\varsigma \to e^- e^+ \gamma)$ in the plot is computed as that with a photon having $E \geq 10^{-3} m_e$. It means that $\Gamma(\varsigma \to e^- e^+ \gamma)$ with a photon having $E \leq 10^{-3} m_e$ is treated as a radiative correction to $\Gamma(\varsigma \to e^- e^+)$.}

\subsubsection{Light scalar WIMP with the scalar mediator}
\label{subsub: SS}

We now introduce a light WIMP to the system discussed above, i.e., the one with the SM particles and the singlet scalar mediator particle. We first consider the case of the WIMP being a scalar particle. From the viewpoint of minimality, we take the WIMP as a real scalar, singlet under the SM gauge symmetry, and odd under the $Z_2$ symmetry, as mentioned at the beginning of this section. The most general and renormalizable Lagrangian is given as
\begin{align}
    {\cal L}_\phi = &{\cal L}_S
    +\frac{1}{2} (\partial_\mu \phi)^2
    -\frac{\mu_\phi^2}{2} \phi^2
    -\frac{\mu_{S\phi}}{2} S \phi^2
    -\frac{\lambda_{\phi S}}{4} S^2 \phi^2
    -\frac{\lambda_{H\phi}}{2} |H|^2 \phi^2
    -\frac{\lambda_\phi}{4!} \phi^4,
    \label{eq: S-phi lagrangian}
\end{align}
where ${\cal L}_S$ is the Lagrangian in eq.\,(\ref{eq: S}), and $\phi$ is the field describing the light WIMP. After the electroweak symmetry is broken, the Lagrangian gives interactions between $\phi$, $\varsigma$, and $h$,
\begin{align}
    {\cal L}_\phi \, \supset \,
    -\frac{C_{h \phi \phi}}{2} h \phi^2
    -\frac{C_{\varsigma \phi \phi}}{2} \varsigma \phi^2
    -\frac{C_{h h \phi \phi}}{4} h^2 \phi^2
    -\frac{C_{h \varsigma \phi \phi}}{2} h \varsigma \phi^2
    -\frac{C_{\varsigma \varsigma \phi \phi}}{4} \varsigma^2 \phi^2
    -\frac{\lambda_\phi}{4!} \phi^4
    \label{eq: S-phi interactions}
\end{align}
The coupling constants are given in appendix\,\ref{app: scalar interactions} as a function of the Lagrangian parameters.

\begin{figure}[t]
    \centering
    \includegraphics[keepaspectratio, scale=0.78]{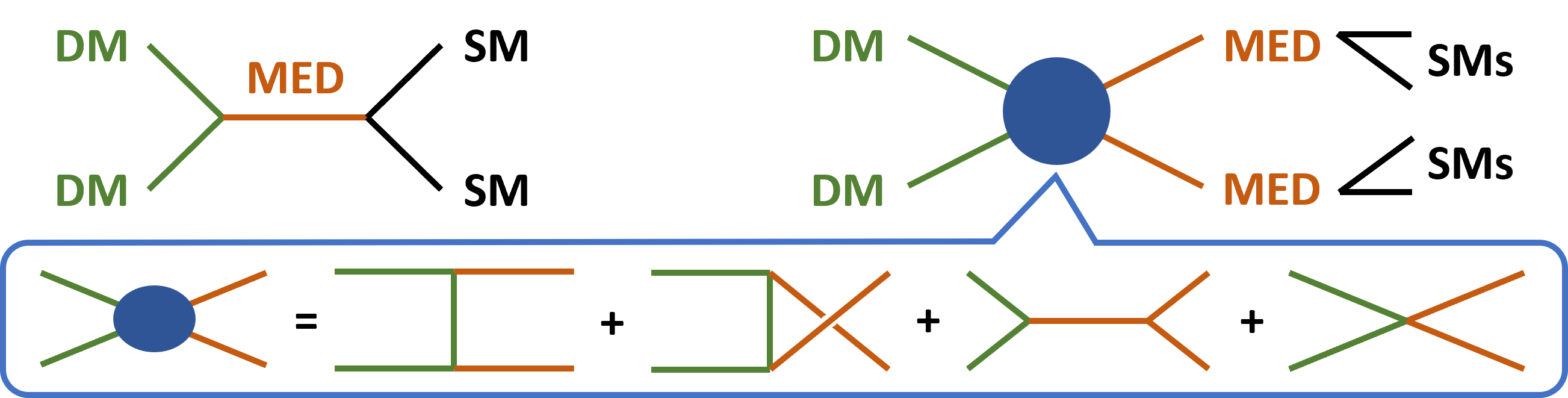}
    \caption{\small \sl Diagrams contributing to the light WIMP annihilation into the SM particles.}
	\label{fig: diagrams S}
\end{figure}

In this model, the light WIMP annihilates into the SM particles in two ways. One is the annihilation through the mediator particle (and the Higgs boson) in the s-channel, as shown in Fig.\,\ref{fig: diagrams S} (left diagram). This is only the annihilation process when the mediator particle is heavier than the WIMP. For instance, the annihilation cross-section into $e^- e^+$ is given by
\begin{align}
    &\sigma_a v\,(\phi \phi \to e^- e^+)
    = \frac{m_e^2}{4 \pi v_H^2}
    \left|
        \frac{\sin\theta\,C_{\varsigma \phi \phi}}{s - m_\varsigma^2 + i\,s^{1/2}\,\Gamma_{\varsigma; \phi}}
        +
        \frac{\cos\theta\,C_{h \phi \phi}}{s - m_h^2}
    \right|^2
    \left( 1 - \frac{4 m_e^2}{s^2} \right)^{3/2},
    \label{eq: S-phi Annihilation cross-section}
\end{align}
with "$s$" being the center-of-mass energy of the annihilation. Since we discuss the light WIMP whose mass is smaller than 100\,MeV, as mentioned at the beginning of this section, we do not consider the imaginary part of the Higgs propagator in the above cross-section. On the other hand, we consider the imaginary part of the propagator of the mediator particle,
{\small
\begin{align}
    &\Gamma_{\varsigma; \phi} (s) \equiv
    \left[
        \Gamma\,(\varsigma \to \phi \phi) + \Gamma\,(\varsigma \to {\rm SMs})
    \right]_{m_\varsigma^2 \to s},
    \quad
    \Gamma(\varsigma \to \phi \phi) = \frac{C_{\varsigma \phi \phi}^2}{32 \pi m_\phi} \left(1 - \frac{4 m_\phi^2}{m_\varsigma^2} \right)^{1/2},
    \label{eq: imaginary part of varsigma}
\end{align}
}\noindent
with $m_\phi = (\mu_\phi^2 + \lambda_{H \phi} v_H^2/2)^{1/2}$ being the physical mass of the scalar WIMP $\phi$. Here, $\Gamma\,(\varsigma \rightarrow \phi \phi)$ is the decay width into a pair of the WIMPs, and $\Gamma\,(\varsigma \to {\rm SMs}) \simeq \Gamma\,(\varsigma \to e^- e^+) + \Gamma\,(\varsigma \to e^- e^+ \gamma) + \Gamma\,(\varsigma \to \gamma \gamma)$ is the (total) decay width into the SM particles. In addition to the above process, the WIMP also annihilates into $e^- e^+ \gamma$ at the next-to-leading order, as well as into a pair of photons, which dominates the WIMP annihilation when $m_\phi < m_e$. Note that the annihilation processes, $\phi \phi \to e^- e^+ \gamma$ and $\phi \phi \to \gamma \gamma$, produce a distinctive MeV gamma-ray signal against backgrounds in the indirect dark matter detection, as seen in Fig.\,\ref{fig: dN/dE}.

Another one is the annihilation into a pair of mediator particles, followed by the decay of the mediator particle into the SM particles, as shown in Fig.\,\ref{fig: diagrams S} (right diagram). Since the coupling constants of the mediator particle to the SM particles are suppressed (due to the suppressed mixing angle $\sin\theta$) while that with the WIMP is not, this process dominates the annihilation when $m_\phi > m_\varsigma$. Its differential annihilation cross-section is given by
{\small
\begin{align}
    \frac{d\sigma_a v\,(\phi \phi \to \varsigma \varsigma)}{d\Omega}
    = \frac{\sqrt{1 - 4 m_\varsigma^2/s}}{32 \pi^2\,s}
    \left|
        C_{\varsigma \varsigma \phi \phi}
        +
        \frac{C_{\varsigma \phi \phi}\,C_{\varsigma \varsigma \varsigma}}{s -m_\varsigma^2}
        +
        \frac{C_{h \phi \phi}\,C_{\varsigma \varsigma h}}{s - m_h^2}
        +
        \frac{C_{\varsigma \phi \phi}^2}{t-m_\phi^2}
        +
        \frac{C_{\varsigma \phi \phi}^2}{u-m_\phi^2}
    \right|^2,
\end{align}
}\noindent
where $t$ and $u$ are the so-called Mandelstam variables that depend on the center-of-mass energy squared $s$ and the angle between the incoming $\phi$ and outgoing $\varsigma$. So, the total annihilation cross-section is obtained by integrating the above differential cross-section over the angle. Since the final state of the process is composed of two $\varsigma$s so that $s$ never hits the pole of the $\varsigma$-propagator, we do not consider the imaginary part of the propagator. Note that this annihilation process also produces a distinctive MeV gamma-ray signal when (one of) the mediator particles in the final state decay(s) into a pair of photons, leading to a box-shaped gamma-ray spectrum seen in Fig.\,\ref{fig: dN/dE} (middle panel). Moreover, when $m_\phi \simeq m_\varsigma$, the process also gives a distinctive signal even if the mediator particle decays into $e^- e^+ \gamma$ instead of $\gamma \gamma$ because the mediator particles in the final state are produced at rest.\footnote{
    The signal from $\varsigma \to e^- e^+ \gamma$ in the final state is still harder than backgrounds' even when $m_\phi \gg m_\varphi$, but the energy spectrum at the threshold $E \simeq m_\phi$ is smeared, giving a softer spectrum than that with $m_\phi \simeq m_\varphi$.}

Finally, we consider the scatterings of the WIMP with an electron and a nucleon in this model, which are relevant to direct dark matter detection at various underground experiments. First, the scattering cross-section between the light WIMP and an electron is
\begin{align}
    & \sigma_s(\phi e \to \phi e) \simeq
    \frac{m_e^4}{4 \pi v_H^2 (m_\phi + m_e)^2}
    \left(
        \sin \theta \frac{C_{\varsigma\phi\phi}}{m_\varsigma^2}
        +\cos \theta \frac{C_{h\phi\phi}}{m_h^2}
    \right)^2.
    \label{eq: phi-e scattering}
\end{align}
The above cross-section is suppressed by the fourth power of the electron mass when $m_\phi \gg m_e$ because the cross-section is proportional to the electron Yukawa coupling squared, $y_e^2 \sim m_e^2/v_H^2$, and the reduced mass squared, $\mu_{\phi e}^2 = m_\phi^2 m_e^2/(m_\phi + m_e)^2$.\footnote{
    Here, it is implicitly assumed in the formula that the mass of the mediator particle is larger than the typical size of the momentum transfer at this scattering, i.e., $\sim \alpha m_e$, with $\alpha$ being the fine structure constant\,\cite{Essig:2011nj}.}
On the other hand, the scattering cross-section between the WIMP and a nucleon is given by the formula:
\begin{align}
    \sigma_s(\phi N \to \phi N) \simeq
    \frac{f_N^2\,m_N^4}{4 \pi v_H^2 (m_\phi+m_N)^2}
    \left(
        \sin \theta \frac{ C_{\varsigma\phi\phi} }{m_\varsigma^2}
        +\cos \theta \frac{ C_{h\phi\phi} }{m_h^2}
    \right)^2,
    \label{eq: phi-N scattering}
\end{align}
where $m_N$ is the nucleon mass, while $f_N$ is the nuclear form factor, which is estimated to be $f_N \simeq 0.284$ and 0.287 when the nucleon $N$ is a neutron and proton, respectively\,\cite{Belanger:2013oya, Thomas:2012tg, Alarcon:2011zs, Alarcon:2012nr}.

\subsubsection{Light fermion WIMP with the scalar mediator}
\label{subsubsec: FS}

We next consider the case of the light WIMP being a fermionic particle and introduce it to the system discussed in section\,\ref{subsec: scalar mediator}.  We take the WIMP as a Majorana fermion from the viewpoint of minimality, which is singlet under the SM gauge symmetry and odd under the $Z_2$ symmetry. The most general and renormalizable Lagrangian of the model is given by
\begin{align}
    {\cal L}_\chi = {\cal L}_S 
    + \frac{1}{2} \bar{\chi} (i\slashed{\partial} - m_{\chi}) \chi
    - \frac{C_s}{2} S \bar{\chi} \chi
    - \frac{C_p}{2} S \bar{\chi} i \gamma_5 \chi,
    \label{eq: S-chi lagrangian}
\end{align}
where ${\cal L}_S$ is the Lagrangian in eq.\,(\ref{eq: S}), and $\chi$ is the field describing the light fermionic WIMP. After the electroweak symmetry breaking, the Lagrangian gives the following interactions:
\begin{align}
    \mathcal{L}_{\rm int} \supset
    -\frac{\cos\theta}{2} (C_s \phi \bar{\chi} \chi + C_p \phi \bar{\chi} i \gamma_5 \chi)
    + \frac{\sin\theta}{2} (C_s h \bar{\chi} \chi + C_p h \bar{\chi} i \gamma_5 \chi).
\end{align}
The interaction with the coupling $C_p$ breaks the CP in the dark matter (and the mediator) sector, i.e., outside the SM sector. Whether the CP is broken or not in the dark matter sector affects the property of the WIMP annihilation, as seen in the following discussion.

As in the previous model with the scalar WIMP, the fermionic WIMP also annihilates into the SM particles in two ways shown in Fig.\,\ref{fig: diagrams S}. The WIMP annihilation cross-section into $e^- e^+$ through the $s$-channel exchange of the mediator particle (and the Higgs boson), which corresponds to the process described by the left diagram in the figure, is given by
\begin{align}
    \sigma_a v\,(\chi \chi \to e^- e^+) =
    &\frac{s_\theta^2 c_\theta^2}{8 \pi v_H^2}
    \left[ C_s^2\,v_\chi^2(s) + C_p^2 \right]
    \left|
        \frac{m_e\,s^{1/2}}{s - m_\varsigma^2 + i\,s^{1/2}\,\Gamma_{\varsigma; \chi}(s)}
        +
        \frac{m_e\,s^{1/2}}{s - m_h^2}
    \right|^2
    v_e^3(s),
\end{align}
where $v_\chi(s) \equiv (1 - 4 m_\chi^2/s)^{1/2}$ and $v_e(s) \equiv (1 - 4 m_e^2/s)^{1/2}$. The imaginary part of the mediator particle propagator is given by $\Gamma_{\varsigma; \chi} (s) \equiv[\Gamma\,(\varsigma \to \chi \chi) + \Gamma\,(\varsigma \to {\rm SMs})]_{m_\varsigma^2 \to s}$, where $\Gamma\,(\varsigma \to {\rm SMs})$ is the (total) decay width into the SM particles mentioned in section\,\ref{subsub: SS}, and $\Gamma\,(\varsigma \to \chi \chi)$, i.e., the decay width into a pair of light fermionic WIMPs, is obtained as follows:
\begin{align}
    \Gamma(\varsigma \to \chi \chi) = \cos^2 \theta\,\frac{m_\varsigma}{16 \pi}
    \left[
        C_s^2 \, v_\chi^3(m_\varsigma)
        + C_p^2\,v_\chi(m_\varsigma).
    \right].
\end{align}
The above annihilation cross-section is suppressed by "p-wave" when $C_p = 0$, i.e., the WIMP does not annihilate when the angular momentum of the incident WIMPs is zero due to the CP and angular momentum conservation. Then, the cross-section is suppressed by $v_\chi^2(s) = (1 - 4m_\chi^2/s) \simeq v^2/4$ at the non-relativistic (NR) limit of the WIMP, with $v$ being the relative velocity between the WIMPs. As in the case of the scalar WIMP, the fermionic WIMP also annihilates into $e^- e^+ \gamma$ at the next-to-leading order and into a pair of photons, where all the processes are "p-wave"-suppressed when $C_p = 0$, as it originates in the conservation.

On the other hand, the fermionic WIMP also annihilates into a pair of mediator particles when $m_\chi \gtrsim m_\varsigma$, which is described by the right diagram in Fig.\,\ref{fig: diagrams S}.\footnote{
    We do not have the contact diagram (the last one in Fig.\,\ref{fig: diagrams S}) due to the fermionic nature of the WIMP.}
When the CP is conserved in the dark matter sector, i.e., $C_p = 0$, the cross-section at the NR limit is obtained as
\begin{align}
    \sigma_a v\,(\chi \chi \to \varsigma \varsigma) \simeq
    \frac{C_s^2 v_\varsigma(s)\,v_\chi^2(s)}{384\pi}
    &
    \left[
        \left(\frac{c_\theta C_{\varsigma\varsigma\varsigma}}{3m_\varsigma^2} + \frac{s_\theta C_{\varsigma\varsigma h}}{m_h^2}\right)^2
        +\frac{16C_s^2 c_\theta^4 m_\chi^2
        (9m_\chi^4 - 8m_\varsigma m_\chi^2 + 2m_\varsigma^4)}{(2m_\chi^2 - m_\varsigma^2)^4}
    \right.
    \nonumber \\
    & \qquad
    \left.
        -\left( \frac{c_\theta C_{\varsigma\varsigma\varsigma}}{3m_\varsigma^2} + \frac{s_\theta C_{\varsigma\varsigma h}}{m_h^2} \right)
        \frac{8C_s c_\theta^2m_\chi(5m_\chi^2 - 2m_\varsigma^2)}{(2m_\chi^2 - m_\varsigma^2)^2}
    \right],
\end{align}
with $v_\varsigma(s)$ being $v_\varsigma(s) \equiv (1 - 4m_\varsigma^2/s)$.
It is seen that the annihilation cross-section is proportional to $v_\chi^2(s)$, i.e., suppressed by "p-wave" because of the same reason as above. On the contrary, when $C_s = 0$ and $C_p \neq 0$, the cross-section at the NR limit is obtained as 
\begin{align}
    \sigma_a v\,(\chi \chi \to \varsigma \varsigma) \simeq
    &
    \frac{C_p^2 v_\varsigma(s)}{32 \pi} \left(\frac{c_\theta C_{\varsigma\varsigma\varsigma}}{3m_\varsigma^2}+ \frac{s_\theta C_{\varsigma\varsigma h}}{m_h^2}\right)^2
    \nonumber \\
    &
    +\frac{C_p^2 v_\varsigma(s)\,v_\chi^2(s)}{128\pi}
    \left[
        \left(\frac{c_\theta C_{\varsigma\varsigma\varsigma}}{3m_\varsigma^2} + \frac{s_\theta C_{\varsigma\varsigma h}}{m_h^2}\right)^2
        +\frac{16 C_p^2 c_\theta^4 m_\chi^2 (m_\chi^2 - m_\varsigma^2)^2}{3(2m_\chi^2 - m_\varsigma^2)^4} \right].
\end{align}
In the first term of the right-hand side, the cross-section is not suppressed by "p-wave," as the co-existence of non-vanishing $C_p$ and $C_{\varsigma\varsigma\varsigma}$ (or $C_{\varsigma\varsigma h}$) breaks CP. On the other hand, as seen in the last term of the right-hand side, the contribution from diagrams depending only on $C_p$ is suppressed by $v_\chi^2(s)$, as only the use of the interaction with $C_p$ preserves the CP.

Finally, the light fermionic WIMP also scatters off an electron and a nucleon in the same manner as the scalar WIMP. Their scattering cross-sections are then obtained as follows:
\begin{align}
    & \sigma_s(\chi e \to \chi e) \simeq
    \frac{m_e^4 m_\chi^2 c_\theta^2 s_\theta^2 C_s^2}{\pi v_H^2 (m_\chi + m_e)^2}
    \left( \frac{1}{m_\varsigma^2}- \frac{1}{m_h^2} \right)^2,
    \label{eq: chi-e scattering}
    \\
    & \sigma_s(\chi N \to \chi N) \simeq
    \frac{f_N^2 m_N^4 m_\chi^2 c_\theta^2 s_\theta^2 C_s^2}{\pi v_H^2 (m_\chi + m_N)^2}
    \left( \frac{1}{m_\varsigma^2}- \frac{1}{m_h^2} \right)^2.
    \label{eq: chi-N scattering}
\end{align}
Note that the pseudo-scalar interaction (with the coupling $C_p$) does not contribute to the cross-sections, as corresponding amplitudes are suppressed by the incident WIMP velocity. The size of the above cross-sections is similar to those of the scalar WIMP discussed in the previous sub-subsection, which is proportional to (effective) Yukawa coupling squared of the electron (nucleon) and reduced mass squared between the WIMP and the particle.

\subsection{Models with a vector mediator}
\label{subsec: vector mediator}

Next, we consider models with a vector mediator particle. Assuming that the mediator particle originates in a gauge boson associated with a new U(1) gauge symmetry, which is spontaneously broken by the condensate of a scalar field $\Sigma$, the renormalizable Lagrangian describing interactions between the mediator and the SM particles is generally given by
\begin{align}
    {\cal L}_V = {\cal L}_{\rm SM}
    -\frac{1}{4} (V_{\mu \nu})^2
    +\left| (\partial_\mu + i g_V q_\Sigma V_\mu) \Sigma \right|^2
    -V(\Sigma, H)
    -\frac{\xi}{2} V_{\mu \nu}\,B^{\mu \nu}
    -g_V \sum_f q_f \bar{f} V_\mu \gamma^\mu f, 
    \label{eq: V}
\end{align}
where $V_{\mu\nu}$ is the field strength tensor of the new U(1)$_V$ gauge boson $V_\mu$, $g_V$ is its gauge coupling, $q_X$ is the U(1)$_V$ charge of a particle `$X$,' and $f$ represents an SM fermion, i.e., $f = Q$, $U$, $D$, $L$ and $E$ with those being the quark doublet, up-type quark singlet, down-type quark singlet, lepton doublet, and charged lepton singlet, respectively. The scalar interactions among $\Sigma$ and the SM Higgs double $H$ are gathered in $V(\Sigma, H)$, including interactions that induce the vacuum condensate of $\Sigma$. We assume in the above Lagrangian for simplicity that the Higgs field does not carry the U(1)$_V$ charge (i.e., $q_H = 0$) and the U(1)$_V$ gauge symmetry is flavor-blind (i.e., $q_x$ for the SM fermions do not depend on the generation). After the EW and U(1)$_Y$ symmetries are spontaneously broken down, with $H = (0, v_H + h)^T/\sqrt{2}$ and $\Sigma = (v_\Sigma + \zeta)/\sqrt{2}$, the quadratic terms of the neutral gauge bosons are obtained as
\begin{align}
    &{\cal L}_V \supset
    \frac{1}{2} (W^3_\mu,\,B_\mu,\,V_\mu)
    \left[ (\Box g^{\mu \nu} - \partial^\mu \partial^\nu)\,{\cal K} + {\cal M}\,g^{\mu \nu} \right]
    (W^3_\nu,\,B_\nu,\,V_\nu)^T,
    \nonumber \\
    & {\cal K} \equiv
    \begin{pmatrix} 1 & 0 & 0 \\ 0 & 1 & \xi \\ 0 & \xi & 1 \end{pmatrix},
    \qquad
    {\cal M} \equiv
    \begin{pmatrix} g^2 v_H^2/4 & -g g' v_H^2/4 & 0 \\ -g g' v_H^2/4 &  g^{\prime 2} v_H^2/4 & 0 \\ 0 & 0 & g_V^2 q_\Sigma^2 v_\Sigma^2 \end{pmatrix},
\end{align}
with $g$ and $g'$ being the gauge couplings of the SU(2)$_L$ and U(1)$_Y$ symmetries in the SM. The redefinition of the vector mediator and the SM gauge fields, as well as the diagonalization of the above mass matrix ${\cal M}$, give mass eigenstates having canonical kinetic terms,
\begin{align}
    &
    X {\cal K} X^T = {\bf 1},
    \qquad
    X {\cal M} X^T = {\rm diag}(m_Z^2,\,0,\,m_{Z'}^2),
    \qquad
    (Z_\mu,\,A_\mu,\,Z'_\mu)^T = (X^{-1})^T (W^3_\mu,\,B_\mu,\,V_\mu)^T,
    \nonumber \\
    & \qquad\qquad
    X \simeq
    \begin{pmatrix}
        \cos \theta_W & -\sin \theta_W & \displaystyle \xi \frac{\sin \theta_W m_Z^2}{m_Z^2 - m_{Z'}^2} \\
        \sin \theta_W & \cos \theta_W & 0 \\
        \displaystyle -\xi \frac{\cos \theta_W \sin \theta_W m_Z^2}{m_Z^2 - m_{Z'}^2} & \displaystyle \xi \frac{m_{Z'}^2 - \cos^2 \theta_W m_Z^2}{m_Z^2 - m_{Z'}^2} & 1
    \end{pmatrix},
\end{align}
assuming the parameter $\xi \ll 1$ with $m_{Z'}$ being the mass of the vector mediator particle $Z'$. With the mass eigenstates $Z_\mu$, $A_\mu$ and $Z'_\mu$ (i.e., the $Z$ boson, the photon, and the vector mediator particle, respectively), the Lagrangian\,(\ref{eq: V}) gives the following interactions:
\begin{align}
    {\cal L}_V \supset
    -\sum_f g_f \bar{f} Z'_\mu \gamma^\mu f + \cdots,
    \label{eq: V interactions}
\end{align}
where "$\cdots$" denotes other interactions among the mediator particle $Z'$, the physical mode in $\Sigma$ denoted by $\zeta$, and the SM particles. Note that some SM interactions are also affected by the new U(1)$_V$ gauge symmetry, such as the interactions between the $Z$ boson and the SM fermions and the self-interactions among the EW gauge bosons. However, those are highly suppressed by the tiny coupling $\xi$. We do not give their explicit forms because these interactions are irrelevant to the following discussion. On the other hand, the interactions given in eq.\,(\ref{eq: V interactions}) play an essential role in the discussion, where the coupling $g_f$ is given by
\begin{align}
    g_f \simeq g_V q_f
    -g' \xi\,Q_{f} \frac{c_W^2 m_Z^2 - m_{Z'}^2}{m_Z^2 - m_{Z'}^2}
    -g' \xi\,T_f\frac{m_{Z'}^2}{m_Z^2 - m_{Z'}^2},
\end{align}
at the leading order of $\xi$. Here, $Q_f$ is the electric charge and $T_f$ is the third component of the weak isospin, i.e., $T_f = 1/2\,(-1/2)$ for the upper\,(lower) components of the quark and lepton doublets, $Q$ and $L$, while $T_f = 0$ for the quark and lepton singlets, $U$, $D$ and $E$.

The charge $q_f$ is fixed once we concretely choose the U(1)$_V$ symmetry. The simplest choice is that U(1)$_V$ is irrelevant to the SM sector, i.e., $q_f = 0$ for any `f, ' which is called the dark photon scenario\,\cite{Fabbrichesi:2020wbt, Holdom:1985ag, Galison:1983pa}: The vector mediator interacts with the SM particles only through the kinetic mixing term proportional to $\xi$. Another well-known choice is that U(1)$_V$ is identified with the so-called U(1)$_{\rm B-L}$\,\cite{Mohapatra:1980qe, Buchmuller:1991ce}. This case is well-motivated because a right-handed neutrino is introduced at each generation to make the theory anomaly-free, which enables us to describe observed neutrino masses via the see-saw mechanism\,\cite{Yanagida:1979as, Gell-Mann:1979vob}. The charge assignment in this case is $q_Q = q_U = q_D= 1/3$ and $q_L = q_E = -1$. The other choice, giving a very different phenomenological prediction from the above cases, is that U(1)$_V$ is identified with U(1)$_{\rm B}$, where the charge assignment is $q_Q = q_U = q_D= 1/3$ and $q_L = q_E = 0$. This case calls for a few more fermions charged under the SM gauge symmetry (instead of the right-handed neutrinos) to make the theory anomaly-free\,\cite{Carone:1995pu, Duerr:2014wra, Feng:2016ysn, Duerr:2013lka, Duerr:2013dza, Duerr:2014wra}. A simple example of such a particle content is composed of fermions that are chiral for the U(1)$_{\rm B}$ symmetry, while vector-like for the SM gauge symmetry\,\cite{Duerr:2013lka, Duerr:2013dza, Duerr:2014wra}: $\psi_{L/R} \sim ({\bf 1}, {\bf 2}, -1/2, B_1/B_2)$, $\eta_{R/L} \sim ({\bf 1}, {\bf 1}, -1, B_1/B_2)$, and $\chi_{R/L} \sim ({\bf 1}, {\bf 1}, 0, B_1/B_2)$, where the numbers in each bracket are SU(3)$_c$, SU(2)$_L$, U(1)$_Y$ and U(1)$_{\rm B}$ charges, respectively, for each Weyl fermion. The cancellation of the $SU(2)_L \otimes U(1)_{\rm B}$ anomaly requires $B_1-B_2 = -3$. So, if the U(1)$_{\rm B}$ charge of the aforementioned scalar field $\Sigma$ is $-3$, these fermions acquire their masses through Yukawa interactions with $\Sigma$, such as $\bar{\psi}_L \Sigma \psi_R$. On the other hand, as seen in section\,\ref{sec: Accelerator Light}, the U(1)$_{\rm B}$ gauge coupling $g_B$ is severely constrained as $g_B \lesssim 10^{-3.5} (m_{Z'}/{\rm GeV})$, so the vacuum expectation value $v_\Sigma \sim m_{Z'}/g_B$ must be greater than several TeV, enabling the fermions to have heavy enough masses and to avoid constraints from high-energy collider experiments.\footnote{
    In this simple scenario, a $Z_2$ symmetry, i.e., $\psi_{L/R} \to -\psi_{L/R}$, $\eta_{L/R} \to -\eta_{L/R}$, and $\chi_{L/R} \to -\chi_{L/R}$, remains even after the U(1)$_{\rm B}$ and EW symmetry breaking, so the lightest particle among them may serve as an additional dark matter component contributing to the dark matter abundance observed today. Such an additional contribution can be avoided if we assume that a reheating temperature of the universe is low enough.}

\begin{figure}[t]
    \centering
    \includegraphics[keepaspectratio, scale=0.8]{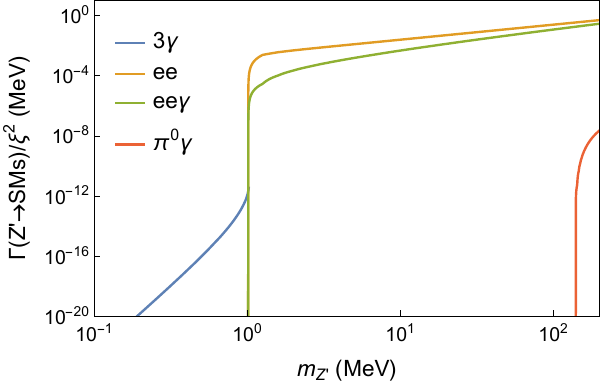}
    \caption{\small \sl The partial decay widths of $Z'$ into the SM particles in the dark photon scenario.}
    \label{fig: Z' to SMs}
\end{figure}

Let us consider the decay of the vector mediator particle in general. When its mass is smaller than twice the electron mass, i.e., $m_{Z'} \leq 2m_e$, it decays mainly into a pair of neutrinos and three photons. The partial decay width into a pair of neutrinos is
\begin{align}
    \Gamma(Z' \to \nu_i \bar{\nu}_i) = \frac{g_L^2 m_{Z'}}{24 \pi},
\end{align}
with $i$ being the generation index. So, the decay width of the mediator particle into all the neutrino generations is three times larger than the above. Next, we utilize the result given in Refs.\,\cite{Pospelov:2008jk, McDermott:2017qcg} for the partial decay width into three photons, which is obtained as
\begin{align}
    \Gamma(Z' \to \gamma \gamma \gamma) \simeq
    \frac{17\alpha^3 (g_L+g_E)^2}{2^{11} 3^6 5^3} \frac{m_{Z'}^9}{m_e^8},
\end{align}
when $m_{Z'} \ll 2 m_e$.\footnote{
    The result obtained by the exact one-loop calculation in Ref\,\cite{McDermott:2017qcg} is used in our analysis and also Fig.\,\ref{fig: Z' to SMs}.}
Here, we assumed that the vector coupling of the mediator particle with the electron ($Z'_\mu \bar{e} \gamma^\mu e$) dominates the axial-vector one ($Z'_\mu \bar{e} \gamma^\mu \gamma_5 e$) with $e$ being the electron field involving both the left- and right-handed chiralities. This assumption is validated in the cases mentioned above, i.e., U(1)$_V$ $=$ U(1)$_{\rm B - L}$ and U(1)$_{\rm B}$, when $m_{Z'} \ll m_Z$. Furthermore, the mediator particle decays into $e^- e^+$ and $e^- e^+ \gamma$ when $m_{Z'} \geq 2 m_e$, as in the scalar mediator case in section\,\ref{subsec: scalar mediator}. The partial decay width into  a pair of electrons is obtained as
{\small
\begin{align}
    &\Gamma(Z' \to e^- e^+) = \frac{m_{Z'}}{48 \pi}
    \left[
        (g_L + g_E)^2 \left(1 + \frac{2 m_e^2}{m_{Z'}^2}\right)
        \left(1 - \frac{4 m_e^2}{m_{Z'}^2}\right)^{1/2}
        + (g_L - g_E)^2 \left(1 - \frac{4 m_e^2}{m_{Z'}^2}\right)^{3/2}
    \right],
    \label{eq: Z'-SM coupling}
\end{align}
}\noindent
The differential partial decay width into a pair of electrons associated with a photon is
{\small
\begin{align}
    &
    \frac{d\Gamma(Z' \to e^- e^+ \gamma)}{dE} =
    \frac{2 \alpha}{\pi m_{Z'}} \Gamma(Z' \to e^- e^+)
    \times {\rm FSRV}(2E/m_{Z'}, m_e/m_{Z'}),
    \label{eq: Z' to eegamma}
    \\
    &
    {\rm FSRV}(x, \mu) =
    \frac{1 + (1 - x)^2 - 4\mu^2 (x + 2 \mu^2)}{x\,\sqrt{1 - 4\mu^2}\,(1+2\mu^2)}
    \log\left[\frac{1 + v_\mu(x)}{1 - v_\mu(x)}\right]
    -\frac{1 + (1 - x)^2 + 4\mu^2 (1 - x)}{x\,\sqrt{1 - 4\mu^2}\,(1 + 2\mu^2)}
    v_\mu(x),
    \nonumber
\end{align}
}\noindent
with $v_\mu(x) \equiv [1 - 4 \mu^2/(1 - x)]^{1/2}$. The partial decay width of the process, $Z' \to e^- e^+ \gamma$, is obtained by integrating the above differential width over an appropriate period of the energy $E$; we compute the width $\Gamma(Z' \to e^- e^+ \gamma)$ in the same manner adopted in section\,\ref{subsec: scalar mediator}, i.e., it is defined as the one with a photon having $E \geq 10^{-3} m_e$. On the other hand, unlike the scalar mediator case, the vector mediator particle decays into a neutral pion and a photon when $m_{Z'} \geq m_{\pi^0}$ with $m_{\pi^0}$ being the neutral pion mass\,\cite{Coogan:2019qpu, Coogan:2021sjs}. Its partial decay width is
\begin{align}
    \Gamma(Z' \to \pi^0 \gamma) =
    \frac{\alpha\,(3g_Q + 2g_U + g_D)^2 (m_{Z'}^2 - m_{\pi^0}^2)^3}{6144\,\pi^4 f_\pi^2 m_{Z'}^3}.
\label{eq: decay pi0gamma}
\end{align}
Here, we again assumed that the vector couplings of the mediator particle with the SM quarks dominate those of the axial-vector couplings, as for the decay width into $3\gamma$.

The partial decay widths of the vector mediator particle discussed above, i.e., $Z' \to \gamma\gamma\gamma$, $e^- e^+$, $e^- e^+ \gamma$, and $\pi^0 \gamma$, are shown in Fig.\,\ref{fig: Z' to SMs}, assuming the dark photon scenario. The partial decay width into 3$\gamma$ is only shown at $m_{Z'} \leq 2 m_e$, as it is much smaller than other decay channels above the threshold, $m_{Z'} \geq 2 m_e$. The decay width into a neutrino pair is added when we consider the U(1)$_{\rm B-L}$ with a large B-L coupling, which should be in the same order as that into $e^- e^+$. When considering the U(1)$_{\rm B}$ scenario, the decay width into $\pi^0 \gamma$ alters respective to that in the dark photon scenario, depending on the U(1)$_{\rm B}$ coupling.

\subsubsection{Light scalar WIMP with the vector mediator}
\label{subsubsec: Scalar WIMP and Vector Mediator}

We introduce a light scalar WIMP to the system discussed above, i.e., the one with the SM particles and the vector mediator particle. Since the WIMP should be charged under this new U(1)$_V$ symmetry to have an interaction with the SM particles, i.e., the WIMP should carry the U(1)$_V$ charge, we take the WIMP as a complex scalar, which is singlet under the SM gauge symmetry and odd under the $Z_2$ symmetry, as discussed at the beginning of this section. The most general and renormalizable Lagrangian is obtained as follows:
\begin{align}
    {\cal L}_\varphi = {\cal L}_V
    +|(\partial_\mu + i g_V q_\varphi V_\mu)\,\varphi|^2
    -M_\varphi^2 |\varphi|^2
    -\lambda_{H \varphi} |H|^2 |\varphi|^2
    -\lambda_{\Sigma \varphi} |\Sigma|^2 |\varphi|^2
    -\frac{1}{4}\lambda_{\varphi} |\varphi|^{4},
\end{align}
where ${\cal L}_V$ is the Lagrangian in eq.\,(\ref{eq: V}), and $\varphi$ is the field describing the light scalar WIMP. In addition to the interactions shown above, there is an additional one, $\varphi^2\Sigma + h.c.$, when $q_\Sigma + 2 q_\varphi = 0$ is satisfied, or $\varphi^2 \Sigma^2 + h.c.$, when $q_\Sigma + q_\varphi = 0$ is satisfied. After the U(1)$_V$ symmetry is broken, these interactions make the complex scalar field $\varphi$ split into two real scalar fields as mass eigenstates; it leads to the so-called in-elastic dark matter scenario, giving viable models for the sub-GeV scale WIMP\,\cite{Izaguirre:2014dua, Brahma:2023psr, Tucker-Smith:2001myb, Finkbeiner:2007kk, Arkani-Hamed:2008hhe, Cheung:2009qd, Chen:2009ab, Batell:2009vb, Graham:2010ca}. We no longer consider such a scenario in this article because we focus on the other type of scenario for the light WIMP using velocity-dependent annihilation, as mentioned in the introductory section. After the breaking of the electroweak and U(1)$_V$ symmetries, the interactions involving $\varphi$ are obtained as
\begin{align}
    {\cal L}_\varphi \, \supset \,
    -i g_\varphi Z^{\prime \mu} (\varphi^* \overleftrightarrow{\partial_\mu} \varphi)
    +g_\varphi^2 Z'_\mu Z^{\prime \mu} |\varphi|^2
    -\frac{1}{4}\lambda_{\varphi} |\varphi|^{4}
    +\cdots,
    \label{eq: Z'-varphi interactions}
\end{align}
where $g_\varphi = g_V q_\varphi$. Here, we implicitly assumed that the physical mode of the scalar field $\Sigma$ and the electroweak bosons (i.e., the Higgs and $Z$ bosons) are so heavy that these bosonic fields do not play significant roles in the physics that we will discuss in the followings.

The light scalar WIMP with the vector mediator particle also annihilates into the SM particles in the same manner as in Fig.\,\ref{fig: diagrams S}. For instance, the WIMP annihilation cross-section into a $e^- e^+$ pair through the $s$-channel exchange of the mediator particle is obtained as
\begin{align}
    \sigma_a v\,(\varphi \varphi^* \to e^- e^+) =
    \frac{g_\varphi^2\,(g_L + g_E)^2}{24\pi}
    v_e(s)\,v_\varphi^2(s)\,(s+2m_e^2)
    \left|
        \frac{1}{s - m_{Z'}^2 + i\,s^{1/2}\,\Gamma_{Z'; \varphi}(s)}
    \right|^2,
    \label{eq: varphi varphi to ee}
\end{align}
where $v_\varphi(s) \equiv (1 - 4 m_\varphi^2/s)^{1/2}$, with $m_\varphi = (M_\varphi^2 + \lambda_{H\varphi} v_H^2/2 + \lambda_{\Sigma \varphi} v_\Sigma^2/2)^{1/2}$ being the physical mass of the scalar WIMP $\varphi$. The imaginary part of the propagator of the mediator particle in the cross-section is defined as $\Gamma_{Z'; \varphi}(s) \equiv [\Gamma\,(Z' \to \varphi \varphi^*) + \Gamma\,(Z' \to {\rm SMs})]_{m_{Z'}^2 \to s}$. Here, the second term of the right-hand-side, $\Gamma\,(Z' \to {\rm SMs})$, is the (total) decay width of $Z'$ into the SM particles, while the first term, i.e., the decay width into a pair of the WIMPs, is
\begin{align}
    \Gamma(Z' \to \varphi \varphi^*) =
    \frac{g_\varphi^2 m_{Z'}}{48\pi}  v_\varphi^3(s)
\end{align}
The above annihilation cross-section is proportional to $v_\varphi^2(s)$, i.e., it is suppressed by "p-wave" because of CP and angular momentum conservation between the initial state (WIMP pair) and the intermediate state (mediator particle $Z'$). As deduced from the $Z'$ decays in section\,\ref{subsec: vector mediator}, the WIMP also annihilates into $e^- e^+ \gamma$, a pair of neutrinos, three photons, and $\pi^0 \gamma$. The last two processes, i.e., those into a pair of neutrinos and three photons, dominate the annihilation when $m_\varphi < 2 m_e$, while that into $\pi^0 \gamma$ is kinematically allowed when $m_\varphi > m_{\pi^0}$. These processes are also suppressed by "p-wave" due to the conservation above. Note that the processes, $\varphi \varphi^* \to \gamma \gamma \gamma$, $e^- e^+ \gamma$ and $\pi^0 \gamma$ produce a hard MeV gamma-ray signal against backgrounds in the indirect dark matter detection, as seen in Fig.\,\ref{fig: dN/dE}.

The WIMP also annihilates into a pair of mediator particles $Z'$, followed by the $Z'$ decay into the SM particles. As for the models with the scalar mediator particle discussed in section\,\ref{subsec: scalar mediator}, this process dominates the WIMP annihilation when $m_\varphi > m_{Z'}$ because the coupling of the mediator particle to the WIMP is not suppressed while those to the SM particles are, as seen in the next sections. Its annihilation cross-section is obtained as\footnote{
    We do not have the $s$-channel diagram (the third one in Fig.\,\ref{fig: diagrams S}) due to the U(1) nature of the $Z'$ particle.}
\begin{align}
    \sigma_a v (\varphi \varphi^* \to Z' Z') \simeq
    \frac{4 g_\varphi^4 m_{Z'}\,
    (3 m_{Z'}^4 - 8 m_{Z'}^2 m_\varphi^2 + 8 m_\varphi^4)}{128 \pi m_\varphi^3\,(2m_\varphi^2 - m_{Z`}^2)^2}
    v_{Z'}(s),
\end{align}
with $v_{Z'}(s) \equiv (1 - 4 m_{Z'}^2/s)^{1/2}$. Here, we take the non-relativistic limit regarding the incident WIMP particles, though we use the annihilation cross-section without taking this limit in our numerical analysis. It is seen that the above cross-section is not suppressed by "p-wave," unlike the case of eq.\,(\ref{eq: varphi varphi to ee}). Note also that this annihilation process produces a distinctive MeV gamma-ray signal when (one of) the mediator particles in the final state decay(s) into $\gamma \gamma \gamma$ and $\pi^0 \gamma$. Moreover, when we consider the case $m_\varphi \simeq m_{Z'}$, the process also gives a distinctive signal even if $Z'$ decays into $e^- e^+ \gamma$, as $Z'$ are produced almost at rest.

The light WIMP scatters off an electron and a nucleon by exchanging $Z'$, as in the models with a scalar mediator particle. Their scattering cross-sections are obtained as follows:
\begin{align}
    & \sigma_s(\varphi e \to \varphi e) \simeq
    \frac{m_e^2 m_\varphi^2\,g_\varphi^2 (g_L + g_E)^2}
    {4\pi m_{Z'}^4 (m_\varphi + m_e)^2},
    \label{eq: varphi-e scattering}
    \\
    & \sigma_s(\varphi N \to \varphi N) \simeq
    \frac{m_N^2 m_\varphi^2\,g_\varphi^2 g_N^2}
    {4\pi m_{Z'}^4 (m_\varphi + m_N)^2},
    \label{eq: varphi-N scattering}
\end{align}
with $g_N \equiv 3g_Q + 2g_U + g_D$ and $3g_Q + g_U + 2g_D$ for the cases that $N$ is a proton and a neutron, respectively. Comparing to those given in eqs.\,(\ref{eq: phi-e scattering}) and (\ref{eq: phi-N scattering}) assuming $C_{\varsigma \phi \phi} = {\cal O}(m_\phi)$, it is found that the above cross-sections are not suppressed by the factors $m_e^2/v_H^2$ and $f_N^2m_N^2/v_H^2$. So, the signal of the scalar dark matter $\varphi$ (with the vector mediator $Z'$) at the direct dark matter detection is expected to be stronger than that of the scalar dark matter $\phi$ (with the scalar mediator $\varsigma$). However, the couplings $g_\varphi$, $g_L$, and $g_E$ are required to be very small to satisfy the relic abundance condition and the constraint from accelerator experiments in the parameter region we focus on, as seen in the next sections. As a result, the strength of the signal becomes below the sensitivity of the present direct detection experiments.

\subsubsection{Light fermion WIMP with the vector mediator}
\label{subsubsec: FV}

Next, we consider a light fermionic WIMP and introduce it to the system discussed in this subsection. The WIMP should carry the U(1)$_V$ charge, as in the case of section\,\ref{subsubsec: Scalar WIMP and Vector Mediator}, so we take the WIMP as a Dirac fermion, which is singlet under the SM gauge symmetry and odd under the $Z_2$ symmetry. The most general and renormalizable Lagrangian of the model is
\begin{align}
    {\cal L}_\psi = {\cal L}_V
    +\bar{\psi} (i\slashed{\partial} - g_V q_\psi \slashed{V} - m_\psi)\,\psi,
\end{align}
where ${\cal L}_V$ is the Lagrangian defined in eq.\,(\ref{eq: V}), and $\psi$ is the field describing the light fermionic WIMP. There is an additional interaction $\overline{\psi^C} \psi\,\Sigma + h.c.$ when $q_\Sigma + 2 q_\psi = 0$ is satisfied, with $\Sigma$ being the scalar field spontaneously breaking U(1)$_V$ defined in eq.\,(\ref{eq: V}). The interaction makes the Dirac fermion field $\psi$ split into two Majorana fields as mass eigenstates, leading to the in-elastic dark matter scenario. We no longer consider such a scenario for the same reason as the aforementioned scalar WIMP. After the electroweak and U(1)$_V$ symmetries are broken, the interactions of the fermionic WIMP are obtained as
\begin{align}
    {\cal L}_\psi \, \supset \,
    -g_\psi Z^\prime_\mu \bar{\psi} \gamma^\mu \psi 
    +\cdots,
    \label{eq: Z'-psi interactions}
\end{align}
where $g_\psi = g_V q_\psi$. We implicitly assumed here that the physical mode of the scalar field $\Sigma$ and the electroweak bosons (i.e., the Higgs and $Z$ bosons) are heavy, so these bosonic fields do not play significant roles in the physics that we will discuss in the followings.

The fermionic WIMP with the vector mediator particle annihilates into the SM particles again in the same manner as in Fig.\,\ref{fig: diagrams S}. The WIMP annihilates into a $e^- e^+$ pair through the $s$-channel exchange of the vector mediator particle, and its cross-section is given by
{\small
\begin{align}
    \sigma_a v (\psi \bar{\psi} \to e^- e^+) =
    \frac{g_\psi^2\,(g_L + g_E)^2}{24\pi s}\,(s+2m_e^2)\,(s+2m_\psi^2)
    \left|
        \frac{1}{s - m_{Z'}^2 + i\,s^{1/2}\,\Gamma_{Z'; \psi}(s)}
    \right|^2
    v_e(s).
\end{align}
}
The imaginary part of the $Z'$ propagator in the above cross-section is defined by $\Gamma_{Z'; \psi}(s) \equiv [\Gamma\,(Z' \to \psi \bar{\psi}) + \Gamma\,(Z' \to {\rm SMs})]_{m_{Z'}^2 \to s}$, where the partial decay width into a WIMP pair is
\begin{align}
    \Gamma\,(Z' \to \psi \bar{\psi}) =
    \frac{g_\psi^2\,(m_{Z'}^2 + 2m_\psi^2)}{12\pi m_{Z'}}
    v_\psi(m_{Z'}),
\end{align}
with $v_\psi \equiv (1 - 4 m_\psi^2/s)^{1/2}$. It is found that, unlike the case in eq.\,(\ref{eq: varphi varphi to ee}), the annihilation cross-section is not suppressed by "p-wave." On the other hand, as in the previous case in section\,\ref{subsubsec: Scalar WIMP and Vector Mediator}, the fermionic WIMP also annihilates into $\nu\bar{\nu}$, $e^- e^+ \gamma$, $\gamma \gamma \gamma$, and $\pi^0 \gamma$, by exchanging $Z'$ in the $s$-channel. Again, all the processes are not suppressed by "p-wave." The last three processes give hard MeV gamma-ray signals at indirect dark matter detection.

The light fermionic WIMP also annihilates into a pair of vector mediator particles, followed by the $Z'$ decay into the SM particles. This process dominates the annihilation of this fermionic WIMP when $m_\psi > m_{Z'}$. Its annihilation cross-section is obtained as follows:\footnote{
    No $s$-channel and contact diagrams in Fig.\,\ref{fig: diagrams S} exist due to the U(1) and fermionic natures of $Z'$ and $\psi$.}
\begin{align}
    \sigma_a v (\psi \bar{\psi} \to Z' Z') \simeq
    \frac{g_\psi^4\,m_{Z'}\,(m_\psi^2-m_{Z'}^2)}
    {8\pi m_\psi\,(2m_\psi^2 - m_{Z'}^2)^2} 
    v_{Z'}(s).
    \label{eq: FS II annihilation}
\end{align}
Here, we take the non-relativistic limit regarding the incident WIMP particles to simplify the expression. On the other hand, we use the annihilation cross-section without taking this limit in our numerical analysis. The above cross-section is not suppressed by "p-wave" and gives a distinctive MeV gamma-ray signal when (one of) the $Z'$s in the final state decay(s) into $\gamma \gamma \gamma$ and $\pi^0 \gamma$. Furthermore, when $m_\psi \simeq m_{Z'}$, it also gives a distinctive signal even if $Z'$ decays into $e^- e^+ \gamma$, though the cross-section is suppressed by the factor $(m_\psi^2 - m_{Z'}^2)$.

The light fermionic WIMP scatters off an electron or a nucleon by exchanging $Z'$, as for the light scalar WIMP in section\,\ref{subsubsec: Scalar WIMP and Vector Mediator}. Their scattering cross-sections are obtained as
\begin{align}
    & \sigma_s(\psi e \to \psi e) \simeq\frac{m_e^2 m_\psi^2\,g_\psi^2 (g_L + g_E)^2}
    {4\pi m_{Z'}^4 (m_\psi + m_e)^2}
    ,
    \label{eq: psi-e scattering}
    \\
    & \sigma_s(\psi N \to \psi N) \simeq \frac{m_N^2 m_\psi^2\,g_\psi^2 g_N^2}
    {4\pi m_{Z'}^4 (m_\psi + m_N)^2}
    .
    \label{eq: psi-N scattering}
\end{align}
It is found that the above scattering cross-sections are the same as those in eqs.\,(\ref{eq: varphi-e scattering}) and (\ref{eq: varphi-N scattering}) for the light scalar WIMP, because of the conservation of the vector currents.

\subsection{Light WIMP models considered in this article}
\label{subsec: light WIMP models}

Among the light WIMP models discussed so far, we focus on the models summarized in table\,\ref{tab: light WIMP models} in this article. As mentioned at the beginning of this section, we consider models with the scalar mediator particle $\varsigma$ with the light WIMP being a real scalar $\phi$ (sections\,\ref{subsub: SS}) or a Majorana fermion $\chi$ (sections\,\ref{subsubsec: FS}), where those are called the {\bf SS} and {\bf FS} models, respectively, in this article. With the light WIMP being a complex scalar $\varphi$ (section\,\ref{subsubsec: Scalar WIMP and Vector Mediator}) or a Dirac fermion $\psi$ (section\,\ref{subsubsec: FV}), we also consider models with the vector mediator particle originating in the U(1)$_{\rm B-L}$ symmetry, assuming the physical mode $\zeta$ in the scalar field $\Sigma$ is heavier enough than the mediator particle and $\xi \gg g_V$, i.e., the vector mediator particle is almost the dark photon. Those are called the {\bf SV} and {\bf FV} models. The reason why we do not consider the models with the "exact" dark photon mediator particle (i.e., $g_V = 0$) is that the existence of the tiny gauge coupling $g_V$ enables to drastically relax the cosmological constraint on the lower limit of the WIMP mass as discussed in section\,\ref{sec: Cosmology} while not affecting other physics of the light WIMP. Furthermore, we also consider models with the vector mediator particle originating in the U(1)$_{\rm B}$ gauge symmetry, with the light WIMP being again a complex scalar $\varphi$ or a Dirac fermion $\psi$. As discussed in the following, these U(1)$_{\rm B}$ models are treated as extra; they predict a very different MeV gamma-ray signal than those of the dark photon models. The models are called the {\bf SV(B)} and {\bf FV(B)} models, assuming additional fermions making the theory anomaly-free are also very heavy.

\begin{table}[t]
    \centering
    \begin{tabular}{r|ccc}
        & Scalar mediator & "Dark photon" mediator & U(1)$_{\rm B}$ mediator \\
        \hline
        Scalar WIMP & {\bf SS} & {\bf SV} & {\bf SV(B)} \\
        Fermionic WIMP & {\bf FS} & {\bf FV} & {\bf FV(B)} \\
        \hline
    \end{tabular}
    \caption{\sl\small The light WIMP models we consider in this article: Those are named {\bf SS}, {\bf FS}, {\bf SV} models, etc., according to the properties of WIMP and mediator particle. See the text for more details.}
    \label{tab: light WIMP models}
\end{table}

\section{Conditions and Constraints on the Light WIMPs}
\label{sec: conditions and constraints}

The light WIMPs discussed in the previous section should satisfy several conditions and constraints to be viable dark matter candidates, which is very different from traditional WIMPs with an electroweak-scale mass. In this section, we discuss the conditions and constraints obtained from cosmological observations, i.e., those concerning the CMB and BBN observations, and from new particle detections, i.e., those from experiments searching for new particles at accelerators, underground laboratories, and astrophysical observations.

\subsection{Cosmology of the light WIMPs}
\label{sec: Cosmology}

The light WIMP, whose mass is smaller than 100\,MeV, is expected to be in equilibrium with the SM particles at or close to the epoch of the neutrino decoupling and Big Bang Nucleosynthesis\,(BBN). Furthermore, though the WIMP decouples from the SM bath after these epochs, it keeps annihilating and thus injecting energy into the bath. So, the WIMP affects the physics explored by the CMB and BBN observations, and it constrains the WIMP models in the previous section because we have not observed such effects so far.

\subsubsection{Limit on the WIMP annihilation from the CMB observation}
\label{subsubsec: the scenrios}

The most striking constraint for the light WIMP is from the CMB observation: The light WIMP keeps injecting electromagnetically interacting particles (electrons, positrons, and photons) into the SM bath during the recombination epoch. It changes the thermal history of the recombination by increasing the residual ionization fraction, and the CMB anisotropy is modified\,\cite{Slatyer:2015jla, Kawasaki:2021etm, Lopez-Honorez:2013cua}. Because the precise CMB observation prefers the anisotropy without the injection, the WIMP annihilation cross-section at the CMB epoch is constrained as
\begin{align}
\label{eq: CMB limit on annihilation}
    f_{\rm eff\,}(m_{\rm DM})\,
    \frac{\langle \sigma_a v \rangle_{\rm CMB}}{m_{\rm DM}} \leq
    4.1 \times 10^{-28}\,[{\rm cm}^3\,{\rm s}^{-1}\,{\rm GeV}^{-1}],
\end{align}
at 95\,\%\,C.L. by the PLANCK collaboration\,\cite{Planck:2018vyg}. Here, $m_{\rm DM}$ is the WIMP mass, and $\langle \sigma_a v \rangle_{\rm CMB}$ is the WIMP annihilation cross-section averaged by the velocity distribution of the WIMP at the recombination epoch. On the other hand, $f_{\rm eff\,}(m_{\rm DM})$ is the efficiency of the deposited energy by the annihilation to be injected into the bath. We used the prescription given in Ref.\,\cite{Slatyer:2015jla} to compute the efficiency $f_{\rm eff\,}(m_{\rm DM})$. The annihilation of the WIMP into $e^- e^+$ is the dominant process contributing to this efficiency in the models discussed in this article.

When the WIMP annihilates dominantly into electromagnetically interacting particles, the annihilation cross-section $\langle \sigma_a v \rangle_{\rm CMB}$ is required to be smaller than 10$^{-30}$--10$^{-28}$\,cm$^3$/s when $m_{\rm DM} \simeq$ 1--100\,MeV. It is much smaller than the canonical value of the annihilation cross-section for the correct relic abundance, $\langle \sigma_a v \rangle_{\rm FO} \sim$ 10$^{-25}$--10$^{-26}$\,cm$^3$/s, with $\langle \sigma_a v \rangle_{\rm FO}$ being the cross-section averaged by the velocity distribution of the WIMP at the freeze-out epoch. So, a naive scenario that the WIMP annihilates into electromagnetically interacting particles with a constant cross-section over the entire region of a non-relativistic WIMP does not work. The following scenarios are considered to overcome this difficulty:
\begin{itemize}
    \item WIMP dominantly annihilates into harmless particles, such as neutrinos\,\cite{Asai:2020qlp, Kelly:2019wow}.
    \item {\bf WIMP annihilates velocity-dependently in the non-relativistic region\,\cite{Feng:2017drg, Bernreuther:2020koj, Binder:2022pmf, Brahma:2023psr, Boehm:2002yz, Boehm:2003hm, DAgnolo:2015ujb}.}
    \item WIMP abundance is determined by a process different from annihilation\,\cite{Izaguirre:2014dua, Hochberg:2014dra, Hochberg:2014kqa, Kaplan:2009ag, Edsjo:1997bg}.
    \item Annihilation determines abundance, but in a non-standard cosmology\,\cite{Davoudiasl:2015vba,Dutra:2018gmv, Compagnin:2022elr}.
\end{itemize}
We consider the second scenario in this article. Thanks to the existence of the mediator particle, a non-trivial velocity dependence can be realized in the WIMP models discussed in the previous section.\footnote{
    The 1st and 3rd scenarios are also discovered in the models by choosing model parameters appropriately.}
Such a velocity dependence appears in the following regions:
\begin{itemize}
    \item The WIMP annihilation cross-section is suppressed by "p-wave."
    \item The WIMP annihilation proceeds via the $s$-channel resonance of the mediator particle.
    \item The WIMP annihilation proceeds via the forbidden channel into the mediator particles.
\end{itemize}

\begin{figure}[t]
    \centering
    \includegraphics[keepaspectratio, scale=0.41]{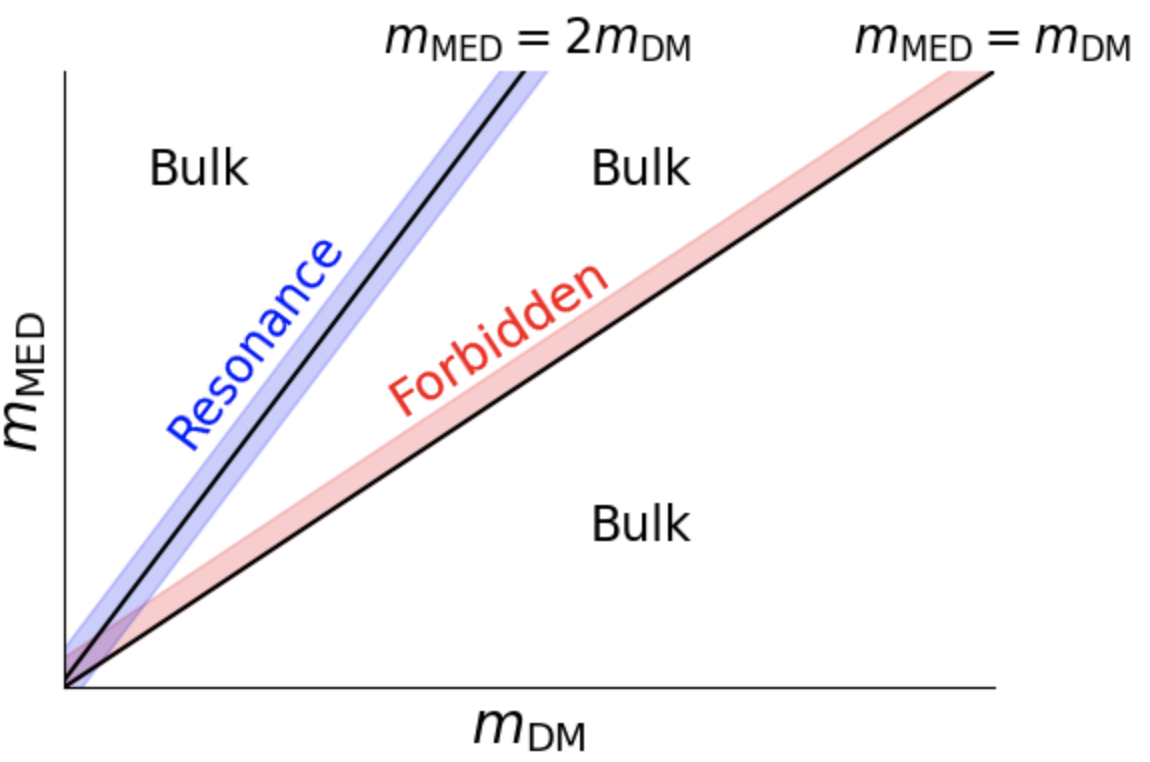}
    \quad
    \caption{\small \sl Classifying the parameter region of the light WIMP models on the $(m_{\rm DM}, m_{\rm MED})$-plane. Each region can predict a velocity-dependent annihilation cross-section. See text for more details.}
    \label{fig: v-dependent regions}
\end{figure}

We also visualize the above parameter regions in Fig.\,\ref{fig: v-dependent regions} with $m_{\rm DM}$ and $m_{\rm MED}$ being the WIMP and mediator particle masses: The first one (p-wave) is found in the bulk parameter region when the WIMP and the mediator are a fermion and a scalar, or a scalar and a vector, respectively. The second one (Resonance) is realized when the mass of the mediator particle is about twice the WIMP mass, and the third one (Forbidden) is the case where the mediator particle is slightly heavier than the WIMP. Therefore, we consider the above parameter regions in each model discussed in section\,\ref{subsec: light WIMP models}, as summarized in table\,\ref{tab: light WIMP scenarios}.

In the {\bf SS} model, the WIMP dominantly annihilates into the SM particles in "s-wave" in all the parameter regions, as discussed in section\,\ref{subsub: SS}. So, no bulk ("p-wave") region exists. In addition, in the resonance region ($m_{\rm MED} \simeq 2 m_{\rm DM}$) of this model, we impose an additional condition, $m_{\rm MED} > 2 m_{\rm DM}$. This is because, in the resonance region with $m_{\rm MED} < 2 m_{\rm DM}$, the desired velocity-dependence of the annihilation cross-section, i.e., $\langle \sigma_a v \rangle_{\rm CMB} < \langle \sigma_a v \rangle_{\rm FO}$, is not obtained. The forbidden region is simply defined by ($m_{\rm MED} \simeq m_{\rm DM})\wedge(m_{\rm MED} > m_{\rm DM}$). 

In the {\bf FS} model, assuming that the CP symmetry is not violated in the dark sector, we consider the parameter region with $C_p =0$, i.e., the region where the pseudo-scalar coupling between the WIMP and the mediator particle vanishes.\footnote{
    It is also possible to discuss the model with both $C_s$ and $C_p$ in eq.\,(\ref{eq: S-chi lagrangian}) being nonzero\,\cite{Chen:2024njd}. The model's prediction is expected to be between the SS and FS ($C_p = 0$) models. We leave this study for future work.}
In such a case, the WIMP annihilates into the SM particles in "p-wave" in all the parameter regions, as discussed in section\,\ref{subsubsec: FS}. Then, in the resonance region ($m_{\rm MED} \simeq 2 m_{\rm DM}$), we do not have to impose the additional condition used in the {\bf SS} model, as $\langle \sigma_a v \rangle_{\rm CMB}$ can be smaller than $\langle \sigma_a v \rangle_{\rm FO}$ due to the "p-wave" nature even at the resonance region with $m_{\rm MED} < 2 m_{\rm DM}$. The forbidden region in this model is defined in the same way as that in the {\bf SS} model, $(m_{\rm MED} \simeq m_{\rm DM})\wedge(m_{\rm MED} > m_{\rm DM})$. The bulk region is defined by neither the resonance nor the forbidden regions.

In the {\bf SV} model, the WIMP annihilates in "p-wave" when $m_{\rm DM} < m_{\rm MED}$, while in "s-wave" when $m_{\rm DM} > m_{\rm MED}$, as discussed in section\,\ref{subsubsec: Scalar WIMP and Vector Mediator}. So, we do not have to impose any further constraints in the resonance region ($m_{\rm MED} \simeq 2 m_{\rm DM}$), as in the {\bf FS} model. The forbidden region is defined in the same way as other models, $(m_{\rm MED} \simeq m_{\rm DM})\wedge(m_{\rm MED} > m_{\rm DM})$. In contrast, the bulk region is defined as $(m_{\rm MED} \slashed{\simeq} 2 m_{\rm DM})\wedge(m_{\rm MED} \slashed{\simeq} m_{\rm DM})\wedge(m_{\rm MED} > m_{\rm DM})$.

In the ${\bf FV}$ model, the WIMP annihilates in "s-wave" in all the parameter regions, as discussed in section\,\ref{subsubsec: FV}. So, no "p-wave" region exists, as in the ${\bf SS}$ model. We also impose an additional condition, $m_{\rm MED} > 2 m_{\rm DM}$, in the resonance region ($m_{\rm MED} \simeq 2 m_{\rm DM}$). The forbidden parameter region is defined in the same way as others, $(m_{\rm MED} \simeq m_{\rm DM})\wedge(m_{\rm MED} > m_{\rm DM})$.

In the {\bf SV(B)} model, which is considered as an extra model as addressed in section\,\ref{subsec: light WIMP models}, we only consider a limited parameter region, namely its resonance region. The region is defined by $m_{\rm MED} \simeq 2 m_{\rm DM}$, as the WIMP annihilates into the SM particles in "p-wave."

We again consider the resonance region in the {\bf FV(B)} model. Since the WIMP annihilates into the SM particle in "s-wave," the region is defined as $(m_{\rm MED} \simeq 2 m_{\rm DM})\wedge(m_{\rm MED} > 2 m_{\rm DM})$.

We consider all the parameter regions discussed above and call those the light WIMP scenarios in this article, as summarized in table\,\ref{tab: light WIMP scenarios}. In our analysis, we impose the CMB constraint on $\langle \sigma_a v \rangle_{\rm CMB}$ given in the eq.\,(\ref{eq: CMB limit on annihilation}) on the parameter space of each scenario.

\begin{table}[t]
    \centering
    \begin{tabular}{r|ccc}
        & Bulk ("p-wave") region & Resonance region & Forbidden region \\
        \hline
        {\bf SS} model & {\footnotesize N/A} & {\bf SS-R} & {\bf SS-F} \\
        {\bf FS} {\sl "}\qquad\,~ & {\bf FS-B} & {\bf FS-R} & {\bf FS-F} \\
        {\bf SV} {\sl "}\qquad\,~ & {\bf SV-B} & {\bf SV-R} & {\bf SV-F} \\
        {\bf FV} {\sl "}\qquad\,~ & {\footnotesize N/A} & {\bf FV-R} & {\bf FV-F} \\
        {\bf SV(B)} {\sl "}\qquad\,~ & -- & {\bf SV(B)-R} & -- \\
        {\bf FV(B)} {\sl "}\qquad\,~ & -- & {\bf FV(B)-R} & -- \\
        \hline
    \end{tabular}
    \caption{\sl\small The light WIMP scenarios we consider in this article: Those are named {\bf SS-F}, {\bf SV-R} scenarios, etc., depending on the WIMP model and the parameter region. See the text for more details.}
    \label{tab: light WIMP scenarios}
\end{table}

\subsubsection{Limit on the WIMP mass from the CMB observation}
\label{subsubsec: mass CMB}

The existence of new light particles may affect the physics of the neutrino decoupling: When the new particles' masses are less than or comparable to the decoupling temperature\,($T_D \sim 2$\,MeV), they eventually change the expansion rate of the universe in the recombination epoch and modifies the CMB anisotropy spectrum\,\cite{Dolgov:2002wy, Ibe:2018juk}. The new particles affect the physics in two ways: First, the entropy that is initially carried by the new particles is injected into the electromagnetic plasma (neutrinos) after the new particles become non-relativistic, making the photon (neutrino) temperature larger and the expansion rate of the universe smaller (larger). Next, when the new particles couple to both the electromagnetic sector (photons, electrons, and positrons) and neutrinos, it lowers the decoupling temperature because the temperature is determined not by the weak interaction but by the new one that the new particles mediate. The precise CMB observation supports the standard expansion without such effects, leading to the lower limits on the new particles' masses\,\cite{Giovanetti:2021izc, Sabti:2021reh, Chu:2022xuh}.

The limit is given in terms of the effective number of relativistic degrees of freedom:
\begin{align}
    N_{\rm eff} =
    3 \left[ \frac{11}{4}\left( \frac{T_\nu}{T_\gamma}\right)^3 \right]^{4/3},
\end{align}
where $T_\gamma$ and $T_\nu$ are the temperatures of the electromagnetic sector and neutrinos at the recombination epoch. The CMB observation insists $N_{\rm eff} = 2.99 \pm 0.17$ at 1$\sigma$ confidence level\,\cite{Planck:2018vyg}, and it is consistent with the SM prediction, i.e., that without such a new particles' effect\,\cite{deSalas:2016ztq, Bennett:2020zkv}. When the new particles couple only to the electromagnetic sector, which is the case of {\bf the light WIMP models with the scalar mediator} in the previous section, the effective number $N_{\rm eff}$ involving the new particles' contribution is estimated as follows:
\begin{align}
    N_{\rm eff} \simeq 3 \left\{ 1 + \frac{45}{11\pi^2 T_D^3} \sum_i s_i (T_D)   \right\}^{-4/3},
    \qquad
    s_i(T_D) = h_i(T_D) \frac{2\pi^2}{45} T_D^3,
    \label{eq: electrophilic Neff}
\end{align}
where $h_i(T_D) = (15 x_i^4)/(4 \pi^4) \int^{\infty}_1 dy \,(4y^2 -1)\sqrt{y^2 -1}/(e^{x_i y} \mp 1)$ with $x_i \equiv m_i/T_D$, $g_i$ is the degree of freedom, and `i' represents a new particle such as the WIMP and the mediator particle.\footnote{
    A similar formula can also be obtained for the case that the new particles couple only to neutrinos.}
The $-$\,($+$) sign is applied when the particle `i' is bosonic\,(fermionic). The new particles' effect appears only through the entropy injection, and the instantaneous neutrino decoupling by the weak interaction is assumed to obtain the above formula with $T_D = 1.7$\,MeV\,\cite{Matsumoto:2018acr, Ibe:2018juk}. The formula gives the result well-agreeing with that obtained by a more careful analysis using coupled Boltzmann equations describing electromagnetic, neutrino, and dark matter sectors\,\cite{Escudero:2018mvt, Sabti:2019mhn}. The comparison between the result of the formula\,(\ref{eq: electrophilic Neff}) and that of the CMB observation gives us the lower limit on the new particles' masses at 95\% C.L. to be ${\cal O}(1$--$10)$\,MeV, which are summarized in table\,\ref{tab: lower mass limits} ({\bf SS} and {\bf FS} models).

\begin{table}[t]
    \centering
    \begin{tabular}{r|cc|cc|cc}
        & \multicolumn{2}{c|}{Bulk region} & \multicolumn{2}{c|}{Resonance region} & \multicolumn{2}{c}{Forbidden region} \\
        & CMB & BBN & CMB & BBN & CMB & BBN \\
        \hline
        {\bf SS} model & {\footnotesize N/A} & {\footnotesize N/A} &  4.7\,MeV & 0.3\,MeV & 6.3\,MeV & 0.4\,MeV \\
        {\bf FS} {\sl "}\qquad\,~ & $\geq$ 6.3\,MeV & $\geq$ 0.4\,MeV &  6.4 {\sl "}~~~~~~ & 0.4 {\sl "}~~~~~~ & 7.5 {\sl "}~~~~~~ & 0.5 {\sl "}~~~~~~ \\
        {\bf SV} {\sl "}\qquad\,~ & {\footnotesize No limit} & $\geq$ 0.4 {\sl "}~~~~~~ & {\footnotesize No limit} & 0.5 {\sl "}~~~~~~ & {\footnotesize No limit} & 0.7 {\sl "}~~~~~~ \\
        {\bf FV} {\sl "}\qquad\,~ & {\footnotesize N/A} & {\footnotesize N/A} & {\footnotesize No limit} & 0.7 {\sl "}~~~~~~ & {\footnotesize No limit} & 0.9 {\sl "}~~~~~~ \\
        {\bf SV(B)} {\sl "}\qquad\,~ & -- & -- &  6.6 {\sl "}~~~~~~ & 0.5 {\sl "}~~~~~~ & -- & -- \\
        {\bf FV(B)} {\sl "}\qquad\,~ & -- & -- & 8.3 {\sl "}~~~~~~ & 0.7 {\sl "}~~~~~~ & --& -- \\
        \hline
    \end{tabular}
    \caption{\sl\small The lower limit on the mass of the WIMP in the light WIMP scenarios defined in table\,\ref{tab: light WIMP scenarios}, which is obtained by the CMB and BBN observations. The lower limit varies in the bulk ("p-wave") scenarios, as it also depends on the mass of the mediator particle.\protect\footnotemark\, See text for more details.}
    \label{tab: lower mass limits}
\end{table}
\footnotetext{The lower limit on the mass of $m_{\rm MED}$ is also obtained when $m_{\rm MED}$ is comparable or less than $m_{\rm DM}$.}

On the other hand, when the new particles couple to both the electromagnetic and neutrino sectors, which is the case of {\bf the light WIMP models with the U(1)$_{\rm B-L}$ vector mediator} in the previous section, the above formula cannot be applied because the new particles' effect appears not only through the entropy injection but also as a new interaction connecting the electromagnetic and neutrino sectors. It has recently been shown that, based on a careful analysis using the coupled Boltzmann equations, the lower limit on the masses is very sensitive to the relative strength of the new particles' interactions with electromagnetic particles and neutrinos.\,\cite{Escudero:2018mvt, Chu:2023jyb}. It is found that the two contributions cancel each other when the relative strength is appropriately chosen; the resultant effective number $N_{\rm eff}$ becomes the same as the SM prediction. In the {\bf SV} and {\bf FV} models, such a cancellation seems to occur when the relative branching ratio of the WIMP annihilation cross-section is $10^{-6}$--$10
^{-7}$, i.e., $\sigma_a({\rm DM\,DM} \to \nu \bar{\nu})/\sigma_a({\rm DM\,DM} \to e^- e^+) = \Gamma(Z' \to \nu \bar{\nu})/\Gamma(Z' \to e^- e^+) \sim 10^{-6}$--$10^{-7}$. Our analysis adopts such relative branching ratios to involve the region with as small a WIMP mass as possible. So, the CMB observation does not put a lower limit on the WIMP mass in the models. Note that having such a relative branching ratio is obtained by considering a tiny U(1)$_{\rm B-L}$ gauge coupling in the models, as discussed in sections\,\ref{subsec: vector mediator} and \ref{subsec: light WIMP models}, and it does not affect the physics of the WIMP models except for the one discussed here.

Unlike the U(1)$_{\rm B-L}$ models, the vector mediator couples dominantly to the electromagnetic particles in {\bf the light WIMP models with the U(1)$_{\rm B}$ vector mediator}. So, the cancellation of the contributions to $N_{\rm eff}$ mentioned above does not occur, and the limit on the new particles' masses is again obtained by the formula in eq.\,(\ref{eq: electrophilic Neff}). The lower limit on the WIMP mass at 95\% C.L. in the {\bf SV(B)-R} and {\bf FV(B)-R} scenarios are also summarized in table\,\ref{tab: lower mass limits}.

\subsubsection{Limit on the WIMP mass from the BBN observation}
\label{subsubsec: mass BBN}

The presence of the new particles with MeV scale masses affects the expansion rate of the universe and the time evolution of the photon and neutrino temperatures before their energy densities get Boltzmann suppressed. These modify the abundance of primordial light elements, e.g., the change of the temperatures' evolution modifies the proton-to-neutron conversion rate, and also, the change of the expansion rate modifies the time at which the deuterium forms. The BBN observation supports the standard thermal history without such new physics effects, leading to lower limits on the new particles' masses\,\cite{Boehm:2013jpa}.

The new particles contribute to the background evolution of the universe behind the BBN as new light degrees of freedom, so their masses are constrained even in the parameter region where the cancellation discussed in section\,\ref{subsubsec: mass CMB} occurs. To draw precise statements about the constraint, i.e., about the new particles' effect on the BBN, we should implement the background evolution obtained by solving the coupled Boltzmann equations mentioned above in an appropriate modern BBN code\,\cite{Pisanti:2007hk, Arbey:2011nf, Pitrou:2018cgg}. The code also enables us to include several uncertainties associated with the computation of the elements' abundances, e.g., those on nuclear reaction rates, observational outputs, etc., for a robust estimate of the limits. According to the latest studies, the limits on the masses are between several hundred keV to a few MeV, depending on the new particles' spins and interactions\,\cite{Boehm:2013jpa, Sabti:2019mhn}, so it is generally less constraining than those from the CMB observation in section\,\ref{subsubsec: mass CMB}.

We, however, include the limit on the new particles' masses from the BBN observation in our analysis for {\bf SV} and {\bf FV} models because the CMB observation does not put lower limits on the masses, as discussed in section\,\ref{subsubsec: mass CMB}. No detailed calculation of the new particles' effect on the BBN has been performed for the models that predict a velocity-dependent annihilation cross-section, except for the bulk ("p-wave") scenarios. So, we adopt the semi-analytical calculation developed in Ref.\,\cite{Escudero:2018mvt} to estimate the new particles' effect as follows: We first consider the Boltzmann equation that governs the number of neutrons,
\begin{align}
    \frac{d X_n}{dt} = \Gamma_{pn} (1 - X_n) - \Gamma_{np} X_n,
    \label{eq: Boltzmann eq neutron}
\end{align}
where $X_n = n_n/(n_n + n_p)$ with $n_n$ and $n_p$ being the number densities of the neutron and the proton in the early universe, respectively. We solve the above equation with the initial condition being the equilibrium one, $X_n^{\rm (eq)} \equiv [1 + e^{(m_n - m_p)/T}]^{-1}$, at $T =$ 100\,MeV. Note that all the species (photons, electrons, neutrinos, and new particles) are in thermal equilibrium and have the same temperature at such a high-temperature era. The Helium abundance is estimated by remembering that almost all the present neutrons at the time at which photons no longer dissociate the Deuterium will form Helium\,\cite{Sarkar:1995dd}: $Y_p \simeq 2X_n|_{T_\gamma = T_{\rm ^2H}}$ with $T_{\rm ^2H}$ being such a photon temperature, $T_{\rm ^2H} \simeq$ 0.073\,MeV. On the other hand, three processes convert protons into neutrons efficiently and the other way around: $n + \nu_e \to p + e^-$, $n + e^+ \to p + \bar{\nu}_e$, and $n \to p + e^- + \bar{\nu}_e$. Hence, the rate for the conversion is obtained as follows\,\cite{Dicus:1982bz}:
\begin{align}
    \Gamma_{np} =
    K \int^\infty_1 d\epsilon
    \frac{(\epsilon - q)^2 (\epsilon^2 - 1)^{1/2}\epsilon}
    {(1 + e^{-\epsilon z_\gamma})[1 + e^{(\epsilon - q) z_\nu}]}
    +K \int^\infty_1 d\epsilon
    \frac{(\epsilon + q)^2 (\epsilon^2 - 1)^{1/2}\epsilon}
    {(1 + e^{\epsilon z_\gamma})[1 + e^{-(\epsilon + q) z_\nu}]},
\end{align}
where $z_\gamma = m_e/T_\gamma$, $z_\nu = m_e/T_\nu$, $q = (m_n - m_p)/m_e$, $m_n - m_p = 1.2933$\,MeV, and $K \simeq (1.636\,\tau_n)^{-1}$ with the neutron lifetime being $\tau_n =$ 878.4\,s\,\cite{ParticleDataGroup:2022pth}. The proton-to-neutron conversion rate is obtained by $\Gamma_{pn} = \Gamma_{np}(-q)$. At sufficiently high temperatures, these rates are related by a detailed balance as $\Gamma_{pn} = \Gamma_{np}\,e^{(m_n - m_p)/T}$, and the equilibrium solution in eq.\,(\ref{eq: Boltzmann eq neutron}) is obtained as addressed above. The time evolution of the photon and neutrino temperatures, $T_\gamma(t)$ and $T_\nu(t)$, is required to estimate the above conversion rates, and the new particles affect the evolution when they are in equilibrium with the SM particles and before their energy densities are Boltzmann suppressed. We use the code developed in Ref.\,\cite{Escudero:2018mvt} to compute the evaluation, including the new particles' effect. The code allows us to include a tiny interaction with neutrinos, in addition to the primary interaction with electrons.

The comparison between the result obtained in the above method and the observation $Y_p = 0.247 \pm 0.0020$\,\cite{Kurichin:2021wlb} gives lower limits on the new particles’ masses at 95\% C.L. to be ${\cal O}(0.1$--$1)$\,MeV, which are summarized in table\,\ref{tab: lower mass limits}. The limits agree well with those obtained in Ref.\,\cite{Sabti:2019mhn}, where the new particle effect was estimated accurately and comprehensively based on the PRIMAT code. Note that the comprehensive calculation gives not only the prediction of the Helium abundance but also those of other light elements such as Deuterium and $^3$He. Comparing such predictions with their observations provides additional severe limits. For instance, in some cases, the limits would be a MeV or so, as shown in Ref.\,\cite{Sabti:2019mhn}. However, we use the limits obtained by the above semi-analytical method because changing the limits on the masses being $\sim$\,MeV does not significantly alter our discussion.

\subsubsection{Limit on the WIMP annihilation from the BBN observation}
\label{subsubsec: BBN on annihiation}

The BBN observations also put a limit on the WIMP annihilation cross-section. High-energy electromagnetically interacting particles, such as electrons, positrons, and photons, emitted by new particles into the cosmic plasma induce electromagnetic showers and produce energetic photons copiously. Though the high-energy particles emitted during the BBN are quickly thermalized through various electromagnetic processes and do not have a significant effect on the light elements being synthesized, those emitted after the BBN may destroy the light elements through so-called photo-disintegration processes\,\cite{Kawasaki:1994af, Ellis:1990nb}. So, the WIMP annihilation cross-section is constrained not to spoil the success of the standard BBN.

\begin{figure}[t]
    \centering
    \includegraphics[keepaspectratio, scale=0.23]{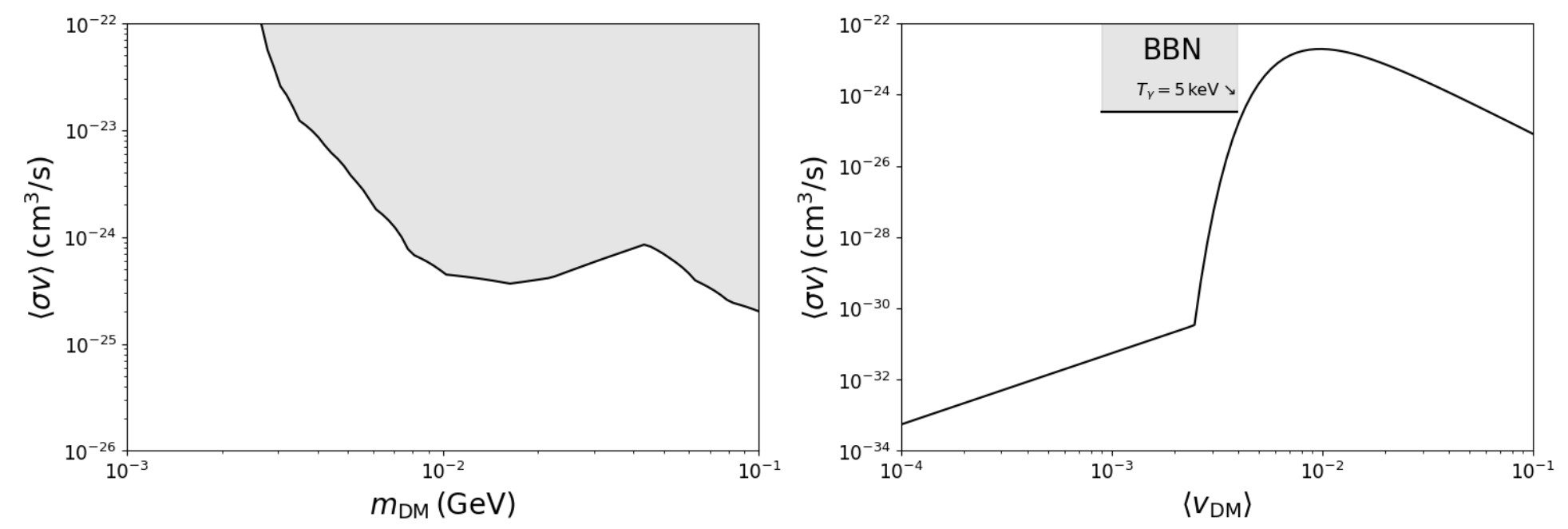}
    \quad
    \caption{\small \sl {\bf Left:} The constraint on the WIMP annihilation cross-section at 95\% C.L. from the BBN observation, assuming that the cross-section is velocity-independent. {\bf Right:} An example showing the constraint on the WIMP annihilation cross-section with the $s$-channel resonance of the mediator particle, with $\langle v_{\rm DM}\rangle$ being the mean value of the WIMP velocity. The cross-section in the panel is obtained in the {\bf SV-R} scenario, assuming model parameters as $m_\varphi = 70$\,MeV, $v_R = 4\,[(m_{Z'} - 2 m_\varphi)/m_\varphi]^{1/2} = 0.012$, $\xi = 7.8 \times 10^{-8}$, and $g_\varphi = 0.1$. According to the calculation concerning the relic abundance condition in section\,\ref{subsubsec: thermal relics}, $\langle v_{\rm DM} \rangle \simeq 4 \times 10^{-3}$ when $T_\gamma = 5$\,keV. See text for more details.}
    \label{fig: BBN-photo}
\end{figure}

A recent study shows that the WIMP annihilation cross-section is constrained at 95\% C.L. as given in the left panel of Fig.\,\ref{fig: BBN-photo}, assuming that the annihilation cross-section is velocity-independent when the WIMP is non-relativistic, i.e., a simple s-wave annihilation cross-section\,\cite{Depta:2019lbe}. Though the constraint is weaker than that given by the CMB observation in section\,\ref{subsubsec: the scenrios} and does not also contradict the condition of the thermal relic abundance in section\,\ref{subsubsec: thermal relics}, it does constrain the light WIMP models when the WIMP annihilates through the $s$-channel resonance of the mediator particle, i.e., the {\bf SS-R}. {\bf FS-R}, {\bf SV-R}, {\bf FV-R}, {\bf SV(B)-R}, and {\bf FV(B)-R} scenarios. This is because the annihilation cross-section may become larger than the limit shown in Fig.\,\ref{fig: BBN-photo} at the epoch relevant to the photo-dissociation.

Hence, we impose the above BBN constraint on the scenarios with the $s$-channel resonance of the mediator particle. According to the result in Ref.\,\cite{Depta:2019lbe}, when the WIMP mass is less than 100\,MeV, the constraint comes from the photo-dissociation (disintegration) of the Deuterium. An important fact is that a photon injected into the cosmic plasma with energy less than $\sim 2$\,MeV does not disintegrate the Deuterium, as the Deuterium has the binding energy of $E_{\rm B}({\rm D}) \simeq 2.22$\, MeV\,\cite{Ando:2005cz}. On the other hand, a photon injected into the cosmic plasma with energy more than $m_e^2/(22T_\gamma)$ quickly loses the energy through interactions with background photons (pair-creation process), with $T_\gamma$ being the temperature of the cosmic plasma\,\cite{Kawasaki:1994af}. So, a photon emitted at a temperature larger than $m_e^2/22/E_{\rm B}({\rm D}) \sim 5$\,keV does not disintegrate the Deuterium. Therefore, we impose the BBN constraint on the WIMP annihilation cross-section in the resonance scenarios as follows: At each $m_{\rm DM}$, the annihilation cross-section averaged by the WIMP velocity distribution function at the time that the temperature of the thermal plasma is $T_\gamma$ does not exceed the limit given in the left panel of Fig.\,\ref{fig: BBN-photo} whenever $T_\gamma \leq 5$\,keV, as exemplified in the right panel of the figure.

The photo-dissociation of the Deuterium by the WIMP annihilation also becomes ineffective at a temperature smaller than $T_{\rm eff} \sim {\cal O}(100)$\,eV because the WIMP density dilutes enough as the universe expands\,\cite{Cyburt:2002uv}. Hence, it is also possible to evade the BBN constraint when the velocity averaged cross-section is below the limit of Fig.\,\ref{fig: BBN-photo} (left panel) at a temperature above $T_{\rm eff}$, even if the cross-section exceeds the limit below $T_{\rm eff}$. However, we numerically confirmed that such a parameter region is ruled out by the indirect dark matter detection constraint, even if we include various uncertainties associated with the constraint. So, we discard such a parameter region by requiring the BBN constraint as above.

\subsubsection{Relic abundance condition on the WIMP}
\label{subsubsec: thermal relics}

The CMB observation also gives precious information about the WIMP's phase-space density (e.g., the abundance). The evolution of the density in the early universe is governed by the Boltzmann equation, $L [ f_{\rm DM} ] = C [ f_{\rm DM} ]$, with $L$ and $C$ being the Liouville operator and the collision terms, respectively. The latter term describes WIMP interactions, such as the scattering with and annihilation into the SM particles. In the Friedman-Robertson-Walker universe, with $H$ being the Hubble parameter, the equation is written as follows\,\cite{Kolb:1990vq, Gondolo:1990dk}:
\begin{align}
    E\,(\partial_t - H\,\vec{p}\cdot\vec{\nabla}_p)\,f_{\rm DM}
    = C_a\left[ f_{\rm DM} \right] + C_s\left[ f_{\rm DM} \right],
    \label{eq: Boltzmann}
\end{align}
with $C_a$ and $C_s$ being collision terms describing the annihilation and scattering of the WIMP. With multi-dimensional integrals in the collision terms, it is time-consuming to solve the equation numerically. An alternative is considering the first few moments of the phase-space density, i.e., using the so-called hydrodynamical approach\,\cite{Binder:2017rgn, Binder:2021bmg}. The zeroth moment is the WIMP number density, $n_{\rm DM}(t) \equiv g_{\rm DM} \int d^3p/(2\pi)^3\,f_{\rm DM}(\vec{p}, t)$. In the thermal equilibrium, the WIMP has the number density, $n_{\rm DM}^{\rm eq} \sim (m_{\rm DM} T)^{3/2}\exp (-m_{\rm DM}/T)$, which is Boltzmann-suppressed compared to those of relativistic SM particles, $n_{\rm SM}^{\rm eq} \sim T^3$ when $T \lesssim m_{\rm DM}$.\footnote{
    When the freeze-out of the WIMP happens after the neutrino decoupling, we must specify what is the WIMP temperature. The WIMP interacts mainly with the electromagnetic sector in the scenarios we focus on in this article; the temperature can be identified with that of the sector, i.e., $T_\gamma$, as verified in Refs.\,\cite{Li:2023puz, Escudero:2018mvt}.} 
Therefore, with the crossing symmetry, the elastic scattering between the WIMP and the SM particles is usually much faster than the WIMP annihilation into the SM particles. So, the WIMP is assumed to be in kinetic equilibrium with the SM particles during the freeze-out epoch, which closes the above Boltzmann hierarchy at the zeroth-moment level,
\begin{align}
    \dot{n}_{\rm DM} + 3 H n_{\rm DM} =
    -\langle \sigma_a v \rangle^T \,\left[n_{\rm DM}^2 - \left( n_{\rm DM}^{\rm eq}\right)^2 \right],
    \label{eq: nBE}
\end{align}
with $\langle \sigma_a v \rangle^T$ being the annihilation cross-section averaged by the WIMP velocity distribution function when the temperature of the universe is $T$, i.e., the thermally averaged one.

The above discussion can be applied to the light WIMP scenarios when the mediator particle is heavier enough than the WIMP, except for the resonance scenarios. Such parameter regions are realized in the {\bf FS-B} and {\bf SV-B} scenarios in section\,\ref{subsubsec: the scenrios}. On the other hand, it cannot be straightforwardly applied when the mediator particle is lighter enough than the WIMP, which is realized in another parameter region of the {\bf FS-B} scenario. This is because the mediator particle exists in the universe at the freeze-out epoch. Hence, we impose a condition on the scenario that the mediator particle must have a larger decay width into the SM particles than the Hubble parameter $H$ at the freeze-out temperature $T_{\rm FO}$,
\begin{align}
    \langle \Gamma ({\rm Med} \to {\rm SMs}) \rangle^{T_{\rm FO}}
    > H(T_{\rm FO}),
    \label{eq: mediator equilibrium}
\end{align}
where $\langle \Gamma(\cdots) \rangle^T$ is the decay width, including the effect of the Lorentz gamma factor at the temperature $T$. Moreover, the freeze-out temperature $T_{\rm FO}$ is defined in the same way as adopted in the micrOMEGAs code, i.e., $n_{\rm DM} = 2.5 n_{\rm DM}^{\rm eq}$ at $T = T_{\rm FO}$, where $n_{\rm DM}$ is obtained by solving eq.\,(\ref{eq: nBE}) assuming that the mediator particle is in equilibrium with the SM particles. Note that the condition\,(\ref{eq: mediator equilibrium}) guarantees that the mediator particle is in equilibrium with the SM particles not only at but also after the freeze-out epoch, as $H(T)$ is more quickly suppressed than $\langle \Gamma(\cdots) \rangle^T$ at $T \leq T_{\rm FO}$. Furthermore, since the WIMP annihilates mainly into a pair of mediator particles in such a parameter region, as discussed in the previous section, the WIMP is also in kinematic equilibrium with the mediator particle via the scattering between the two particles thanks to the crossing symmetry. So, we can safely use the Boltzmann equation\,(\ref{eq: nBE}) to calculate the thermal relic abundance of the light WIMP.

The mass of the mediator particle is almost the same as that of the WIMP in the {\bf SS-F}, {\bf FS-F}, {\bf SV-F}, and {\bf FV-F} scenarios in section\,\ref{subsubsec: the scenrios}, so the number densities of the particles are also almost the same in equilibrium. Then, the scattering between the WIMP and the mediator particle does not automatically guarantee the kinetic equilibrium between them, unlike the above case. Hence, we impose a condition on the scenarios that, in addition to the one in eq.\,(\ref{eq: mediator equilibrium}), the scattering rate between the WIMP and the mediator particle and/or the rate between the WIMP and the SM particles are larger than $H$ at the freeze-out:
\begin{align}
    &\langle \sigma_s v ({\rm DM\,MED} \to {\rm DM\,MED}) \rangle^{T_{\rm FO}}\,n_{\rm MED}^{\rm eq}(T_{\rm FO}) > H(T_{\rm FO})
    \\
    &\qquad\qquad\qquad\qquad {\rm and/or}
    \nonumber \\
    &\langle \sigma_s v ({\rm DM\,SM} \to {\rm DM\,SM})\rangle^{T_{\rm FO}}\,n_{\rm SM}^{\rm eq}(T_{\rm FO}) > H(T_{\rm FO}), 
\end{align}
where $\langle \sigma_s v (\cdots) \rangle^{T_{\rm FO}}$ denotes the scattering cross-section of the WIMP with the mediator or with the SM particles averaged by the WIMP velocity distribution function at the freeze-out temperature, and $n_{\rm MED}^{\rm eq}(T_{\rm FO})$ is the number density of the mediator particle in the thermal equilibrium. Then, the above condition enables us to use the Boltzmann equation\,(\ref{eq: nBE}) to calculate the relic abundance of the light WIMP even in the forbidden scenarios.

Unlike all the above cases, the kinetic equilibrium between the WIMP and the SM particles is not guaranteed in the {\bf SS-R}, {\bf FS-R}, {\bf SV-R}, {\bf FV-R}, {\bf SV(B)-R}, and {\bf FV(B)-R} scenarios in section\,\ref{subsubsec: the scenrios}, where the annihilation of the WIMP is enhanced by the $s$-channel resonance of the mediator particle. This is because the mass of the mediator particle is twice the WIMP mass in the scenarios, so the scattering between the WIMP and the mediator particle does not guarantee the kinematical equilibrium. Moreover, the scattering between the WIMP and the SM particles, which proceeds by exchanging the mediator particle in the $t$-channel, is also highly suppressed due to the small couplings of the mediator particle to the WIMP and the SM particles. So, the above resonance scenarios result in the so-called early kinetic decoupling, and we must go beyond the zeroth moment equation in the Boltzmann hierarchy\,\cite{Binder:2017rgn, Binder:2021bmg}. The first moment of the phase-space density is the WIMP temperature, $T_{\rm DM} = (g_{\rm DM}/n_{\rm DM}) \int d^3p/(2\pi)^3\,p^2/(3E_p)\,f_{\rm DM}(\vec{p}, t)$. We implicitly assume in the scalar WIMP scenarios\,(i.e., the {\bf SS-R}, {\bf SV-R}, and {\bf SV(B)-R} scenarios) that, in addition to that for the mediator particle in eq.\,(\ref{eq: mediator equilibrium}), the couplings of the self-interactions, $\lambda_\phi$ and $\lambda_\varphi$, are large enough to ensure that the WIMP is in kinetic equilibrium during the freeze-out epoch and obeys the Maxwell-Boltzmann distribution with the temperature of $T_{\rm DM}$. 
In the fermionic WIMP scenarios\,(i.e., the {\bf FS-R}, {\bf FV-R}, and {\bf FV(B)-R} scenarios), because the WIMP does not have a self-interaction, we impose a condition on its self-scattering cross-section as follows:
\begin{align}
    &\langle \sigma_s v ({\rm DM\,DM} \to {\rm DM\,DM}) \rangle^{T_{\rm DM}(T_{\rm FO})}\,n_{\rm DM}^{\rm eq}[T_{\rm DM}(T_{\rm FO})] > H(T_{\rm FO}),
    \label{eq: self-scattering condition}
\end{align}
where $T_{\rm DM}(T)$ is the WIMP temperature when the cosmic plasma temperature is $T$, and $T_{\rm FO}$ is defined as $n_{\rm DM}[T_{\rm DM}(T)] = 2.5\,n_{\rm DM}^{\rm eq}[T_{\rm DM}(T)]$ at $T = T_{\rm FO}$, where $n_{\rm DM}[T_{\rm DM}(T)]$ is obtained by solving the following Boltzmann equations assuming that it obeys the Maxwell-Boltzmann distribution with $T_{\rm DM}$. This condition is satisfied when the WIMP-WIMP-MED coupling is larger than the WIMP-WIMP-SM one, as the WIMP self-scattering cross-section is also enhanced by exchanging the mediator particle in the $s$-channel. On the other hand, in the resonance scenarios, the annihilation cross-section keeps increasing, and the net WIMP number in the universe keeps decreasing until a certain low temperature $T <  T_{\rm FO}$. We numerically confirmed that the self-scattering rate is larger than the expansion rate of the universe even during this period when the condition given in eq.\,(\ref{eq: self-scattering condition}) is satisfied. So, the WIMP obeys the Maxwell-Boltzmann distribution during the whole freeze-out epoch.

Then, the Boltzmann hierarchy is closed up to the first-moment level. Integrating the equation\,(\ref{eq: Boltzmann}) for the 0th and 1st moments of $f_{\rm DM}$, we obtain the equations\,\cite{Binder:2017rgn, Binder:2021bmg}:
{\small
\begin{align}
\label{eq: cBE}
    \frac{Y_{\rm DM}'}{Y_{\rm DM}} &=
    \frac{s Y_{\rm DM}}{x \tilde{H}}
    \left[
        \frac{(Y_{\rm DM}^{\rm eq})^2}{Y_{\rm DM}^2}
        \langle \sigma_a v \rangle^{T_\gamma}
        -\langle \sigma_a v \rangle^{T_{\rm DM}}
    \right],
    \\
    \frac{y_{\rm DM}'}{y_{\rm DM}} &=
    \frac{\langle C_s \rangle_2}{x \tilde{H}}
    +\frac{s Y_{\rm DM}}{x \tilde{H}}
    \left[
        \langle \sigma_a v \rangle^{T_{\rm DM}}
        -\langle \sigma_a v \rangle_2^{T_{\rm DM}}
    \right]
    +\frac{s\,(Y_{\rm DM}^{\rm eq})^2}{x \tilde{H} Y_{\rm DM}}
    \left[
        \frac{y_{\rm DM}^{\rm eq}}{y_{\rm DM}}
        \langle \sigma_a v \rangle_2^{T_\gamma}
        -\langle \sigma_a v \rangle^{T_\gamma}
    \right]
    +2 (1 - w) \frac{H}{x \tilde{H}},
    \nonumber
\end{align}
}\noindent
where $x \equiv m_{\rm DM}/T_\gamma$, and the prime " $'$ " is the derivative with respect to $x$. Here, $Y_{\rm DM}(x) \equiv n_{\rm DM}(x)/s(T_\gamma)$, $y_{\rm DM}(x) \equiv m_{\rm DM} T_{\rm DM}(x)/s(T_\gamma)^{2/3}$, while $Y_{\rm DM}^{\rm eq}(x) \equiv n_{\rm DM}^{\rm eq}(T_{\rm DM})/s(T_\gamma)$, $y_{\rm DM}^{\rm eq}(x) \equiv m_{\rm DM} T_\gamma/s(T_\gamma)^{2/3}$, with $s(T_\gamma)$ being the entropy density of the universe, $\tilde{H}(T_\gamma) \equiv H(T_\gamma)/[1 + (1/3)\,d(\log g_s(T_\gamma))/d(\log T_\gamma) ]$ with $g_s(T_\gamma)$ being the entropy degrees of freedom of the background cosmic plasma, and $w(x) \equiv 1 - \langle p^4/E_p^3 \rangle_{T_{\rm DM}}/(6 T_{\rm DM})$, which characterizes the deviation from the highly non-relativistic case, i.e., $w(x) = 1$. The thermally averaged annihilation cross-section $\langle \sigma_a v \rangle_2^T$, which is a variant of $\langle \sigma_a v \rangle^T$, and $\langle C_{\rm el} \rangle_2$ are defined as follows:
\begin{align}
    \langle \sigma_a v \rangle_2^T &\equiv
    \frac{g_{\rm DM}^2}{T [n_{\rm DM}^{\rm eq}(T)]^2}
    \int\frac{d^3p}{(2\pi)^3}\,\frac{d^3p'}{(2\pi)^3}
    \frac{p^2}{3E_p} (\sigma_a v)
    f_{\rm DM}^{\rm eq}(\vec{p}, T)\,
    f_{\rm DM}^{\rm eq}(\vec{p}', T),
    \\
    \langle C_s \rangle_2 &\equiv
    \frac{g_{\rm DM}}{T_{\rm DM}\,n_{\rm DM}^{\rm eq}(T_{\rm DM})}
    \int \frac{d^3p}{(2\pi)^3} \frac{p^2}{3 E_p^2}
    C_s[f_{\rm DM}^{\rm eq}(\vec{p}, T_{\rm DM})],
\end{align}
where $f_{\rm DM}^{\rm eq}(\vec{p}, T)$ is the WIMP's phase-space density in the thermal equilibrium with the temperature $T$. When the WIMP is non-relativistic and scatters with a lighter particle in the cosmic plasma (thermal bath), the typical momentum transfer per scattering is smaller than the average of the momentum that the WIMP has in the thermal equilibrium. Then, the collision term mentioned above, $C_s[f_{\rm DM}]$, can be cast to the so-called Fokker–Planck operator with an appropriate moment transfer rate (denoted by $\gamma$ below) as follows\,\cite{Binder:2016pnr, Binder:2017rgn}:
\begin{align}
    C_s &\simeq
    \gamma \frac{E_p}{2}
    \left[
        T_{\rm DM} E_p \partial_p^2 +
        \left(
            2 T_{\rm DM} \frac{E_p}{P} + p + T_{\rm DM} \frac{P}{E_p} 
        \right) \partial_p +
        3
    \right] f_{\rm DM}(\vec{p}, T_{\rm DM}).
    \nonumber \\
    \gamma &\equiv
    \frac{1}{3 g_{\rm DM} m_{\rm DM} T_{\rm DM}}
    \int \frac{d^3k}{(2\pi)^3} f_{\rm SM} (\omega)
    \left[ 1 \pm f_{\rm SM}(\omega) \right]
    \int_{-4k_{\rm cm}^2}^0 dt\,(-t)\,\frac{d}{dt}(\sigma_s v),
    \label{eq: momentum transfer rate}
\end{align}
with $\omega \equiv (k^2 + m_{\rm SM}^2)^{1/2}$ and $k_{\rm cm} \equiv m_{\rm DM}^2 k^2/(m_{\rm DM}^2 + 2 \omega m_{\rm DM} + m_{\rm SM}^2)$. The differential scattering cross-section between the WIMP and SM particle is evaluated at the center-of-mass energy $s \simeq m_{\rm DM}^2 + 2 \omega m_{\rm DM} + m_{\rm SM}^2$. We use the DRAKE code\,\cite{Binder:2022pmf}, which takes account of the early kinetic decoupling effect discussed above, to calculate the relic abundance of the WIMP in the resonance scenarios, with the condition for the mediator particle in eq.\,(\ref{eq: mediator equilibrium}).

The CMB observation precisely measures the dark matter abundance as $\Omega_{\rm DM} h^2 \simeq 0.12$ with ${\cal O}(1)$\,\% level observational uncertainty. On the other hand, there are theoretical uncertainties in calculating the dark matter abundance; those are from the relativistic degrees of freedom in the SM plasma, the use of the Boltzmann equation\,(\ref{eq: nBE}) to describe the freeze-out of the WIMP around the neutrino decoupling, the use of the approximated collision term\,(\ref{eq: momentum transfer rate}). We use the relativistic degrees of freedom evaluated in Ref.\,\cite{Saikawa:2020swg} instead of that initially implemented in the code, resulting in ${\cal O}(0.1)$\,\% uncertainty. The uncertainty associated with the freeze-out around the neutrino decoupling is, at most, 10\,\% level when the WIMP almost exclusively interacts with electrons and photons\,\cite{Li:2023puz}, which is the case we focus on in this article.\footnote{
    To more accurately estimate the WIMP abundance when the freeze-out is around the neutrino decoupling, three coupled Boltzmann equations for the WIMP, EM particles\,($e^\pm$ and $\gamma$), and neutrinos should be solved.}
The uncertainty associated with the use of the approximated collision term is less than 10\,\%, as we numerically confirmed that the result using the approximation agrees with that using the full scattering term at less than 10\,\% level even if the WIMP mass is close to the electron mass\,\cite{Aboubrahim:2023yag}. Therefore, we assume 10\,\% theoretical uncertainty in our analysis associated with evaluating the WIMP relic abundance.

\subsection{Detection of the light WIMPs}
\label{sec: Light Thermal DM: Detection}

The basic strategy to detect the light WIMP is, more or less, the same as that of the traditional EW-scale WIMP: detection at accelerator and collider experiments, at underground laboratories\,(i.e., direct detection), at cosmological and astrophysical observations\,(i.e., indirect detection). On the other hand, its detail differs from that of the EW-scale WIMP because of the different energy scales we should search for and the existence of the mediator particle. Below, we discuss the present status and future prospects for detecting the light WIMP.

\subsubsection{Direct dark matter detection}
\label{sec: Direct Detection Light}

Direct detection is an experiment observing recoil energy caused by the dark matter scattering off a usual material, such as a nucleus and an electron, in a deep underground detector. To date, numerous direct detection experiments have been conducted. Thanks to coherence, the so-called spin-independent scattering between dark matter and a nucleus is enhanced for a heavier nucleus. Hence, a noble liquid such as xenon is often used as a target material to detect the traditional WIMP\,\cite{XENON:2023cxc, XENON:2018voc}. However, such an experiment quickly loses sensitivity when the dark matter is much lighter than the nucleus. This is because the recoil energy is proportional to the square of the reduced mass of the dark matter and the nucleus, and the recoil energy falls below the detector threshold. Hence, we need a new strategy to detect a light WIMP. One potential avenue is using detectors with lower energy thresholds, e.g., semiconductor ionization detectors such as germanium and silicon, requiring only tiny energy of ${\cal O}(1)$\,eV to create an electron-hole pair, and crystalline cryogenic bolometers utilize phonons, allowing for low thresholds of ${\cal O}(10)$\,eV. Moreover, the Migdal effect on the scattering has recently been proposed to extend the reach of direct detection experiments: The scattering between dark matter and nucleus leads to atomic excitation and ionization, as the electron cloud associated with the nucleus does not follow the recoiling nucleus instantaneously\,\cite{Ibe:2017yqa, Dolan:2017xbu}. Several experiments, including EDELWEISS\,\cite{EDELWEISS:2019vjv}, CRESST\,\cite{CRESST:2019jnq}, SuperCDMS\,\cite{SuperCDMS:2014cds} and SENSEI\,\cite{SENSEI:2020dpa} have succeeded to yield constraints on light dark matter. On the other hand, the most compelling strategy to detect a light WIMP is to search for the scattering between the dark matter and an electron bound in a material\,\cite{Essig:2011nj, Essig:2015cda}. The scattering makes use of nearly all of the kinetic energy of the dark matter, with the threshold to detect its recoil energy ranging from ${\cal O}(1)$\,meV band gaps in a target with a spin-orbit coupling\,\cite{Chen:2022pyd} to ${\cal O}(10)$\,eV required to ionize electrons in atomic targets\,\cite{Graham:2012su}.

\begin{figure}[t]
    \centering
    \includegraphics[keepaspectratio, scale=0.4]{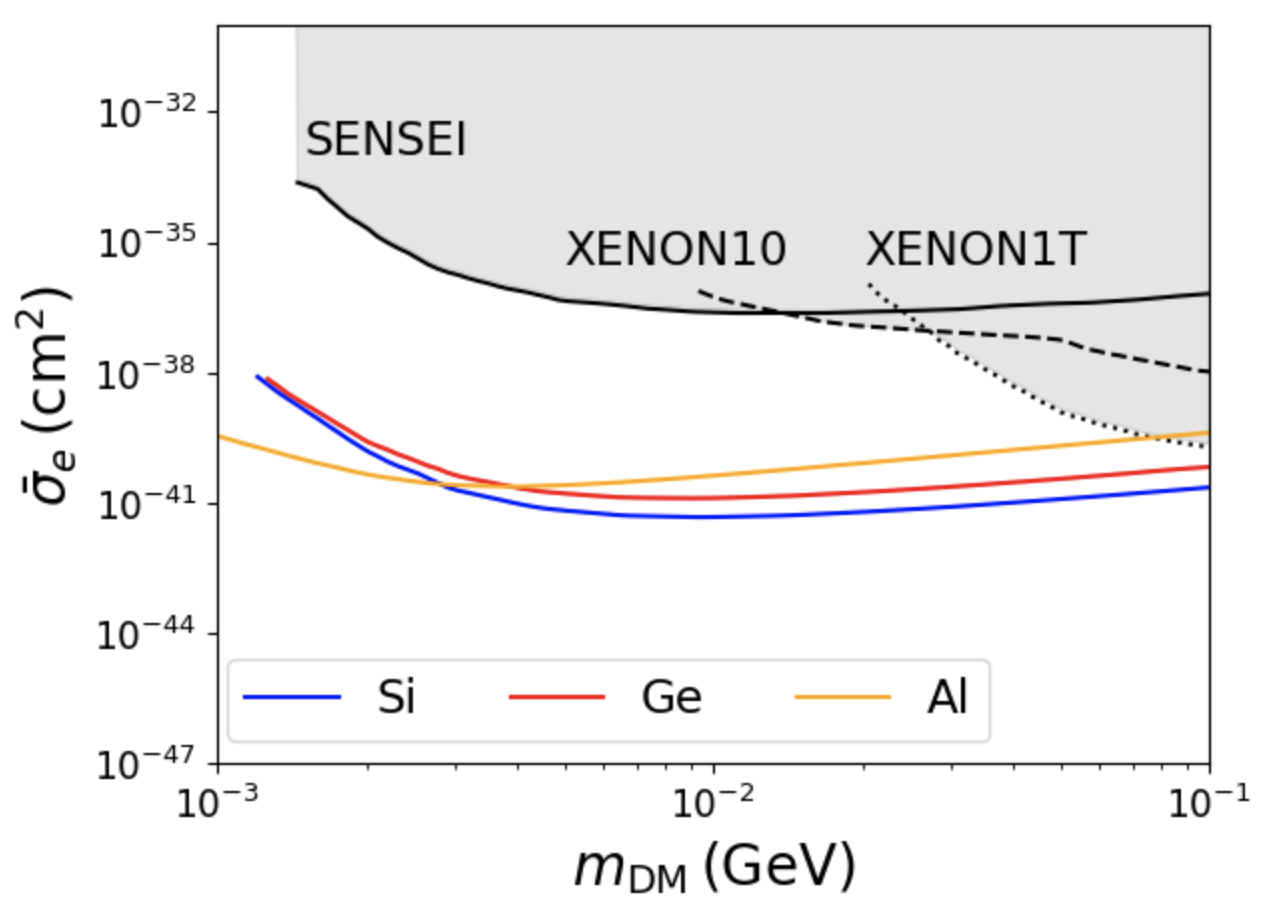}
    \caption{\small \sl Constraints on the elastic scattering between the light WIMP and an electron at 95\,\% C.L. The shaded region is excluded by direct detection experiments performed so far, while colored lines are expected (referential) sensitivities of future detection experiments. See text for more details.}
    \label{fig: direct detection}
\end{figure}

The direct dark matter detection through the WIMP-nucleus scattering has yet to yield a viable constraint for the light WIMP scenarios where the WIMP mass is less than 100\,MeV. So, we focus on the constraint derived from the detection through the WIMP-electron scattering. In the above WIMP mass range, XENON\,10\,\cite{Essig:2017kqs}, XENON\,1T\,\cite{XENON:2019gfn}, and SENSEI\,\cite{SENSEI:2023zdf} experiments currently give the most stringent limits, as shown in Fig.\,\ref{fig: direct detection}. It is worth noting that these constraints are subject to uncertainties related to the local density and the local velocity distribution of dark matter—the recommended values summarized in Refs.\,\cite{Baxter:2021pqo, Lacroix:2020lhn} are used in the figure, but the constraint can be weakened by an order of magnitude\,\cite{Baxter:2021pqo, Wu:2019nhd}. With this fact in mind, it turns out that the present limits do not constrain the light WIMP scenarios we analyze in the next section, so we do not include those in our analysis. Instead, we confirm in the next section that the viable parameter regions of the models that survive after applying other constraints and conditions are consistent with the present direct dark matter detection limits. On the other hand, the sensitivity of the detection searching for the scattering between a WIMP and an electron will be significantly improved in the future. Among several proposals, we consider those discussed in Ref.\,\cite{Knapen:2021run}, where expected (referential) sensitivities of future direct detection experiments at the 95\% C.L. are given, assuming 1\,year$\cdot$kg exposure of silicon\,(Si), germanium\,(Ge), and aluminum \,(Al) detectors with the same dark matter distribution configuration as mentioned above. Those are shown by colored lines in the figure. In the subsequent section, we found that future direct detection experiments will cover certain portions of the viable parameter regions.

\subsubsection{Detection at accelerators}
\label{sec: Accelerator Light}

The models we focus on comprise the light WIMP, the light mediator particle, and the SM particles. While the WIMP only couples to the mediator particle in the energy region much below the EW scale, the mediator particle also couples to the SM particles. Meanwhile, depending on its spin, the mediator particle mixes with SM bosons, such as the photon, mesons, $Z$, and Higgs bosons. Hence, the mixing modifies SM interactions involving such SM bosons, and it also introduces new interactions involving both the SM bosons and the new particles. Therefore, we must consider various accelerator constraints, with the most stringent being experiments searching for direct mediator particle productions at high-luminosity accelerators. We first consider those and then summarize other constraints.
\vspace{0.35cm}

\noindent
\underline{\bf Scalar mediator in the SS and FS models}
\vspace{0.25cm}

\begin{figure}[t]
    \centering
    \includegraphics[keepaspectratio, scale=0.385]{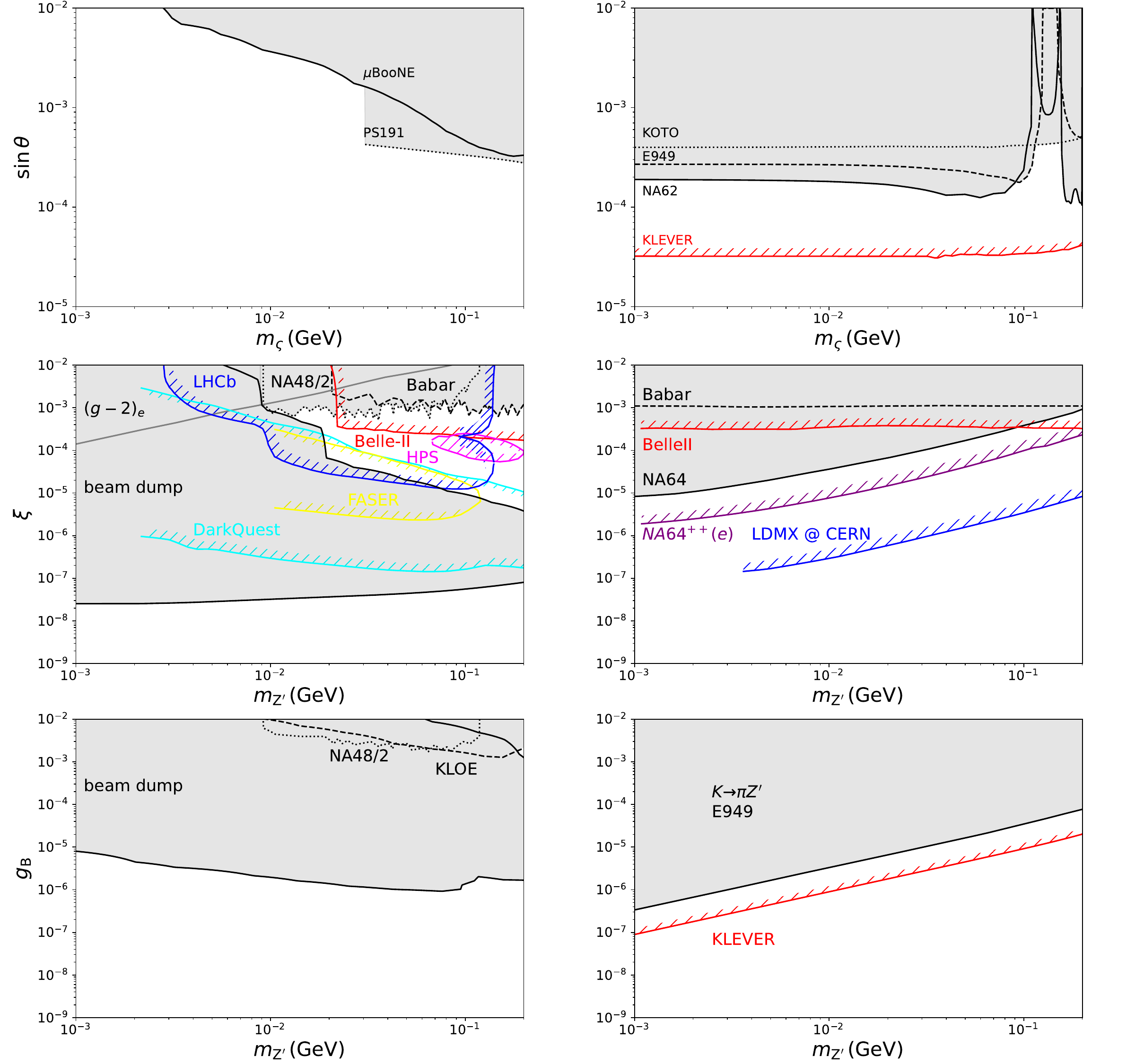}
    \caption{\small \sl Constraints on a visible (invisible) mediator particle at 90\% C.L. from various accelerator experiments are shown at the left (right) panels for the cases of the scalar mediator scenario (upper row), the dark photon scenario (middle row), and the U(1)$_B$ scenario (bottom row). See text for more details. In the middle left panel, we also give a constraint from the observation of the electron magnetic dipole moment, which constrains the region with $m_{Z'} <$ 10\,MeV and $\xi > 10^{-3}$.}
    \label{fig: accelerator constraint}
\end{figure}

\noindent
Once the mediator particle is produced, it decays eventually. Decay channels of the light mediator particle are classified into two cases: visible and invisible decays, which offer two different search strategies to detect the mediator particle at the accelerator experiments. Constraints on the visible scalar mediator particle are summarized in the upper left panel of Fig.\,\ref{fig: accelerator constraint}, assuming the mediator particle decays into a pair of electrons with a 100\,\% branching fraction. When the mediator particle is lighter than 200\,MeV, which is the mass region of our interest, the scalar mediator particle is mainly produced by the rare decay of K mesons via the FCNC process, $s \to d \varsigma$\,\cite{Krnjaic:2015mbs}. As shown in the figure, among various experiments, the PS191 and $\mu$Boone collaborations provide leading constraints on the scalar mediator particle via the search for the processes, $K^0_L \to \pi^0 \varsigma$\,\cite{Gorbunov:2021ccu} and $K^+ \to \pi^+ \varsigma$\,\cite{MicroBooNE:2021usw}.

On the other hand, constraints on the invisible scalar mediator particle are summarized in the upper right panel of Fig.\,\ref{fig: accelerator constraint}, assuming again that the mediator particle decays into a pair of electrons with a 100\,\% branching fraction. In this case, events that only the mediator particle decays outside a detector contribute to the signal.\footnote{
    It is also possible to depict the figure assuming that the mediator particle decays dominantly into a pair of WIMPs. However, as seen in the following sections, the figure shown in Fig.\,\ref{fig: accelerator constraint} is more relevant to our discussion, so we decided to show the figure assuming the mediator particle decays dominantly into $e^- e^+$.
} As shown in the figure, E949\,\cite{BNL-E949:2009dza}, NA62\,\cite{NA62:2021zjw}, and KOTO\,\cite{KOTO:2020prk} collaborations presently provide leading constraints on the "invisible" scalar mediator particle, such as $\sin\theta \lesssim {\cal O}(10^{-4})$ via the processes $K^\pm \to \pi^\pm \varsigma \to \pi^\pm$ + missing (E949, NA62) and $K^0_L \to \pi^0 \varsigma \to \pi^0$ + missing (KOTO). In the near future, as shown by a colored line, the KLEVER collaboration\,\cite{Beacham:2019nyx} will explore the parameter region via the decay process $K^0_L \to \pi^0 \varsigma$, with the sensitivity on the branching fraction measurement of more than an order of magnitude better than the present.
\vspace{0.35cm}

\noindent
\underline{\bf Dark photon mediator in the SV and FV models}
\vspace{0.25cm}

\noindent
Constraints on the visible dark photon mediator particle appearing in our ${\bf SV}$ and ${\bf FV}$ models are summarized in the middle left panel of Fig.\,\ref{fig: accelerator constraint}. To depict this figure, we assume that the branching fraction of the mediator particle decaying into a pair of electrons is 100\,\%. Among various experiments, NA48/2 and BaBar collaborations provide constraints as $\xi \lesssim 10^{-3}$ via the searches for the processes $K^+ \to \pi^+ \pi^0 \to \pi^+ Z' \gamma$\,\cite{NA482:2015wmo} and $e^-e^+ \to \gamma Z'$, respectively, with $Z'$ being the dark photon mediator particle. On the other hand, electron beam-dump experiments\,(E137\,\cite{Bjorken:1988as}, NA64\,\cite{NA64:2018lsq}, etc.) and proton beam-dump experiments\,($\nu$-CAL\,I\,\cite{Blumlein:2011mv, Blumlein:2013cua}, CHARM\,\cite{Gninenko:2012eq}, etc.) provide constraints as $\xi \sim 10^{-3}$--$10^{-7}$ depending on the mediator particle mass via the search for the electron bremsstrahlung process, proton bremsstrahlung process\,\cite{Blumlein:2013cua}, and meson decays\,($\pi^0 \to Z'\gamma$\,\cite{Blumlein:2011mv}, $\eta$\,,$\eta' \to Z'\gamma$\,\cite{CHARM:1985anb}, etc.). Because the beam-dump experiments employ electromagnetic calorimeters placed downstream of their substantial shielding, the length of the decay volume, i.e., the distance between the end-edge of the shielding and the detector, determines a hard limit on the flight distance of the mediator particle. Since the lifetime of the mediator particle scales as $(\xi^2 m_{Z'})^{-1}$, they provide wedge-shaped constraints. The surviving region at $\xi \sim 10^{-4}$ and $m_{Z'} \gtrsim 10$\,MeV will be explored by future experiments (LHCb\,\cite{Ilten:2016tkc}, Belle\,II\,\cite{Belle-II:2018jsg}, FASER\,\cite{FASER:2018eoc}, HPS\,\cite{Moreno:2013mja}, DarkQuest\,\cite{Tsai:2019buq}), as shown by colored lines in the figure.

Constraints on the invisible dark photon, which also appears in our ${\bf SV}$ and ${\bf FV}$ models, are summarized in the middle right panel of Fig.\,\ref{fig: accelerator constraint}. To depict the figure, we assume that the mediator particle decays into a pair of WIMPs with almost 100\,\% branching fraction. As seen in the panel, the NA64 collaboration\,\cite{Banerjee:2019pds} presently provides a leading constraint via the search for the invisible decay of the dark photon that is produced by the bremsstrahlung process from an injected electron beam interacting with an active target. In the near future, the Belle\,II\,\cite{Belle-II:2018jsg} and NA64$^{++}$\,\cite{gninenko2018addendum} collaborations explore a new parameter region with the sensitivity on the mixing parameter $\xi$ of about an order of magnitude better than the present limit. Moreover, in the future, the LDMX collaboration\,\cite{LDMX:2018cma} may explore the parameter region with a sensitivity of a few orders of magnitude better than the present's.

So far, we have considered accelerator constraints on a mediator particle that decays only visibly or invisibly. However, in some parameter regions of the ${\bf SV}$ and ${\bf FV}$ models, the mediator particle decays both visibly and invisibly. We adopt the method developed in Ref.\,\cite{Ilten:2018crw} to put an accelerator constraint on the mediator particle in such a case, which is commonly employed in various past phenomenological studies\,\cite{Brahma:2023psr, Bernreuther:2020koj, Amrith:2018yfb, Flores:2020lji}.
\vspace{0.35cm}

\noindent
\underline{\bf U(1)$_{\rm B}$ vector mediator in the SV(B) and FV(B) models}
\vspace{0.25cm}

\noindent
Constraints on the visible mediator particle in the {\bf SV(B)} and {\bf FV(B)} models are summarized in the lower left panel of Fig.\,\ref{fig: accelerator constraint}, assuming that the mediator particle decays into a pair of electrons with a 100\,\% branching fraction. This decay comes from a small mixing effect caused by one-loop diagrams (or the mixing term to the hyper-charged gauge boson with a coefficient suppressed enough at the same level), leading to the effective mixing parameter of $\xi_{\rm eff} \sim g_B\,e/(4\pi)^2$. Moreover, new additional heavy fermions are introduced to make the theory anomaly-free, as discussed in section\,\ref{subsec: vector mediator}, and it causes the Wess-Zumino type interaction between the vector mediator particle and SM gauge bosons in its low-energy theory. Then, the longitudinal mode of the vector mediator particle is produced with a $E^2/m_{Z'}^2$ enhancement with $E$ being the energy of a produced $Z'$\,\cite{Dror:2017nsg, Dror:2017ehi}. Therefore, as seen in the panel, the mediator particle is longer-lived, and the constraints from proton fixed-target experiments (including beam-dump ones) are more severe than the dark photon case, which closes the surviving parameter region seen in the dark photon case.

Meanwhile, constraints on the invisible mediator particle in the {\bf SV(B)} and {\bf FV(B)} models are summarized in the lower right panel of Fig.\,\ref{fig: accelerator constraint}, assuming that the mediator particle decays into a pair of WIMPs with almost 100\,\% branching fraction. As seen in the panel, KTeV\,\cite{KTeV:2003sls} and E949\,\cite{E949:2008btt, E949:2004uaj} collaborations presently provide leading constraints via the search for the $K \to \pi Z'$ process. Remembering that the couplings between the mediator particle and SM fermions are $\sim \xi$ and $\sim g_B$ for the dark photon and the U(1)$_{\rm B}$ scenarios, respectively, the constraint on the U(1)$_{\rm B}$ mediator particle is found to be more severe than the dark photon's. This is because the production of the longitudinal mode of the U(1)$_{\rm B}$ mediator particle is enhanced, as mentioned above. Similar to the scalar mediator case, the KLEVER collaboration explores a new parameter region in the near future, with the sensitivity of measuring the branching fraction of an order of magnitude better than the present's.
\vspace{0.35cm}

\noindent
\underline{\bf Other accelerator constraints}
\vspace{0.25cm}

\noindent
Modifying SM interactions involving SM bosons that mix with a mediator particle has a negligible impact on constraining the light WIMP models because the mixing is already severely constrained by the abovementioned constraints. New interactions involving the SM bosons and the mediator particle, generated via the mixing above, also do not put any severe constraints for the same reason. On the other hand, the interaction involving the scalar mediator particle (or the scalar WIMP) and the Higgs boson is not suppressed even if the mixing is highly suppressed. This interaction is constrained by the invisible Higgs width measurement at the LHC experiment, resulting in ${\rm Br}(h \to {\rm inv.}) \leq 0.190$ at 95\,\% C.L.\,\cite{CMS:2018yfx}.

\subsubsection{Indirect dark matter detection}
\label{sec: Indirect Detection Light}

Indirect detection is an experiment that observes products produced by WIMP annihilation in space. Depending on the kind of product, the detection is classified into two cases: observing electrically charged particles (i.e., electron, positron, (anti-)proton, etc.) and neutral particles (i.e., photons and neutrinos). The strategy of detecting the light WIMP through indirect detection is qualitatively the same as that of the traditional WIMP. However, the difference between their mass scales causes some quantitative differences. We consider the two cases in the following and discuss constraints obtained by the experiments.
\vspace{0.35cm}

\noindent
\underline{\bf Indirect detection observing charged particles}
\vspace{0.25cm}

\noindent
Given our emphasis on the light WIMP, whose mass is less than 100\,MeV, the observable signal consists of electrons or positrons with energies of, at most, the mass. On the experimental side, however, such particles cannot enter the heliosphere due to the solar magnetic field\,\cite{Boudaud:2016mos}. Only Voyager\,I, i.e., the cosmic-ray detector located outside the heliosphere since 2012, can detect such a light WIMP signal. On the theoretical side, the signal flux is estimated by considering the injected $e^\pm$ spectra from the WIMP annihilation and its propagation in the Milky Way; both suffer from significant systematic uncertainties.

\begin{figure}[t]
    \centering
\includegraphics[keepaspectratio, scale=0.38]{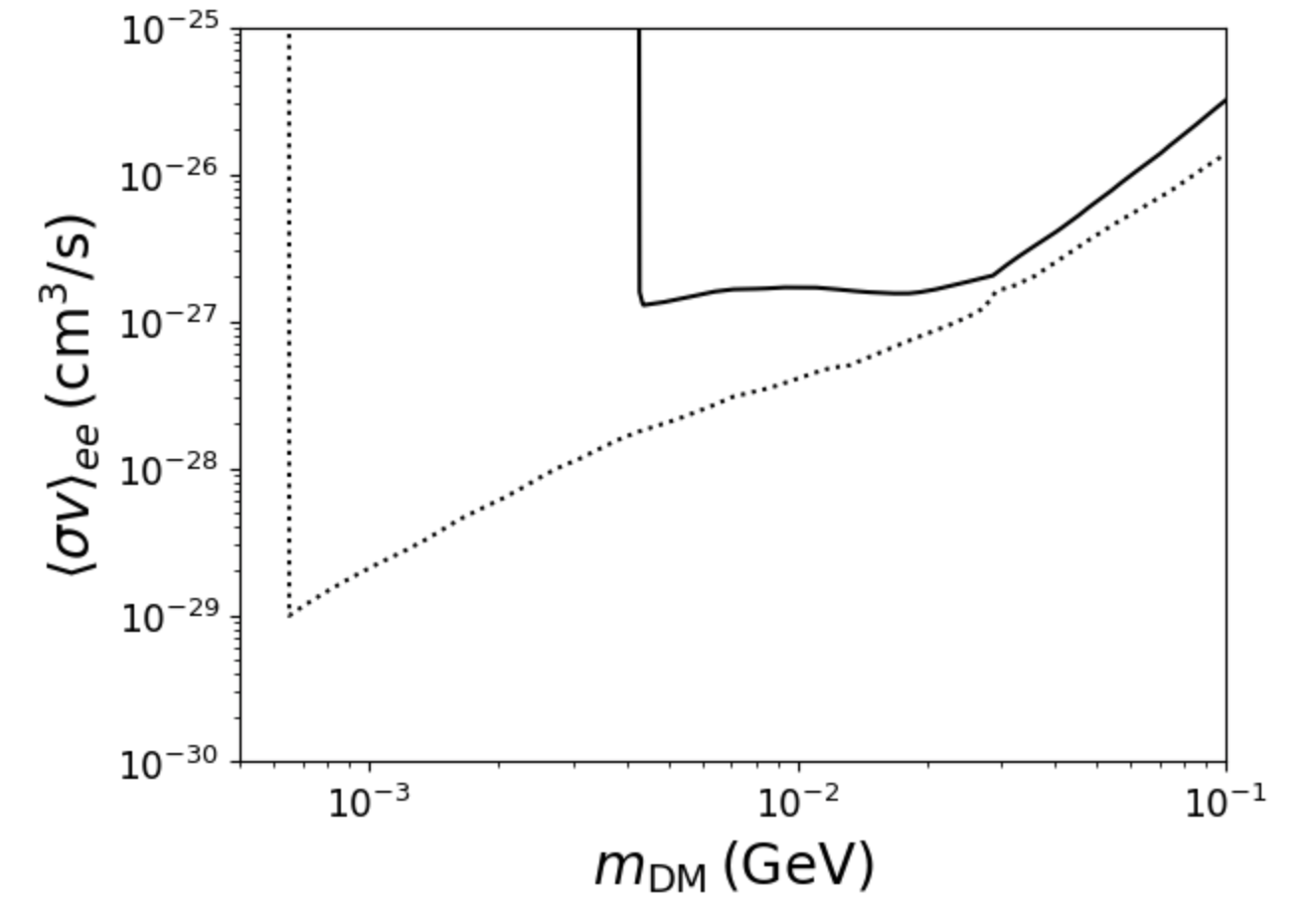}
    \caption{\small\sl Constraints on the WIMP annihilation cross-section at 95\,\% C.L. from Voyager\,I data, assuming the WIMP annihilates into $e^-e^+$. The dotted (solid) line is the constraint assuming the (non-)existence of the $e^\pm$ reacceleration during the propagation. See text for more details.}
    \label{fig: Voyager}
\end{figure}

We adopt the models discussed in Ref.\,\cite{DelaTorreLuque:2023olp} to consider the propagation uncertainty in our analysis. One is the model that best fits the latest AMS-02 data for B, Be, and Li ratios to C and O, with a non-zero reacceleration effect. Another one is the model that is also consistent with the AMS-02 data but without the reacceleration effect. Since the reacceleration pushes the positrons to those with energies more than the WIMP mass, it significantly strengthens the constraint on the WIMP annihilation cross-section. So, the reacceleration gives the leading propagation uncertainty. The constraint on the annihilation cross-section times a relative velocity, averaged by the local WIMP velocity distribution, at 95\,\% C.L. from the Voyager\,I data is shown in Fig.\,\ref{fig: Voyager} by a solid (dotted) line, where the WIMP is assumed to annihilate into $e^- e^+$ that propagate without (with) the reacceleration effect.

It is also worth noting that the above constraint is further subject to uncertainties related to other propagation parameters and the local WIMP density and distribution. First, it is also possible to find other propagation models that are consistent with the AMS-02 data\,\cite{Donato:2003xg, Boudaud:2016mos, Kappl:2015bqa, Reinert:2017aga}, and the difference among the models show the existence of an additional order of magnitude uncertainty. Next, the signal flux is proportional to the WIMP density squared and the WIMP annihilation cross-section averaged over the WIMP velocity distribution. Since energetic electrons and positrons quickly lose energy during propagation due to interactions with the interstellar medium, those produced near the solar system are expected to contribute to the observed signal. Therefore, we must also consider the uncertainty originating in the local WIMP density and distribution. To depict the constraint in Fig.\,\ref{fig: Voyager}, the referential values for the density and distribution\,\cite{Baxter:2021pqo, Lacroix:2020lhn}, also used in Fig.\,\ref{fig: direct detection}, are adopted. However, the local density and distribution are not precisely measured, leading to additional uncertainty regarding the constraint on the WIMP annihilation cross-section.\footnote{
    The local WIMP density is obtained not by global measurements assuming axisymmetry\,\cite{Read:2014qva} but by local ones\,\cite{Bovy:2012tw, Garbari:2012ff, Zhang:2012rsb, Bovy:2013raa}. The uncertainty from the local WIMP velocity distribution is smaller than the density's.}

With the various uncertainties mentioned above, the limit on the WIMP annihilation cross-section presented in Fig.\,\ref{fig: Voyager} can be weakened by, at least, an order of magnitude. With this fact in mind, it turns out that the limit does not constrain the light WIMP models we analyze in the next section, so we do not include those in our analysis. Instead, in the next section, we confirm that the viable parameter regions of the models that survive after applying other constraints and conditions are indeed consistent with the Voyager-I limit.

\vspace{0.35cm}

\noindent
\underline{\bf Indirect detection observing neutral particles}
\vspace{0.25cm}

The light WIMP also contributes to the indirect detection signal composed of neutral particles, such as photons and neutrinos. Gamma-ray observations impose more stringent constraints on WIMP models compared to neutrino observations, unless the WIMP annihilates predominantly into neutrinos. This is because the efficiency of detecting photon signals is significantly higher than that of neutrino signals. In the light WIMP scenarios we focus on, the light WIMP predominantly annihilates into photons (as well as electrons and positrons). Therefore, we consider only the constraints from gamma-ray observations.

On the experimental side, various indirect detection experiments have been dedicated to searching for gamma rays from WIMP annihilation. Such gamma rays interact with matter via three distinct processes: absorption, Compton scattering, and pair creation. Gamma-ray energy influences their efficiency, so each interaction calls for a different search strategy. For gamma rays with energy over 100\,MeV, the primary interaction is pair creation, which is efficiently detected by pair-conversion telescopes such as Fermi-LAT\,\cite{Fermi-LAT:2009ihh}. Conversely, gamma rays with energy less than 100\,keV are predominantly absorbed and efficiently searched for by spectrometers and imagers like INTEGRAL\,\cite{Winkler:2003nn}. On the other hand, gamma rays with the energy of ${\cal O}$(1-10)\,MeV primarily undergo Compton scattering, which is difficult to detect because it produces long-range secondaries requiring massive detectors to stop. Moreover, the background against the signal in this energy range contains large and complicated components, such as atmospheric albedo, activation within the instrument, and diffuse gamma-ray components. Hence, the sensitivity to detect gamma rays in the MeV energy range is not as good as in other energy ranges, known as the 'MeV gap.' However, despite this difficulty, the COMPTEL collaboration\,\cite{schonfelder1993instrument} has observed this energy range, enabling us to explore MeV gamma-ray signals from the light WIMP annihilation. So, we consider the COMPTEL experiment observing the center of our galaxy to put a constraint on the light WIMP scenarios based on the method developed in Refs.\,\cite{Coogan:2021sjs, Coogan:2021rez}.\footnote{
    We also consider the constraint on the light WIMP scenarios from the INTEGRAL experiment observing the galactic center because it puts the most stringent constraint when the WIMP mass is ${\cal O}(1)$\,MeV\,\cite{Siegert:2024hmr}.}

On the theoretical side, the gamma-ray signal, i.e., the gamma-ray flux from the WIMP annihilation at the Galactic center, is estimated by the formula,
\begin{align}
    \frac{d{\cal F}_\gamma}{dE_\gamma} \simeq
    \Bigg[
        \frac{\langle \sigma v \rangle_{v_{\rm GC}}}{8 \pi m_{\rm DM}^2} \sum_f\,{\rm Br}\,({\rm DM\,DM} \to f)\,\left.\frac{d N_\gamma}{dE_\gamma}\right|_f
    \Bigg]
    \times
    \Bigg[
        \int_{\Delta \Omega} d\Omega \int_{\rm l.o.s} ds\,\rho_{\rm DM}^2
    \Bigg],
    \label{eq: flux}
\end{align}
where $\langle \sigma v \rangle_{v_{\rm GC}}$ is the total annihilation cross-section (times the relative velocity) of the light WIMP averaged by the relative-velocity distribution among WIMPs, with the velocity dispersion parameter at the Galactic center, $v_{\rm GC} = 400$\,km/s\,\cite{Lacroix:2020lhn}. The branching fraction of the annihilation channel into the final state `$f$' is denoted by ${\rm Br}\,({\rm DM\,DM} \to f)$, while $d N_\gamma/dE_\gamma|_f$ is called the fragmentation function describing the number of produced photons with energy $E_\gamma$ at a given final state `$f$.' The function originates from the mediator particle decaying into, e.g., $e^- e^+ \gamma$ and $\pi^0 \gamma$, in the WIMP scenarios we focus on in this article, as discussed in section\,\ref{sec: candidates}. The term in the second parenthesis on the right-hand side of the formula is called the J-factor, which is determined by the WIMP density profile at the Galactic center.

\begin{figure}[t]
    \centering
    \includegraphics[keepaspectratio, scale=0.46]{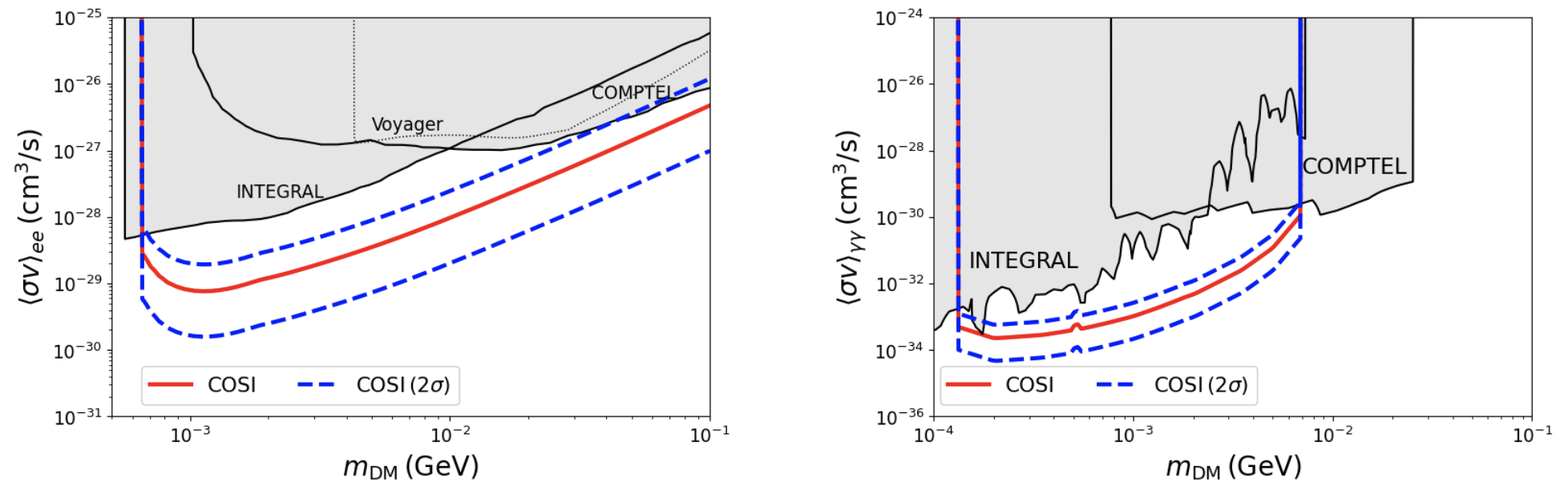}
    \caption{\small \sl 
    The left and right panels show the present COMPTEL and INTEGRAL constraints and projected COSI sensitivities on the light WIMP annihilation into $e^+ e^-$ and $\gamma\gamma$. The red (thick) solid line represents the COSI sensitivity with a median value of the J-factor\,\cite{deSalas:2019pee}, while the blue (thick) dashed lines are those with the uncertainty at the 2$\sigma$ level. The Voyager limit, the same limit without the reacceleration in Fig.\,\ref{fig: Voyager}, is also shown by a thin dotted line in the left panel for comparison.}
    \label{fig: indirect detection}
\end{figure}

By comparing the gamma-ray flux in eq.\,(\ref{eq: flux}) to the MeV gamma-ray observation\,\cite{Coogan:2021rez, Siegert:2024hmr}, we obtain constraints on the WIMP annihilation cross-section (times the relative velocity) averaged by the WIMP velocity function (having the velocity dispersion $v_{\rm GC}$), which is shown in the left panel of Fig.\,\ref{fig: indirect detection} assuming the WIMP annihilates into a pair of electrons associated with a photon, i.e., $\langle \sigma ({\rm DM}\,{\rm DM} \to e^- e^+,\,e^- e^+ \gamma)\,v\rangle_{\rm GC}$, and in the right panel of the figure assuming the annihilation into a pair of photons, i.e., $\langle \sigma ({\rm DM}\,{\rm DM} \to \gamma \gamma)\,v\rangle_{\rm GC}$. Here, we adopted the central value given in Table\,III of Ref.\,\cite{deSalas:2019pee} for the J-factor to depict the lines in the figure. We imposed the constraint on the WIMP scenarios in our analysis; the annihilation cross-sections should not exceed those shown by the black solid lines.

The detection of the MeV gamma-ray will be upgraded in the future; several Compton telescopes are under consideration to overcome the 'MeV gap,' such as COSI\,\cite{Tomsick:2021wed, Aramaki:2022zpw, Tomsick:2023aue}, e-ASTROGAM\,\cite{e-ASTROGAM:2016bph}, and GECCO\,\cite{Orlando:2021get}. We particularly focus on COSI, which will launch in the near future. Based on the method developed in Refs\,\cite{Negro:2021urm, Caputo:2022dkz, Tomsick:2023aue}, we estimate the projected sensitivity on the WIMP annihilation cross-section, $\sigma v$, averaged by the WIMP velocity distribution function with the velocity dispersion $v_{\rm GC}$, assuming two years observations of the Galactic center\,(10\,$\times$\,10 degrees). The sensitivity for the WIMP annihilation into a pair of electrons (associated with a photon) is given by the red (thick) solid line in the left panel and into a pair of photons in the right panel of Fig.\,\ref{fig: indirect detection}. The same J-factor used for the COMPTEL and INTEGRAL constraints is adopted to depict the sensitivity lines. COSI is expected to improve the sensitivities by more than an order of magnitude compared to the present limits. Here, it is worth noting that all the limits from the gamma-ray observations suffer from the uncertainty related to the dark matter density at the Galactic center; various estimates of the dark matter density\,\cite{Benito:2020lgu, deSalas:2019pee, Cautun:2019eaf} suggest that an order of magnitude uncertainty exists; the uncertainty on the COSI sensitivity line is also shown by blue (thick) dashed lines in the figure, based on the J-factor estimation in Ref.\,\cite{deSalas:2019pee}. With this in mind, in the next section, we discuss whether the viable parameter regions of the light WIMP scenarios, which survive after applying all constraints and conditions, can be explored in the future by COSI.

Indirect detection observing lower energy photons is another exciting possibility for detecting the light WIMP, as pointed out in the literature\,\cite{Cirelli:2020bpc, Cirelli:2023tnx, DelaTorreLuque:2023olp}. Many electrons and positrons are emitted from the light WIMP annihilation; these secondary products produce radio waves via synchrotron radiation\,\cite{Bertone:2001jv, Gondolo:2000pn, Aloisio:2004hy}, X-rays via inverse Compton radiation\,\cite{Baltz:2004bb, Cholis:2008wq, Zhang:2008tb} and Bremsstrahlung\,\cite{Cirelli:2013mqa} processes, and 511 keV line emission and corresponding orthopositronium continuum\,\cite{Finkbeiner:2007kk, Boehm:2003bt, Siegert:2023wus}. In particular, X-ray observations are found to give constraints compatible with those discussed above, even if various astrophysical ambiguities suggested in the literature are incorporated.\footnote{
    Observations of the 511\,keV line emission by INTEGRAL/SPI may also give stringent constraints on the light WIMP models, especially when its morphological data is included\,\cite{Vincent:2012an, Cappiello:2023qwl}. However, the dark matter profile and the propagation of low-energy positrons suffer from systematic uncertainties, weakening the constraints.}
Meanwhile, an accurate estimate of the ambiguities is now under debate, and using the observations to put a robust constraint on the light WIMP scenarios is not easy at present, so we do not include the constraint in our analysis. On the other hand, because many future projects observing photons in the lower energy ranges are scheduled with good sensitivities\,\cite{eROSITA:2012lfj}, observing such photons will be crucial for detecting the light WIMP once the ambiguities are evaluated well. We leave the discussion for future study. 

\subsubsection{Other constraints}
\label{subsubsec: other constraints}

We comment on another constraint arising due to the peculiarity of the light WIMP models: the constraint on the mediator particle obtained by observing neutrinos from the supernova (SN) 1987A\,\cite{Kamiokande-II:1987idp}. It is known that the collapse at an SN phenomenon heats the core, and the neutrino emission from the core during the SN explosion is the only mechanism to cool the core within the SM framework. On the other hand, in models with a new light particle weakly interacting with the SM particles, such as the mediator particles, the new particle may be emitted from the core and travel a macroscopic distance without transferring energy back to the stellar material, providing a new cooling mechanism. Since the observed neutrino burst from the SN 1987A quantitatively agrees with the SM predictions\,\cite{Burrows:1987zz}, such an anomalous energy loss should be suppressed enough\,\cite{Turner:1987by, Burrows:1988ah}. So, it constrains the nature of the mediator particle in the light WIMP scenarios so that its coupling to the SM particles is too weak to be emitted or strong enough to be trapped inside the core.

On the other hand, SN physics is very complicated, and we must account for various uncertainties in order to place a robust constraint on the mediator particle\,\cite{Chang:2016ntp, Chang:2018rso, Mahoney:2017jqk, Fischer:2016cyd, Sung:2019xie}. For example, in the physics of the proto-neutron star core, as well as the primary driver of shock revival, the nature (temperature profile, density profile, and equation of state) of the progenitor star, the cross-sections of various QCD processes with soft radiation, and the environmental effects on these processes are all known to introduce uncertainties. To our knowledge, no study has been conducted that places a constraint on a new particle while accounting for all of the above uncertainties. So, we do not include the SN constraint in our analysis and leave the task of robustly incorporating the constraint for future work.\footnote{
    Using the SN constraint from Refs.\,\cite{Chang:2016ntp, Krnjaic:2015mbs}, we found that, although the constraint limits part of the parameter space in some light WIMP scenarios, it does not significantly alter the conclusions of this article.}

\section{Viable parameter regions of the light WIMPs}
\label{sec: Analysis}

In this section, we figure out the viable parameter regions of the light WIMP scenarios defined in section\,\ref{sec: candidates} by imposing the conditions and constraints discussed in section\,\ref{sec: conditions and constraints}. The viable parameter regions are then compared with the detection sensitivities of the future dark matter and mediator particle experiments addressed in section\,\ref{sec: Light Thermal DM: Detection}. Below, we first present our analysis strategy and then discuss the viable parameter regions, focusing on the regions that would be explored in the near-future indirect dark matter detection, the COSI.

\subsection{Analysis strategy}
\label{subsec: analysis}

\begin{figure}[t]
    \centering
    \includegraphics[keepaspectratio, scale=0.45]{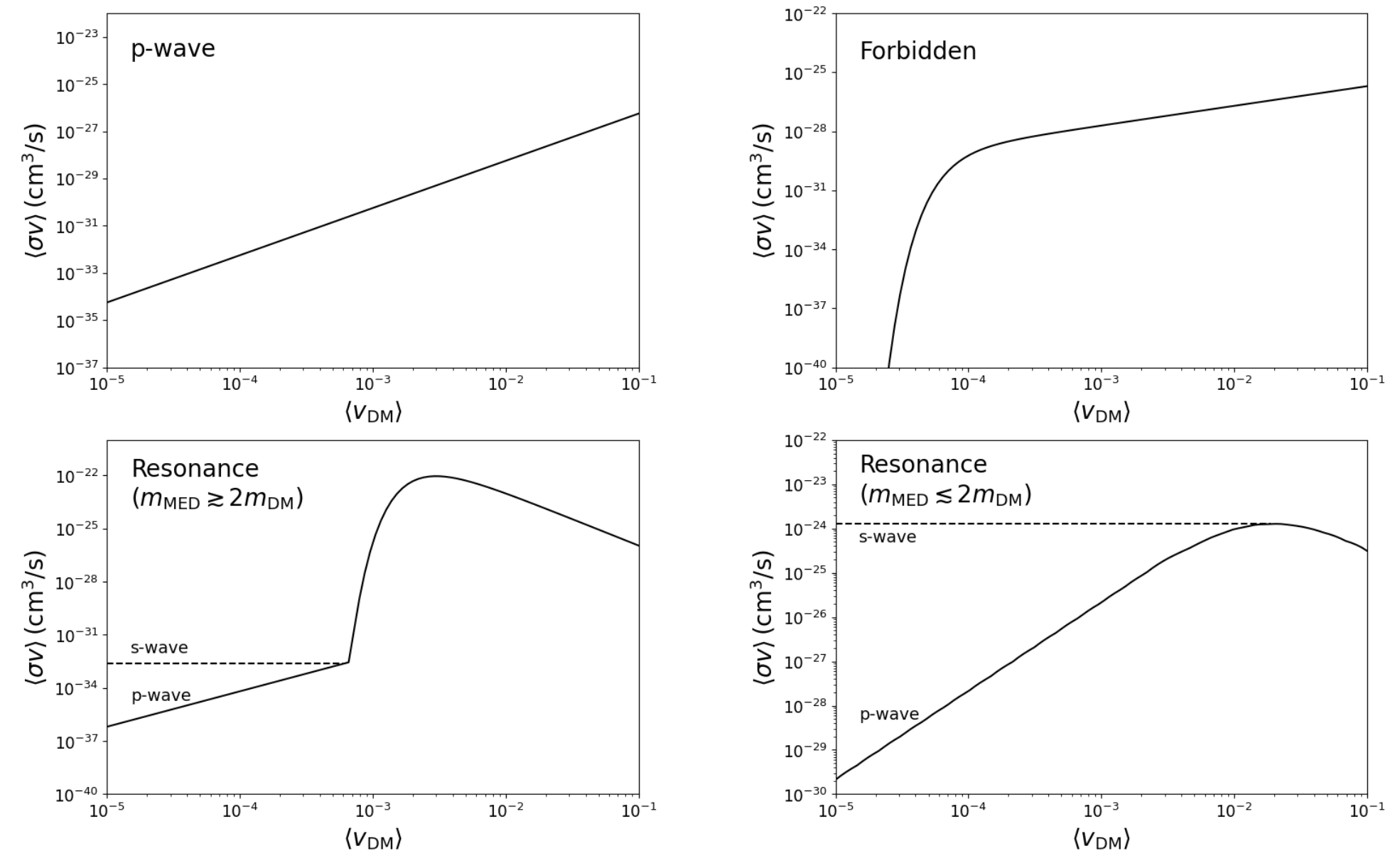}
    \caption{\small \sl Typical velocity-dependencies of the light WIMP annihilation cross-section in the "p-wave"-suppressed, forbidden channel, $s$-channel resonance with an invisible mediator (i.e., $m_{\rm MED} \gtrsim 2 m_{\rm DM}$), and that with a visible mediator (i.e., $m_{\rm MED} \lesssim 2 m_{\rm DM}$) scenarios are shown in the top left, top right, bottom left, and bottom right panels, respectively. The solid lines in the lower panels are resonant cross-sections in the p-wave annihilation, and the dotted ones are those in the s-wave.}
	\label{fig: sigmav_velocity_dependent}
\end{figure}

Among the light WIMP scenarios shown in Table\,\ref{tab: light WIMP scenarios}, we consider those in the resonance and the forbidden regions in the following analysis, i.e., we do not consider the bulk ("p-wave") scenarios. As seen in the top left panel of Fig.\,\ref{fig: sigmav_velocity_dependent}, which illustrates the typical behavior of the light WIMP annihilation cross-section in p-wave, its velocity-dependence is $\sigma v \propto v^2$ with $v$ being the relative velocity between the incidents WIMPs. Because the cross-section is required to be ${\cal O}(10^{-26})$\,cm$^3$/s at $v^2 \sim {\cal O}(0.1)$ to satisfy the relic abundance condition, it is expected to be ${\cal O}(10^{-31})$\,cm$^3$/s in the present universe (i.e., at $v^2 \sim {\cal O}(10^{-6})$), which is an order of magnitude below the projected sensitivity of the COSI, $10^{-30}\,{\rm cm^3/s}$.\footnote{
    This is the COSI sensitivity for a continuum gamma-ray, as shown in the left panel of Fig.\,\ref{fig: indirect detection}, which is from, e.g., the light WIMP annihilation into $e^- e^+ \gamma$. On the other hand, the COSI sensitivity for a line gamma-ray is more sensitive, as seen in the right panel of the figure. However, the cross-section of the light WIMP producing a line gamma-ray is below the sensitivity again in the p-wave scenarios. See appendix\,\ref{app: p-wave} for more details.}
Table\,\ref{tab: conditions & constraints} summarizes the conditions and the constraints imposed in the resonance and forbidden scenarios. First, the relic abundance condition is imposed on all the scenarios according to the strategy discussed in section\,\ref{subsubsec: thermal relics}. Next, we also impose the lower limit on the WIMP mass obtained from cosmological observations in all the scenarios, based on the discussion in sections\,\ref{subsubsec: mass CMB} and \ref{subsubsec: mass BBN}. Moreover, the CMB limit on the annihilation cross-section in the recombination epoch, known to be the most drastic one for the light WIMPs, is imposed on all the scenarios, as addressed in section\,\ref{subsubsec: the scenrios}. On the other hand, the BBN limit on the annihilation cross-section is imposed only on the resonance scenarios because the limit is always weaker than the CMB limit for the forbidden scenarios, as seen in section\,\ref{subsubsec: BBN on annihiation}.

\begin{table}[t]
    \centering
    {\footnotesize
    \begin{tabular}{l|cccccc:cccc}
        & SS-R & FS-R & SV-R & FV-R & SV(B)-R & FV(B)-R
        & SS-F & FS-F & SV-F & FV-F \\
        \hline
        Relic abundance cond.
        & $\circ$ & $\circ$ & $\circ$ & $\circ$ & $\circ$ & $\circ$
        & $\circ$ & $\circ$ & $\circ$ & $\circ$ \\
        CMB \& BBN on $m_{\rm DM}$
        & $\circ$ & $\circ$ & $\circ$ & $\circ$ & $\circ$ & $\circ$
        & $\circ$ & $\circ$ & $\circ$ & $\circ$ \\
        CMB on $\langle \sigma v \rangle$
        & $\circ$ & $\circ$ & $\circ$ & $\circ$ & $\circ$ & $\circ$
        & $\circ$ & $\circ$ & $\circ$ & $\circ$ \\
        BBN on $\langle \sigma v \rangle$
        & $\circ$ & $\circ$ & $\circ$ & $\circ$ & $\circ$ & $\circ$
        & & & & \\
        Accelerator detection
        & $\circ$ & $\circ$ & $\circ$ & $\circ$ & $\circ$ & $\circ$
        & $\circ$ & $\circ$ & $\circ$ & $\circ$ \\
        Indirect detection
        & $\circ$ & $\circ$ & $\circ$ & $\circ$ & $\circ$ & $\circ$
        & $\circ$ & $\circ$ & $\circ$ & $\circ$ \\
        Unitarity limit
        & $\circ$ & $\circ$ & $\circ$ & $\circ$ & $\circ$ & $\circ$
        & $\circ$ & $\circ$ & $\circ$ & $\circ$ \\
        Vacuum stability
        & $\circ$ & $\circ$ & & & &
        & $\circ$ & $\circ$ & & \\
        \hline
    \end{tabular}
    }
    \caption{\sl\small Conditions and constraints imposed in the resonance and forbidden scenarios. Here, those with (without) `$\circ$' are (not) conditions/constraints imposed. See the main text for more details.}
    \label{tab: conditions & constraints}
\end{table}

In addition to the conditions and constraints concerning the early universe cosmology mentioned above, we impose several constraints on the scenarios from dark matter detections and field theoretical consistencies. First, the constraint obtained by accelerator experiments is imposed on all the scenarios based on the strategy discussed in section\,\ref{sec: Accelerator Light}. We also impose the constraint on the scenarios from the indirect dark matter detection, which is from gamma-ray observations addressed in section\,\ref{sec: Indirect Detection Light}. Meanwhile, according to the unitarity of the light WIMP models discussed in appendix\,\ref{app: unitarity}, we impose upper limits on the coupling constants in the models: The coupling constant should be less than $4\pi$ ($\sqrt{4\pi}$) for a dimensionless four-point (three-point) interaction, while it should be less than $\sqrt{4\pi}$ times the mass of the heaviest particle participating in the interaction for a dimensionful three-point interaction. Moreover, we also impose an additional constraint on the scenarios with a scalar mediator (i.e., SS-R, FS-R, SS-F, FS-F), requiring that the scalar potentials in the models should be bounded from below and the SM vacuum should be the global minimum, as discussed in appendix\,\ref{app: vacuum stability}. On the other hand, we do not impose the constraint from the direct dark matter detection because the signal strength is smaller than the present detection sensitivity in the resonance and forbidden scenarios, as mentioned in section\,\ref{sec: Direct Detection Light}.

The main goal of this article is to find viable parameter sets of the light WIMP models that will be explored in the future MeV gamma-ray observatory, such as the COSI. To avoid a complicated interpretation of the statistical analysis, we impose the shape 2$\sigma$ level cut for each condition/constraint in Table\,\ref{tab: conditions & constraints} on the light WIMP scenarios.\footnote{
    Refer to Ref.\,\cite{Balan:2024cmq} to see the difference in the results using various statistical analysis strategies, where the frequentist and Bayesian analyses were performed on the light WIMP scenarios and discuss their results.}
We performed MCMC sampling using the emcee code\,\cite{Foreman-Mackey:2012any} to correct viable parameter sets efficiently and focus on several physical observables to discuss the future prospect of accelerator, direct, and indirect detections for the light WIMP. In the following, we first discuss the viable parameter regions in the forbidden scenarios in section\,\ref{subsec: Forbidden scenarios}, and next the resonance scenarios in section\,\ref{subsec: Resonance scenarios}, and clarify which parts of the regions will be explored in the future experiments.

\subsection{The forbidden scenarios}
\label{subsec: Forbidden scenarios}

We first consider the forbidden (i.e., SS-F, FS-F, SV-F, and FV-F) scenarios, in which the light WIMP primarily undergoes annihilation into a pair of mediator particles, where the mediator particle is slightly heavier than the WIMP, and the mediators subsequently decay into various SM particles.
So, the mass of the mediator particle is parameterized as
\begin{align}
    m_{\rm MED} \equiv
    m_{\rm DM} \left( 1 + v_{\rm th}^2/8 \right),
\end{align}
where $v_{\rm th}$ represents the position of the threshold on the relative velocity between the incident WIMPs. The typical velocity dependence of the annihilation cross-section in the forbidden scenarios is illustrated in the upper right panel of Fig.\,\ref{fig: sigmav_velocity_dependent}; when the relative velocity of the WIMPs $\langle v_{\rm DM} \rangle$ is less than $v_{\rm th}$, the annihilation cross-section is highly suppressed, because the WIMPs cannot annihilate into a pair of mediator particles. It is worth notifying that the cross-section is always suppressed by $v_{\rm Med} \equiv (v^2 - v_{\rm th}^2)^{1/2}$ in the vicinity of the threshold (i.e., $v \sim v_{\rm th}$) due to the phase space factor of the final state, with $v$ being the relative velocity.

Considering the relic abundance condition ($\langle \sigma v \rangle = {\cal O}(10^{-26})$\,cm$^3$/s at $v^2 = {\cal O}(0.1)$), the cross-section at the present universe ($v^2 = 10^{-6}$) is expected to be $\langle \sigma v \rangle \lesssim {\cal O}(10^{-28})$\,cm$^3$/s (due to the $v_{\rm Med} \lesssim 10^{-2}$ suppression) when the WIMP annihilates in s-wave, while $\langle \sigma v \rangle \lesssim {\cal O}(10^{-33})$ cm$^3$/s when annihilating in p-wave; the latter is below the sensitivity of the COSI, so we do not consider the {\bf FS-F} scenarios in the following analysis. Meanwhile, although the annihilation proceeds in s-wave, the cross-section is suppressed by $v_{\rm th}^2$ in the FV-F scenario, as seen in eq.\,(\ref{eq: FS II annihilation}). Since $v_{\rm th}$ must be smaller than ${\cal O}(10^{-3})$ to make the annihilation occur in the present universe, the cross-section is expected to be $\langle \sigma v \rangle \lesssim {\cal O}(10^{-34})$\,cm$^3$/s; it is less than the projected sensitivity of the COSI so that we also do not consider the {\bf FV-F} scenario in the analysis. Hence, we discuss the {\bf SS-F} and {\bf SV-F} scenarios in this subsection.

\subsubsection{The SS-F scenario}

In addition to the WIMP and mediator particle masses (i.e., $v_{\rm th})$, the independent model parameters describing the scenario are summarized in Table\,\ref{tab: parameters}. The SS-F scenario features a plethora of free parameters, including couplings among the WIMP, mediator particle, and the Higgs boson. We vary the eight parameters \mbox{\boldmath $(m_\phi,\,v_{\rm th},\,C_{\varsigma \phi \phi},\,C_{\varsigma \varsigma \phi \phi},\,C_{h \phi \phi},\,\sin\theta,\,C_{\varsigma \varsigma h},\,C_{\varsigma \varsigma \varsigma})$} in the MCMC sampling, as physical observables related to the conditions and constraints discussed in the previous section\,\ref{sec: conditions and constraints} depend on these parameters. The remaining two parameters \mbox{\boldmath $\lambda_\phi$} and \mbox{\boldmath $\lambda_S$} are the couplings of self-interactions among the WIMPs and the mediator particles, respectively, and those are only relevant for the vacuum stability condition. So, given that the former eight parameters pass the conditions and constraints in the previous section, we impose the vacuum stability condition on the latter two parameters. 

\begin{table}[t]
    \centering
    {\begin{tabular}{l|llllllllll}
    & \multicolumn{2}{l}{Masses}
    & \multicolumn{2}{l}{DM-MED Int.}
    & DM-SM Int.
    & \multicolumn{2}{l}{MED-SM Int.}
    & \multicolumn{3}{l}{Self Int.} \\
    \hline
    SS-F &
    $m_\phi$ & $v_{\rm th}$ &
    $C_{\varsigma \phi \phi}$ & $C_{\varsigma \varsigma \phi \phi}$ &
    $C_{h \phi \phi}$ &
    $\sin\theta$ & $C_{\varsigma \varsigma h}$ &
    $\lambda_\phi$ & $C_{\varsigma \varsigma \varsigma}$ & $\lambda_S$ \\
    SV-F &
    $m_\varphi$ & $v_{\rm th}$ &
    $g_{\varphi}$ & &
    $\lambda_{h\varphi\varphi}$ &
    $\xi$ &  &
    $\lambda_\varphi$ & & \\
    SV-R &
    $m_\varphi$ & $\delta$ &
    $g_\varphi$ & &
    $\lambda_{h\varphi\varphi}$ &
    $\xi$ & &
    $\lambda_\varphi$ & & \\
    FV-R &
    $m_\psi$ & $\delta$ &
    $g_\psi$ & &
    &
    $\xi$ & &
    & & \\
    SV(B)-R &
    $m_\varphi$ & $\delta$ &
    $g_\varphi$ & &
    $\lambda_{h\varphi\varphi}$ &
    $\xi$ & $g_{\rm B}$ &
    $\lambda_\varphi$ & & \\
    FV(B)-R &
    $m_\psi$ & $\delta$ &
    $g_\psi$ & &
    &
    $\xi$ & $g_{\rm B}$ &
    & & \\
    \hline
    \end{tabular}
    }
    \caption{\sl\small Model parameters describing the light WIMP scenarios. See the text for more details.}
    \label{tab: parameters}
\end{table}

In the MCMC sampling, we vary the WIMP mass in the range of $0 < m_\phi \leq 100$\,MeV, while the mediator particle mass in the range of $10^{-5} \leq v_{\rm th} \leq 10^{-1}$. Note that a too small $v_{\rm th}$ is not compatible with the CMB limit given in section\,\ref{subsubsec: the scenrios}, while a large $v_{\rm th} \gg 10^{-3}$ prevents the WIMP from annihilating in the present universe. The mixing parameter is varied in the range of $10^{-10} \leq \sin\theta \leq 10^{-1}$. Other "$C$" couplings are varied according to the perturbative unitarity limits discussed in appendix\,\ref{app: unitarity}. The result of the sampling is presented in Fig.\,\ref{fig:  scan_SS-F}. First, it is found that the lower limit on the WIMP mass is given about 6\,MeV due to the CMB limit addressed in section\,\ref{subsubsec: mass CMB}. Next, the mixing angle $\sin\theta$ is restricted to be in a narrow range, i.e., between 10$^{-4}$ and 10$^{-5}$; the upper limit is from the accelerator constraint in section\,\ref{sec: Accelerator Light}, while the lower limit is from the kinematic equilibrium condition in section\,\ref{subsubsec: thermal relics}. Third, upper limits on the couplings $C_{h \phi \phi}$ and $C_{\varsigma \varsigma h}$ are obtained by the accelerator constraint in section\,\ref{sec: Accelerator Light}, which concerns the invisible decay width of the Higgs boson. Forth, the other "C" couplings, $C_{\varsigma \phi \phi}$, $C_{\varsigma \varsigma \phi \phi}$, and $C_{\varsigma \varsigma \varsigma}$, are constrained by the relic abundance condition, so some correlations among the parameters are seen in the figure. Finally, the region with small $m_\varphi$ and $v_{\rm th}$ is excluded because of the indirect dark matter detection constraint utilizing gamma-ray observations discussed in section\,\ref{sec: Indirect Detection Light}.

\begin{figure}[t]
    \centering
    \includegraphics[keepaspectratio, scale=0.17]{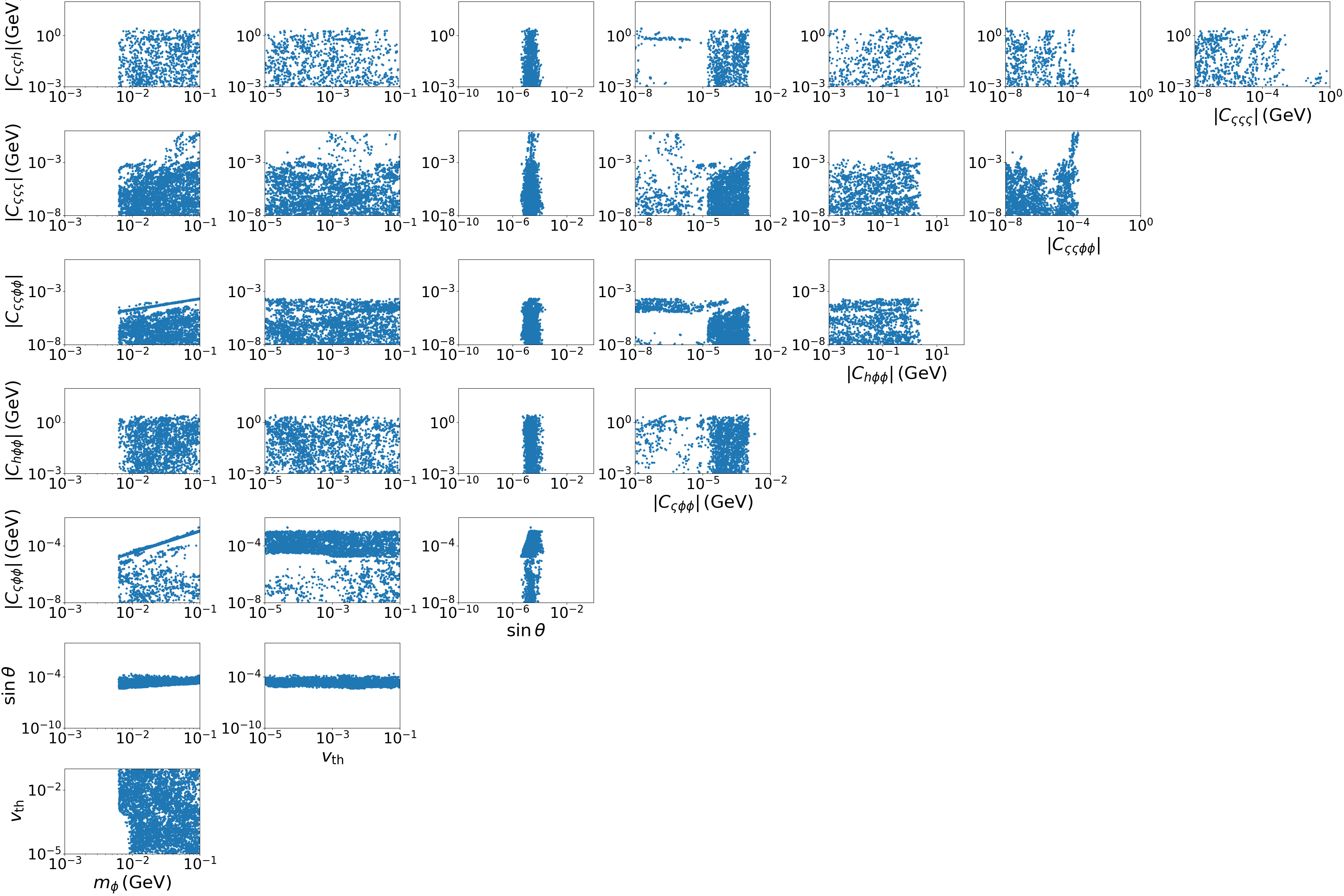}
    \caption{\small \sl The viable parameter region of the SS-F scenario obtained in the MCMC sampling by imposing the conditions and constraints shown in Table\,\ref{tab: conditions & constraints}. See the main text for more details.}
    \label{fig: scan_SS-F}
\end{figure}

\begin{figure}[t]
    \centering
    \includegraphics[keepaspectratio, scale=0.42]{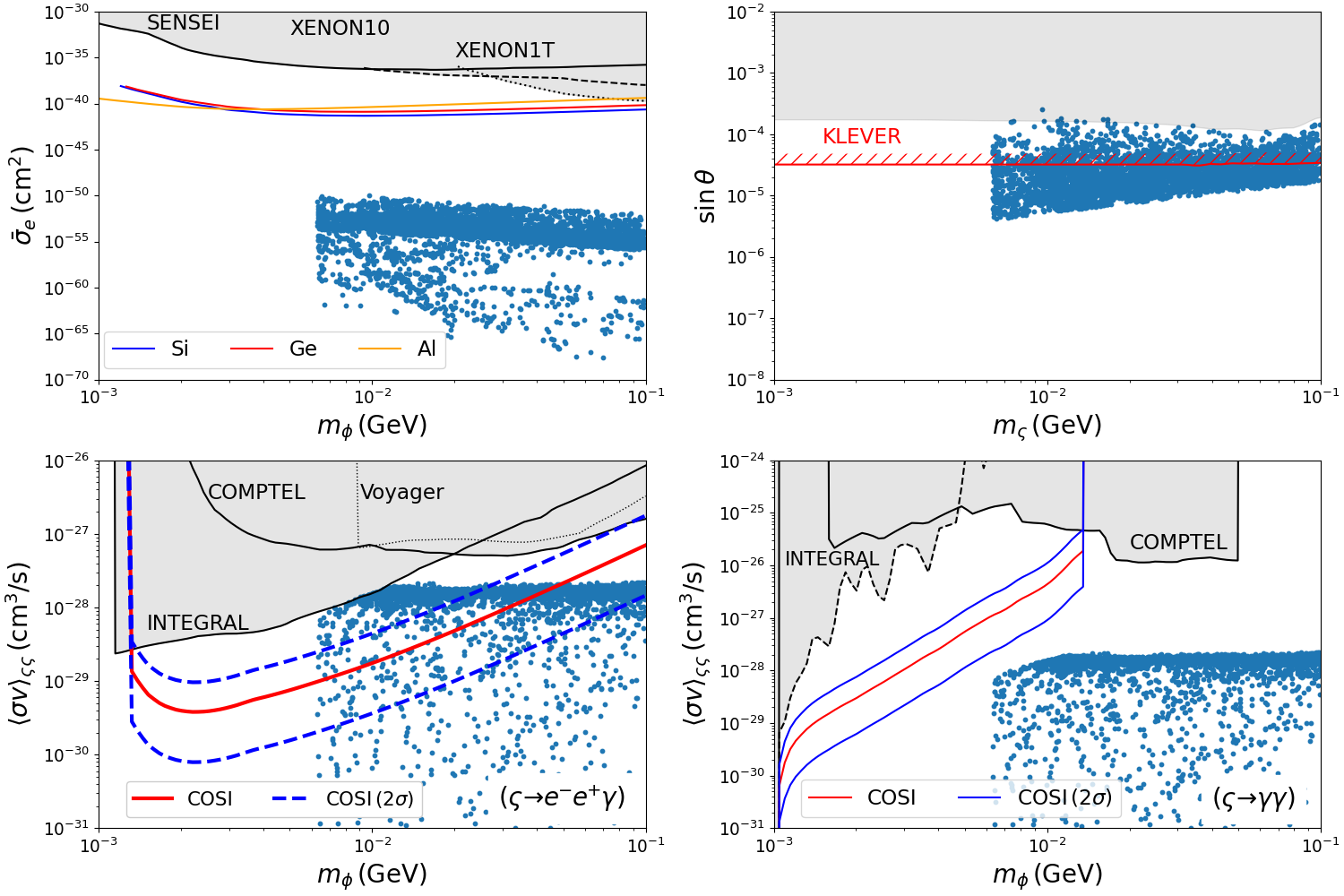}
    \caption{\small \sl The prospect of light WIMP detection in the {\bf SS-F} scenario at various future dark matter and mediator particle search experiments. The prospect at the direct and accelerator detections are given in the top left and right panels, respectively, while the prospect at the indirect detection is given in the bottom panels; the prospect at the continuum gamma-ray observation is given in the bottom left panel, and that at the line gamma-ray observation is in the bottom right panel.}
    \label{fig: experiments_SS-F}
\end{figure}

The viable parameter region in Fig.\,\ref{fig: scan_SS-F} is compared with the projected sensitivities of various future dark matter and mediator particle detection experiments in Fig.\,\ref{fig: experiments_SS-F}. The top left panel shows the prospect for light WIMP detection (in the SS-F scenario) at the underground experiments (direct dark matter detection). As mentioned in section\,\ref{sec: Direct Detection Light}, due to the small mixing angle and the small electron Yukawa coupling, the scattering cross-section between the WIMP and an electron is highly suppressed, rendering the detection difficult. The top right panel shows the prospect for the detection at the accelerator experiments (accelerator dark matter detection), indicating that the KLEVER experiment will explore a large part of the viable parameter region in the future. The bottom panels show the prospect of detection at the astrophysical observations (indirect dark matter detection). The bottom left panel is those observing continuum gamma-rays, originating in the WIMP annihilation $\phi \phi \to \varsigma \varsigma$ \& $\varsigma \to e^- e^+ \gamma$. Meanwhile, the bottom right is those observing line gamma-rays, which originate in $\phi \phi \to \varsigma \varsigma$ \& $\varsigma \to 2 \gamma$. It is found that the COSI observation will explore a substantial part of the viable parameter region via the continuum gamma-ray search, while it is not obvious that the line gamma-ray signal can be detected at the observation. The line gamma-ray signal, however, would be detected if the dark matter profile at the galactic center is (slightly) more cuspy than that of the NFW, as deduced by the prospect in the bottom right panel. It is also worth notifying that the COSI observation will cover the region not explored by the accelerator experiment, as the (former observation) latter experiment is (in)sensitive to the mixing angle $\sin \theta$. So, the two detections play complementary roles.

\subsubsection{The SV-F scenario}
\label{subsubsec: SV-F}

The independent model parameters describing the scenario are summarized in Table\,\ref{tab: parameters}, and we vary the four parameters \mbox{\boldmath $(m_\varphi,\,v_{\rm th},\,\xi,\,g_{\varphi})$} in the MCMC sampling. The self-interaction coupling \mbox{\boldmath $\lambda_\varphi$} is only relevant for the vacuum stability, so we treat it in the same manner as $\lambda_\phi$ and $\lambda_S$ in the SS-F scenario. The other remaining parameter \mbox{\boldmath $\lambda_{h \varphi \varphi}$} is also not included in the sampling, as physical observables related to the conditions and constraints in the previous section depend very weekly on it, unlike the parameter $C_{h \phi \phi}$ in the SS-F scenario.

In the MCMC sampling, the WIMP and mediator particle masses are varied in the same ranges as those in the SS-F scenario, $0 < m_\varphi \leq 100$\,MeV, and $10^{-5} \leq v_{\rm th} \leq 10^{-1}$, while the kinetic mixing parameter is varied in the range of $10^{-11} \leq \xi \leq 10^{-1}$. The coupling between the WIMP and the mediator particle $g_\varphi$ is varied according to the perturbative unitarity limits discussed in appendix\,\ref{app: unitarity}. The result of the sampling is shown in Fig.\,\ref{fig: scan_SV-F}. First, the lower limit on the WIMP mass is found to be around twice the electron mass, $m_\varphi \sim 2 m_e$, which is from the kinematic equilibrium condition in section\,\ref{subsubsec: thermal relics}; the mediator particle can decay into an electron pair and have a enough large width. It is worth notifying that we implicitly assume that the mediator particle in the SV-F scenario also couples to neutrinos via a tiny U(1)$_{\rm B-L}$ gauge coupling, which does not affect physics of the WIMP except that of $N_{\rm eff}$ and alleviates the lower bound on the WIMP mass from the CMB observation, as discussed in section\,\ref{subsubsec: mass CMB}. Next, the mixing parameter $\xi$ is constrained in a non-trivial way, as the accelerator constraint from the beam dump experiments gives a wedge-shaped constraint as shown in section\,\ref{sec: Accelerator Light}, resulting in two separated regions survived. The lower limit on the mixing parameter at $\sim 10^{-10}$ stems from the same reason as the SS-F scenario, the kinematic equilibrium condition. Third, the relic abundance condition determines the coupling between the WIMP and the mediator particle $g_{\varphi}$. Finally, as seen in the SS-F scenario, the region with small $m_\varphi$ and $v_{\rm th}$ is excluded because of the indirect dark matter detection constraint utilizing the MeV gamma-ray and X-ray observations discussed in section\,\ref{sec: Indirect Detection Light}.

\begin{figure}[t]
    \centering
    \includegraphics[keepaspectratio, scale=0.19]{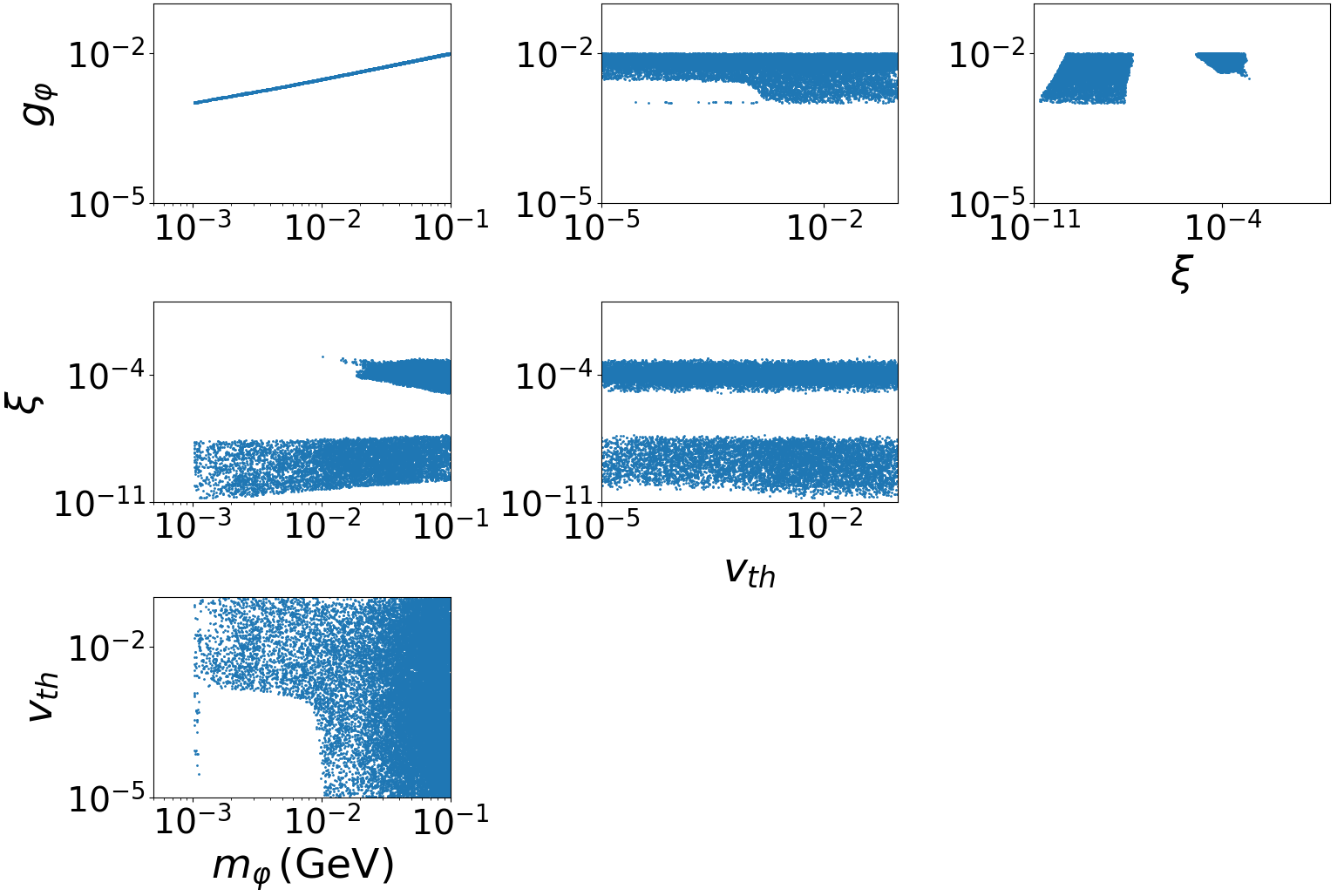}
    \caption{\small \sl The viable parameter region of the SV-F scenario obtained in the MCMC sampling by imposing the conditions and constraints shown in Table\,\ref{tab: conditions & constraints}. See the main text for more details.}
    \label{fig:  scan_SV-F}
\end{figure}

\begin{figure}[t]
    \centering
    \includegraphics[keepaspectratio, scale=0.42]{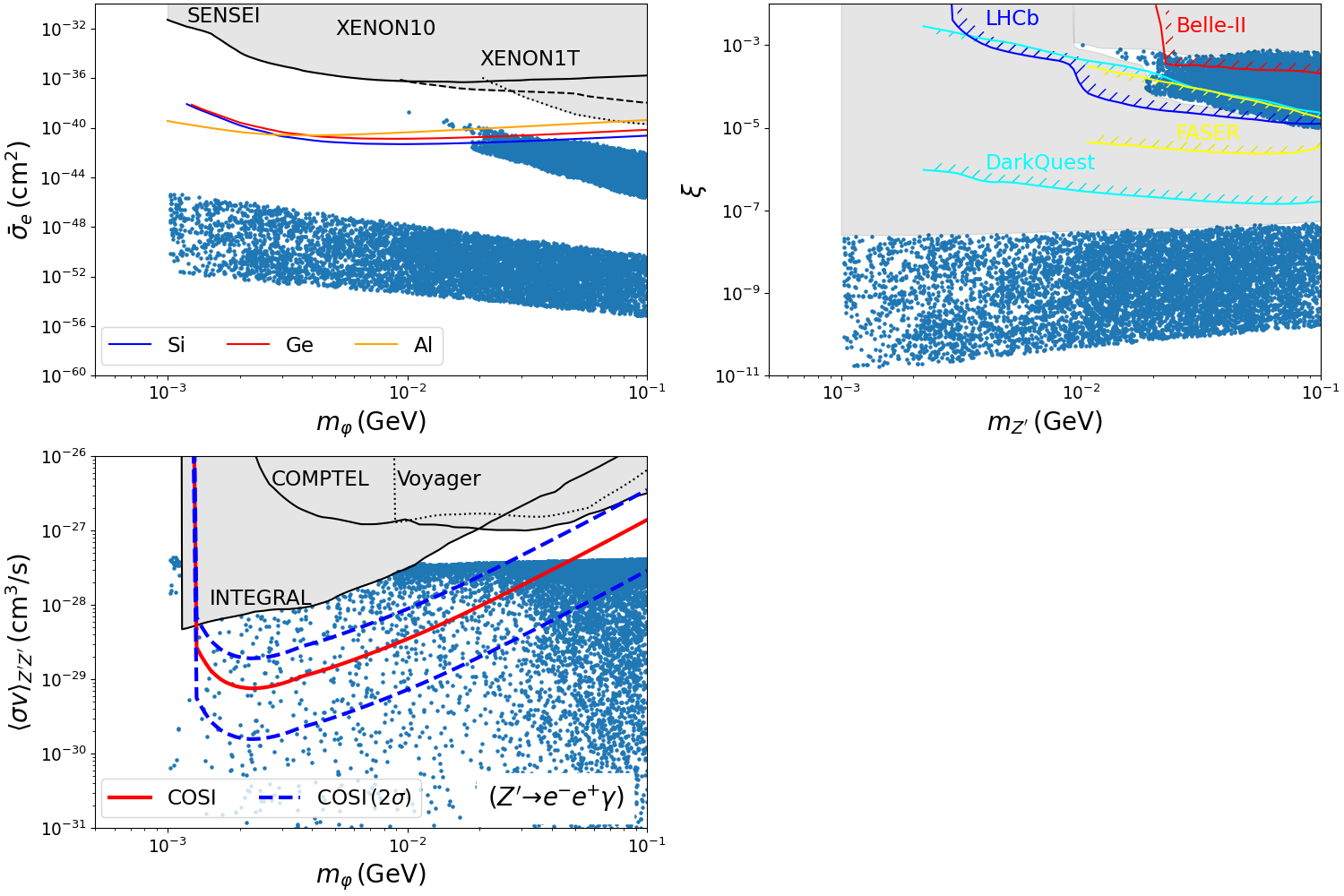}
    \caption{\small \sl The prospect of light WIMP detection in the {\bf SV-F} scenario at various future dark matter and mediator particle search experiments. The prospect at the direct and accelerator dark matter detections are given in the top left and right panels, respectively, while that at the indirect detection is in the bottom left panel; the prospect at the continuum gamma-ray observation is shown.}
    \label{fig:  experiments_SV-F}
\end{figure}

The viable parameter region in Fig.\,\ref{fig:  scan_SV-F} is compared with the projected sensitivities of the future experiments in Fig.\,\ref{fig:  experiments_SV-F}. The top left panel shows the prospect for light WIMP detection (in the SV-F scenario) at the underground experiments (direct dark matter detection). Unlike the previous SS-F scenario, the scattering cross-section between the WIMP and an electron is not suppressed by the small electron Yukawa coupling, so it will be possible to search for the WIMP in future experiments if the WIMP is light enough and the mixing parameter $\xi$ is not highly suppressed. The top right panel shows the prospect for the detection at accelerator experiments (accelerator dark matter detection), indicating that near-future experiments will fully explore the parameter region with $\xi \sim 10^{-4}$. The bottom left panel shows the prospect at the astrophysical observations (indirect dark matter detection), originating in the WIMP annihilation $\varphi \varphi^* \to Z' Z'$ \& $Z' \to e^- e^+ \gamma$. The COSI observation will explore a substantial part of the viable parameter region, which involves those not explored by the accelerator dark matter detection, as the signal strength of the indirect dark matter detection does not depend on $\xi$. So, the two detections play complementary roles. It is also worth notifying that the light WIMP in the SV-F scenario does not predict a line gamma-ray signal when its mass is below 100\,MeV, as deduced from the discussed in section\,\ref{subsubsec: Scalar WIMP and Vector Mediator}.

\subsection{The resonance scenarios}
\label{subsec: Resonance scenarios}

We next consider the resonance (i.e., SS-R, FS-R, SV-R, FV-R, SV(B)-R, and FV(B)-R) scenarios, in which the light WIMP primarily undergoes annihilation into the SM particles through the $s$-channel resonance of the mediator particle. So, based on a similar philosophy to the forbidden scenarios, we parameterize the mass of the mediator particle as follows:
\begin{align}
    m_{\rm MED} \equiv
    2m_{\rm DM}\left( 1 + \delta/8 \right).
\end{align}
Here, when the parameter $\delta$ is {\bf positive}, $\delta^{1/2}$ represents the relative velocity between the incident WIMPs hitting the pole of the mediator particle propagator. In such a case, the mediator particle decays invisibly into a pair of the WIMP and visibly into the SM particles. The typical velocity dependence of the annihilation cross-section (in the resonance scenarios with $\delta > 0$) is illustrated in the bottom left panel of Fig.\,\ref{fig: sigmav_velocity_dependent}; when the relative velocity of the WIMPs is larger than $\delta^{1/2}$, the annihilation cross-section is enhanced thanks to the resonance. On the other hand, the cross-section is exponentially suppressed (compared to that with the resonance) when the relative velocity is much smaller than $\delta^{1/2}$, as it becomes hard to hit the pole by the WIMPs, which leads to the traditional annihilation in s-wave or p-wave. This behavior suggests, even after the freeze-out, the relic density of the WIMP keeps decreasing due to the resonant enhancement of the annihilation cross-section until the era that the relative velocity of the WIMPs becomes enough smaller than $\delta^{1/2}$.

On the other hand, when $\delta$ is {\bf negative}, the incident WIMPs never hit the pole of the mediator particle propagator. In this case, the mediator particle is slightly lighter than twice the WIMP mass and exclusively decays into the SM particles. Interestingly, the annihilation cross-section can have a non-trivial velocity dependence even in such a case. The typical velocity dependence of the annihilation cross-section (in the resonance scenarios with $\delta < 0$) is depicted in the bottom right panel of Fig.\,\ref{fig: sigmav_velocity_dependent}; when the relative velocity of the WIMPs is large enough, the annihilation cross-section is enhanced thanks to the resonance, as in the case of $\delta > 0$. Subsequently, the cross-section stops increasing when the velocity becomes as small as $(-\delta)^{1/2}$, for the WIMPs no longer approach the pole of the propagator. After that, i.e., when the velocity is smaller enough than $(-\delta)^{1/2}$, the cross-section obeys the traditional annihilation in s-wave or p-wave. Since the annihilation in s-wave is not suppressed in the recombination epoch (compared to that in the freeze-out epoch), it is ruled out by the CMB observation discussed in section\,\ref{subsubsec: the scenrios}. Hence, when $\delta < 0$, we only consider the WIMP scenarios that the WIMP annihilates in p-wave, i.e., the {\bf FS-R} and {\bf SV-R} scenarios.

Let us first consider the scenarios with the scalar mediator ({\bf SS-R} and {\bf FS-R} scenarios), where the WIMP primarily annihilates into $e^- e^+$. The annihilation cross-section is highly suppressed both by the small electron Yukawa coupling and the mixing angle, as discussed in section\,\ref{subsec: scalar mediator}. So, the cross-section needs to be highly enhanced for an extended period to satisfy the relic abundance condition; we numerically confirmed that $|\delta|^{1/2}$ should be as small as $10^{-3}$--$10^{-5}$. Such a small $\delta^{1/2}$, however, makes the cross-section at the present universe highly boosted, and no viable parameter region can be found because of the indirect dark matter detection constraint addressed in section\,\ref{sec: Indirect Detection Light}. Hence, we discuss the scenarios with a vector mediator (i.e., {\bf SV-R}, {\bf FV-R}, {\bf SV(B)-R}, and {\bf FV(B)-R} scenarios) in this subsection.

\subsubsection{The SV-R scenario}
\label{subsubsec: SV-R}

The independent model parameters describing the scenario are the same as those in the SV-F scenario, except that we use $\delta$ (instead of $v_{\rm th}$) for the mass of the mediator particle, as shown in Table\,\ref{tab: parameters}. So, we take the strategy for the sampling as that in section\,\ref{subsubsec: SV-F}.

\begin{figure}[t]
    \centering
    \includegraphics[keepaspectratio, scale=0.187]
    {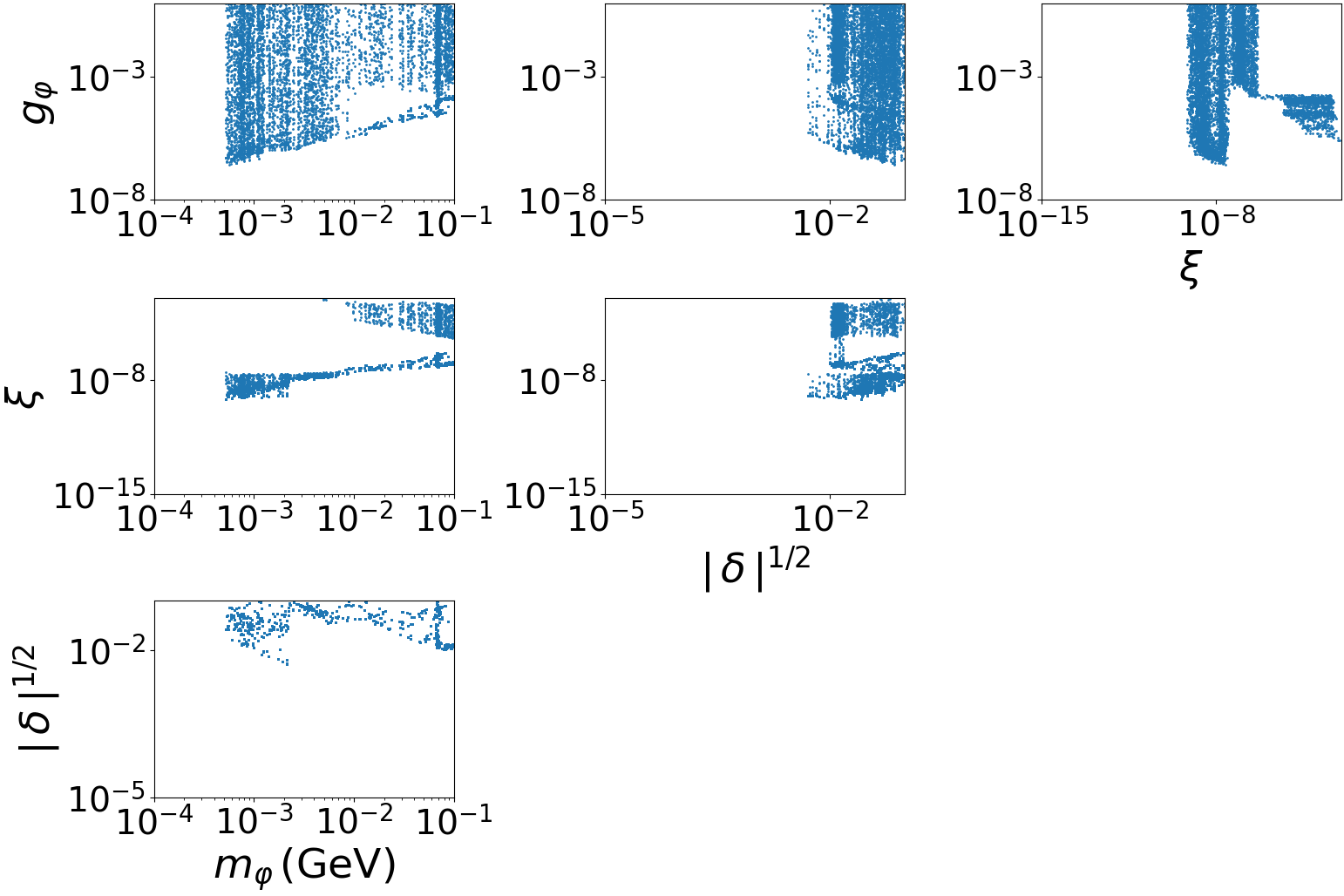}
    \qquad
    \includegraphics[keepaspectratio, scale=0.187]{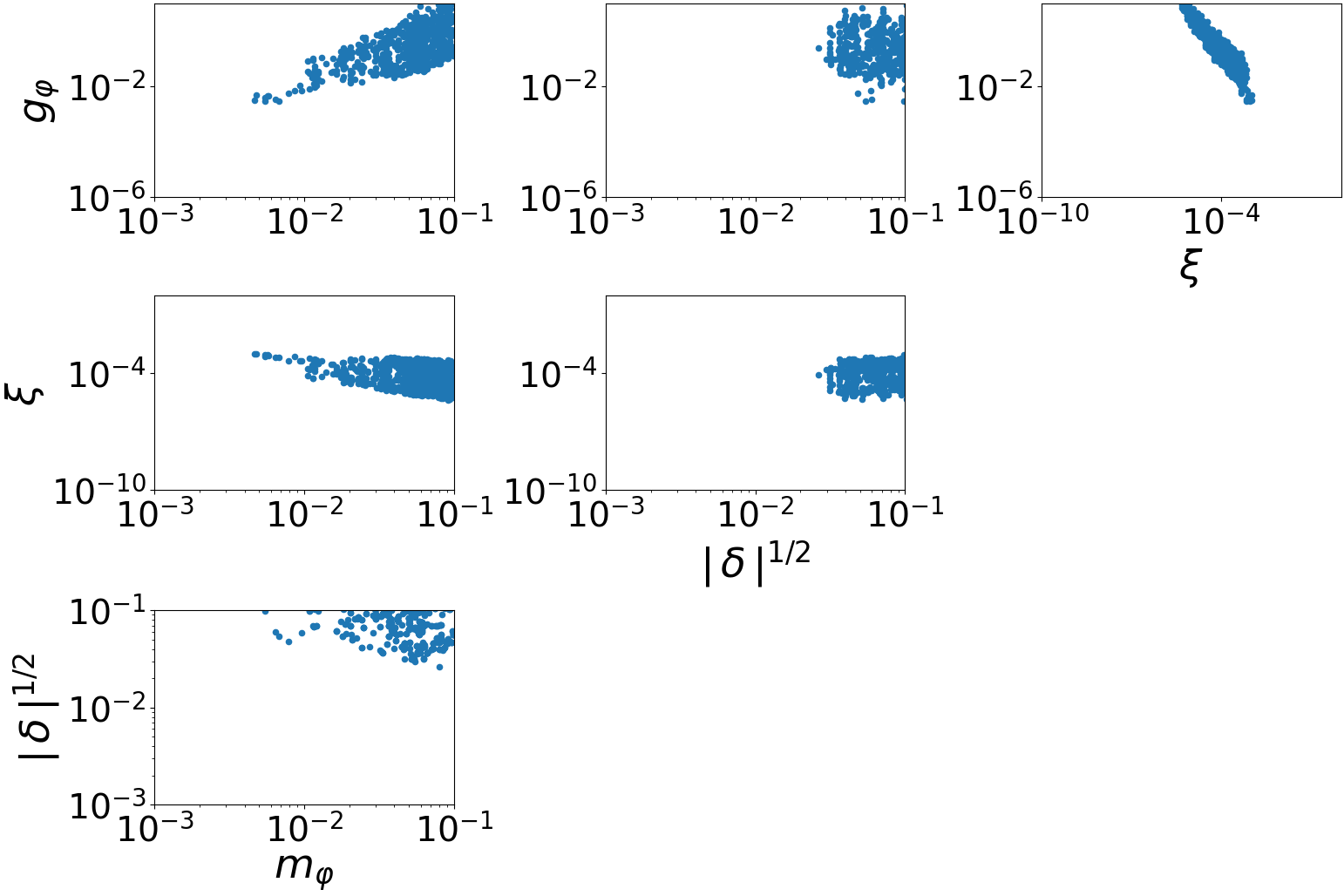}
    \caption{\small \sl 
    The viable parameter region of the SV-R scenario obtained in the MCMC sampling by imposing the conditions and constraints shown in Table\,\ref{tab: conditions & constraints}. The region with $\delta > 0$ is shown in the left 6 panels, while that with $\delta <0$ is in the right 6 panels. See the main text for more details.}
    \label{fig: scan_SV-R-inv-DP}
\end{figure}

In the MCMC sampling, the WIMP and the mediator particle masses are varied in the ranges of $0 < m_\varphi \leq 100$\,MeV and $10^{-3} \leq |\delta|^{1/2} \leq 10^{-1}$, respectively, while the kinetic mixing angle parameter is varied in the range of $10^{-11} \leq \xi \leq 10^{-1}$. The coupling between the WIMP and the mediator particle $g_\varphi$ is varied according to the perturbative unitary limit addressed in appendix\,\ref{app: unitarity}. The result of the sampling with $\delta > 0$ is shown in the left six panels of Fig.\ref{fig: scan_SV-R-inv-DP}. First, it is found that the lower limit on the WIMP mass is around the electron mass, which is from the BBN constraint in section\,\ref{subsubsec: mass BBN} and the kinematic equilibrium condition in section\,\ref{subsubsec: thermal relics}. Next, the lower limit on $\delta^{1/2}$ is found to be around $10^{-2}$ when $m_\varphi \gtrsim$ a few MeV due to the BBN constraint on the annihilation cross-section in section\,\ref{subsubsec: BBN on annihiation}, while it is around 10$^{-3}$ when $m_\varphi \lesssim$ few MeV, which is from the indirect dark matter detection constraint in section\,\ref{sec: Indirect Detection Light}. Finally, the mixing parameter $\xi$ and the coupling $g_\varphi$ are constrained in a non-trivial way because of the accelerator constraint in section\,\ref{sec: Accelerator Light}, resulting in two separated viable parameter regions and giving a void structure on several panels in the figure. The two regions have distinct characteristics: The annihilation cross-section behaves as $\sigma v (\varphi \varphi^* \to e^- e^+) \propto [\Gamma (Z' \to e^- e^+)^{-1} + \Gamma (Z'\to \varphi \varphi^*)^{-1}]^{-1}$ in a narrow width approximation, hence the relic abundance is determined by the smaller width among the two widths $\Gamma (Z' \to e^- e^+)$ and $\Gamma(Z' \to \varphi \varphi^*)$. The former (latter) width dominates in the region with $\xi \gtrsim 10^{-5}$ ($\xi \lesssim 10^{-8}$), and the mediator particle decays visibly (invisibly). It is also worth notifying that lower bounds on the parameters $\xi$ and $g_\varphi$ are from the relic abundance condition as well as the lower limit on $\delta^{1/2}$ because a small $\delta^{1/2}$ leads to a highly enhanced WIMP annihilation cross-section and it requires small couplings to obtain the correct relic abundance.

The result of the MCMC sampling with $\delta <0$ is shown in the right six panels in Fig.\,\ref{fig: scan_SV-R-inv-DP}. First, it is found that the kinetic mixing parameter $\xi$ takes a value of around $10^{-4}$ because of the relic abundance condition and the accelerator constraint: Since the mediator particle always decays visibly when $\delta < 0$, the accelerator constraint requires $\xi$ being around $10^{-4}$ or below about $10^{-8}$, as addressed in section\,\ref{subsubsec: thermal relics}. On the other hand, the WIMP annihilation cross-section is not much enhanced by the resonance compared to the $\delta > 0$ case, so only the region $\xi \sim 10^{-4}$ survives after the relic abundance condition in section\,\ref{sec: Accelerator Light} is imposed. In addition, the accelerator constraint gives a lower limit on the mediator particle mass, as seen in Fig.\,\ref{fig: accelerator constraint}, and it leads to a lower limit on the WIMP mass because we focus on the resonance region $m_{Z'} \simeq 2 m_\varphi$. Next, the lower limit on $|\delta|$ stems from the BBN constraint on the annihilation cross-section discussed in section\,\ref{subsubsec: BBN on annihiation}. Finally, the relic abundance condition determines the coupling between the WIMP and the mediator particle $g_\varphi$.

\begin{figure}[t]
    \centering
    \includegraphics[keepaspectratio, scale=0.42]{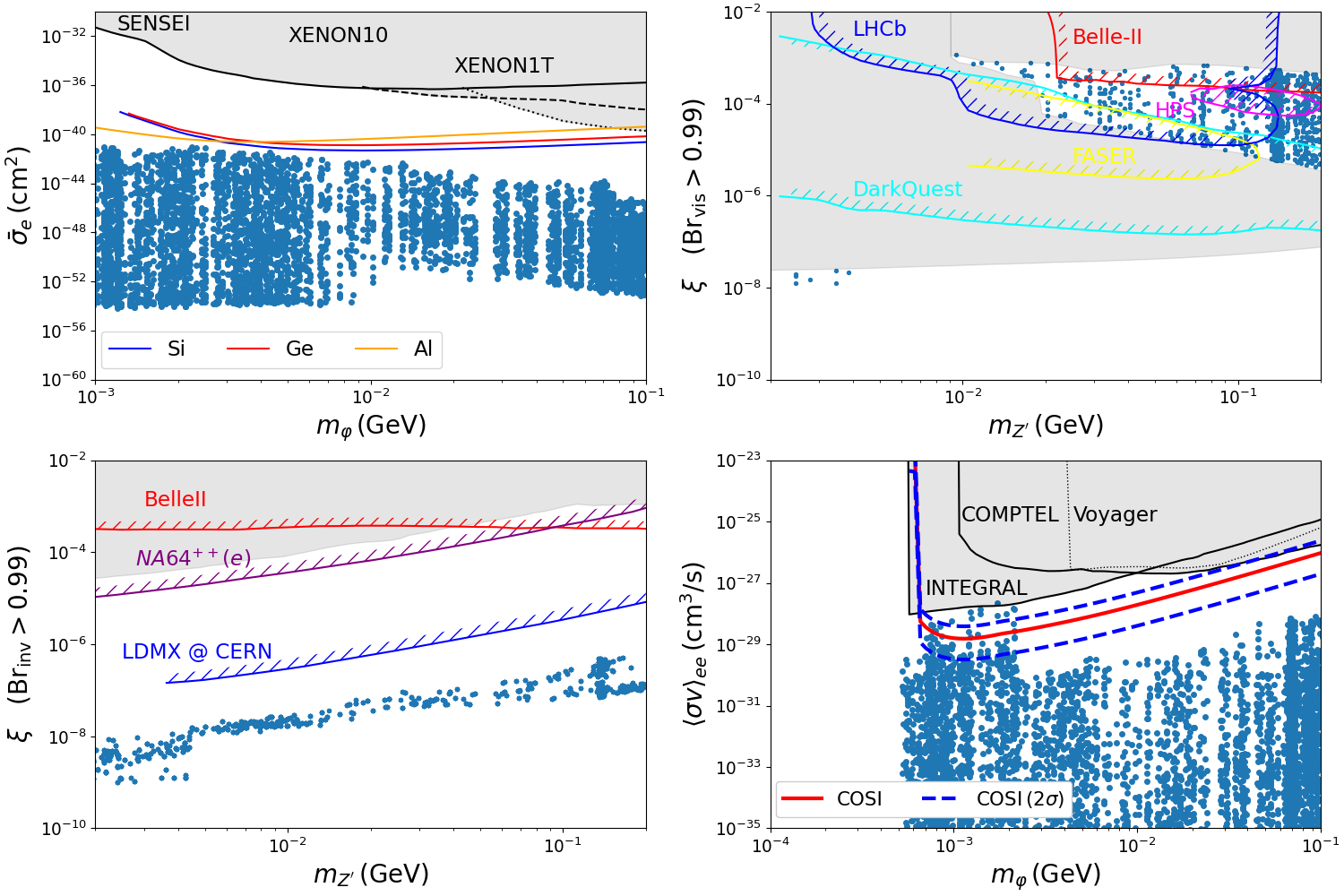}
    \caption{\small \sl The prospect of light WIMP detection in the {\bf SV-R} scenario with \mbox{\boldmath $\delta > 0$} at various future dark matter and mediator particle search experiments. The prospect at the direct detection is in the top left panel, while those at the accelerator detection (utilizing the visible and invisible decay modes of the mediator particle) are in the top right and bottom left panels, respectively. The prospect at the indirect detection (utilizing the continuum gamma-ray observation) is in the bottom right panel.}
    \label{fig:  experiments_SV-R-inv-DP}
\end{figure}

The viable parameter region with $\delta > 0$ in Fig.\,\ref{fig: scan_SV-R-inv-DP} is compared with the projected sensitivities of the future experiments in Fig.\,\ref{fig: experiments_SV-R-inv-DP}. The top left panel shows the prospect for the light WIMP detection (in the SV-R scenario with $\delta >0$) at underground experiments (direct dark matter detection). The scattering cross-section between the WIMP and an electron is suppressed compared to the one in the SV-F scenario because the couplings $\xi$ and $g_\varphi$ are more suppressed to obtain the correct relic abundance via the resonant enhanced annihilation cross-section. So, it is challenging to detect the signal even in future experiments. The top right (bottom left) panel shows the prospect for the detection at accelerator experiments (accelerator dark matter detection) when the mediator particle decays visibly (invisibly); a large portion of the visible region will be explored by the near-future experiments, while the invisible region will not because of the smallness of the $\xi$ parameter. The bottom right panel shows the prospect at astrophysical observations (indirect dark matter detection), originating in the WIMP annihilation $\varphi \varphi^* \to e^- e^+ \gamma$. The COSI observation will explore appearing parameter sets with $m_\varphi = {\cal O}(1)$\,MeV that have survived the severe conditions and constraints. Such a parameter region predicts the invisible decay of the mediator particle, so the accelerator and the indirect dark matter detections play complementary roles.

\begin{figure}[t]
    \centering
    \includegraphics[keepaspectratio, scale=0.42]{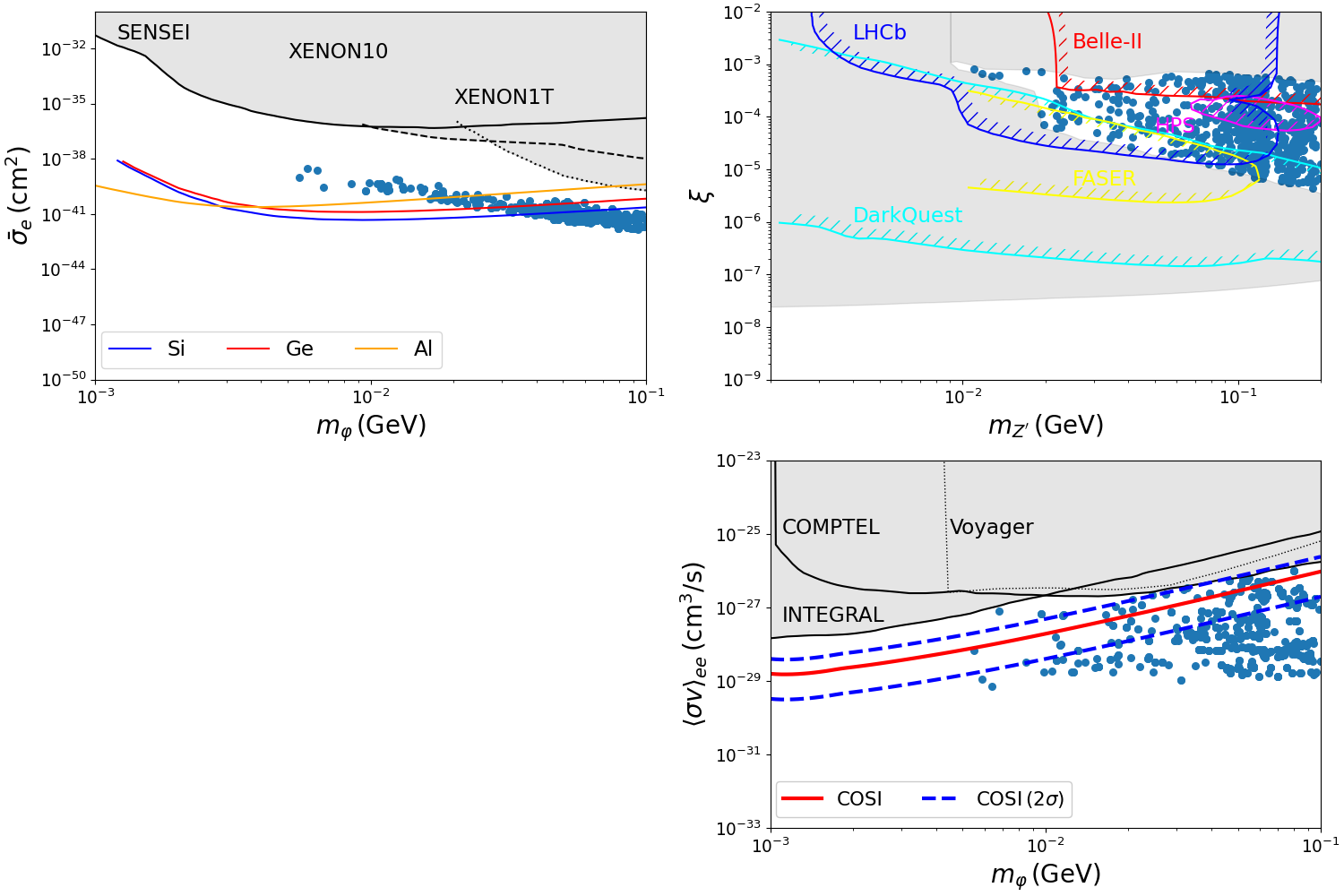}
    \caption{\small \sl The prospect of light WIMP detection in the {\bf SV-R} scenario with \mbox{\boldmath $\delta < 0$} at various future dark matter and mediator particle search experiments. The prospect at the direct dark matter detection is shown in the top left panel, and that at the accelerator detection (utilizing the visible decay mode of the mediator particle) is in the top right panel. The prospect at the indirect dark matter detection (utilizing the continuum gamma-ray observation) is shown in the bottom right panel.}
    \label{fig:  experiments_SV-R-vis}
\end{figure}

The viable parameter region with $\delta < 0$ is compared with the projected sensitivities of the future experiments in Fig.\,\ref{fig: experiments_SV-R-vis}. The top left panel shows the prospect for the light WIMP detection (in the SV-R scenario with $\delta < 0$) at underground experiments (direct dark matter detection). Since both the kinetic mixing parameter $\xi$ and the WIMP coupling with the mediator particle $g_\varphi$ are not suppressed compared to those in the same scenario with $\delta > 0$ (due to the smaller resonance enhancement on the WIMP annihilation), the scattering cross-section between the WIMP and an electron is not suppressed; the future experiments will explore a large part of the viable parameter region. The top right panel shows the prospect at accelerator experiments (accelerator dark matter detection) utilizing the visible decay of the mediator particle; a significant portion of the viable parameter region will be probed in the near future. The bottom right panel shows the prospect at astrophysical observations (indirect dark matter detection), originating in the process $\varphi \varphi^* \to e^- e^+ \gamma$. The COSI observation will also explore a large part of the parameter region in the near future.

\subsubsection{The FV-R scenario}

The independent model parameters describing the scenario are essentially the same as those in the SV-R scenario, as shown in Table\,\ref{tab: parameters}, except that the fermionic WIMP does not have some interactions that the scalar WIMP has, i.e., a direct interaction with the SM Higgs boson and the self-interaction. So, we take the strategy for the sampling as that in section\,\ref{subsubsec: SV-R}.

\begin{figure}[t]
    \centering
    \includegraphics[keepaspectratio, scale=0.187]
    {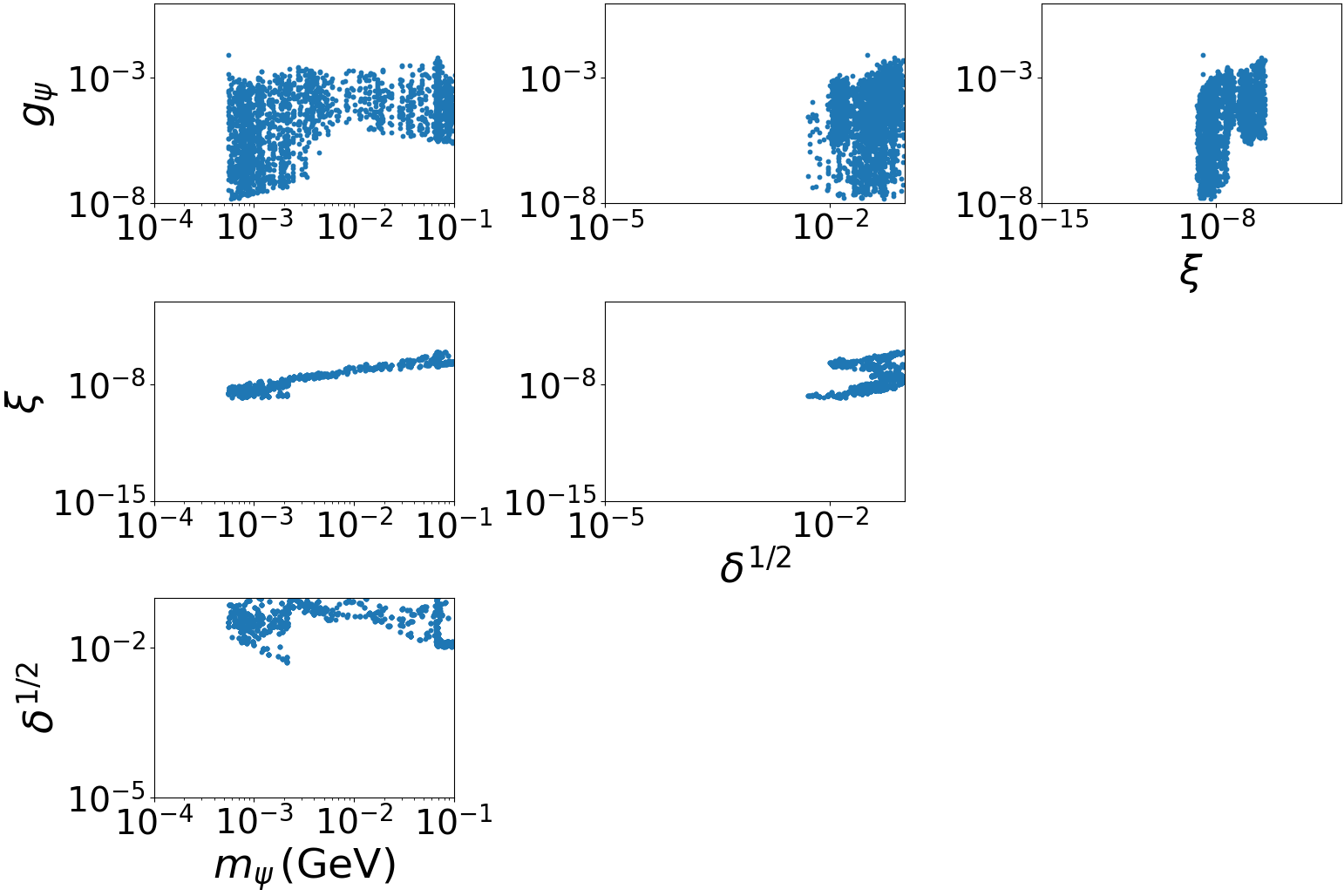}
    \caption{\small \sl 
    The viable parameter region of the FV-R scenario obtained in the MCMC sampling by imposing the conditions and constraints shown in Table\,\ref{tab: conditions & constraints}. See the main text for more details.}
    \label{fig: scan_FV-R-inv-DP}
\end{figure}

In the MCMC sampling, the WIMP mass, the mediator particle mass, and the kinetic mixing parameter are varied in the ranges of $0 < m_\psi \leq 100$\,MeV, $10^{-3} \leq \delta^{1/2} \leq 10^{-1}$, and $10^{-11} \leq \xi \leq 10^{-1}$, respectively. We only consider the parameter region $\delta > 0$, as addressed at the beginning of section\,\ref{subsec: Resonance scenarios}. The coupling between the WIMP and the mediator particle $g_\psi$ is varied according to the perturbative unitary limit in appendix\,\ref{app: unitarity}. The result of the sampling is more or less the same as that of the SV-R scenario with $\delta >0$, as shown in Fig.\ref{fig: scan_FV-R-inv-DP}. First, lower limits on the WIMP mass and $\delta^{1/2}$ are obtained for the same reason as those in the SV-R scenario. Next, unlike the SV-R case, the region with a large kinetic mixing parameter ($\xi \gtrsim 10^{-5}$, hence, a small $g_\psi$ because of the relic abundance condition) is forbidden, for $g_\psi$ in the FV-R scenario must be large enough to have a large self-scattering cross-section among the fermionic WIMPs, as addressed in section\,\ref{subsubsec: thermal relics}. Finally, the region with a large $g_\psi$ is also forbidden, as the fermionic WIMP in this scenario annihilates in s-wave. So, its cross-section is large even after the resonant regime (i.e., at the low WIMP velocity regime) when $g_\psi \gtrsim 10^{-3}$, contradicting with the cosmological constraint in section\,\ref{subsubsec: the scenrios}.

\begin{figure}[t]
    \centering
    \includegraphics[keepaspectratio, scale=0.42]{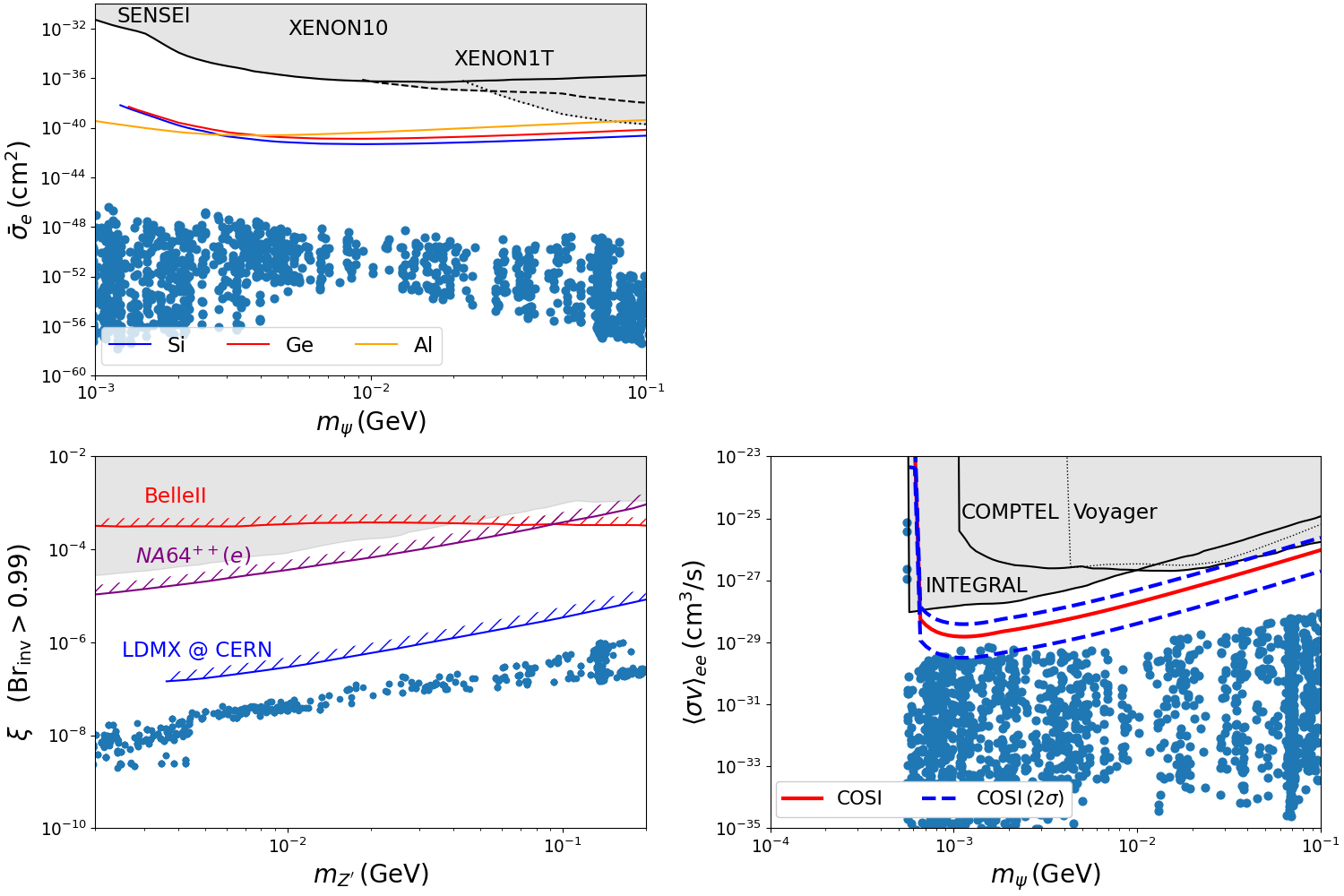}
    \caption{\small \sl The prospect of light WIMP detection in the {\bf FV-R} scenario at various future dark matter and mediator particle search experiments. The prospect at the direct dark matter detection is shown in the top left panel, and that at the accelerator detection (utilizing the invisible decay mode of the mediator particle) is in the bottom left panel. The prospect at the indirect dark matter detection (utilizing the continuum gamma-ray observation) is shown in the bottom right panel.
    }
    \label{fig:  experiments_FV-R-vis}
\end{figure}

The viable parameter region in Fig.\,\ref{fig: scan_FV-R-inv-DP} is compared with the projected sensitivities of the future experiments in Fig.\,\ref{fig:  experiments_FV-R-vis}. The top left panel shows the prospect for light WIMP detection (in the FV-R scenario) at underground experiments (direct dark matter detection). The scattering cross-section between the WIMP and an electron is more suppressed than that of the SV-R scenario because of the smaller kinetic mixing parameter $\xi$, which is well below the sensitivity of future experiments. The bottom left panel shows the prospect for the detection at accelerator experiments (accelerator dark matter detection) utilizing the invisible decay of the mediator particle; it is again challenging to detect the signal, as the mixing parameter $\xi$, hence, the production of the mediator particle, is suppressed. The bottom right panel shows the prospect at astrophysical observations (indirect dark matter detection), originating in the WIMP annihilation $\psi \bar{\psi} \to e^- e^+ \gamma$.
The COSI observation will explore some appearing parameter sets with $m_\varphi = {\cal O}(1)$\,MeV, with a somewhat less strong signal than that of the SV-R scenario; it originates in the s-wave annihilation of the fermionic WIMP, being constrained by the CMB observation (discussed in section\,\ref{subsubsec: the scenrios}) more severely than the p-wave annihilation of the scalar WIMP. So, indirect dark matter detection will be only the probe searching for the light WIMP in the FV-R scenario in the near future.

\subsubsection{The SV(B)-R scenario}
\label{subsubsec: SV(B)-R}

Next, we consider the models with the U(1)$_{\rm B}$ vector mediator particle discussed in section\,\ref{subsec: vector mediator}, i.e., the SV(B) and FV(B) models. We first consider the model with a scalar WIMP focusing on the resonance region, i.e., the SV(B)-R scenario. The independent model parameters describing the scenario are \mbox{\boldmath $(m_\varphi,\,\delta,\,\xi,\,g_{\rm B},\,g_\varphi,\,\lambda_{h \varphi \varphi},\,\lambda_\varphi)$}, which is almost the same as those in the SV-R scenario, as shown in Table\,\ref{tab: parameters}. So, we take the same strategy as in section\,\ref{subsubsec: SV-R}, except for the following two things. First, the WIMP annihilates mainly into $\pi^0 \gamma$ when $\xi \ll g_{\rm B-L}$. However, since the WIMP annihilation cross-section is highly suppressed when $m_\varphi \leq 100$\,MeV as deduced from eq.\,(\ref{eq: decay pi0gamma}), such a scenario is ruled out for the same reason as those of the resonant scenarios with the scalar mediator. On the other hand, the scenario becomes almost the same as the SV-R scenario when $\xi \gg g_{\rm B-L}$, with reduced viable parameter region due to the $N_{\rm eff}$ constraint in section\,\ref{subsubsec: mass CMB}. Therefore, we vary the parameters $\xi$ and $g_{\rm B}$ in appropriate ranges so that the WIMP annihilates primarily into $e^+ e^-$ through the kinetic mixing $\xi$, while also into $\pi^0 \gamma$ through the U(1)$_{\rm B}$ interaction with a certain branching fraction; it gives an interesting signal for the future MeV gamma-ray observation such as the COSI. Moreover, we only consider the parameter region $\delta > 0$; the effect of the resonance in the $\delta < 0$ region is much weaker than that in the $\delta > 0 $ region, so the scenario with a certain branching fraction into $\pi^0 \gamma$ cannot be realized when $\delta < 0$.

\begin{figure}[t]
    \centering
    \includegraphics[keepaspectratio, scale=0.19]{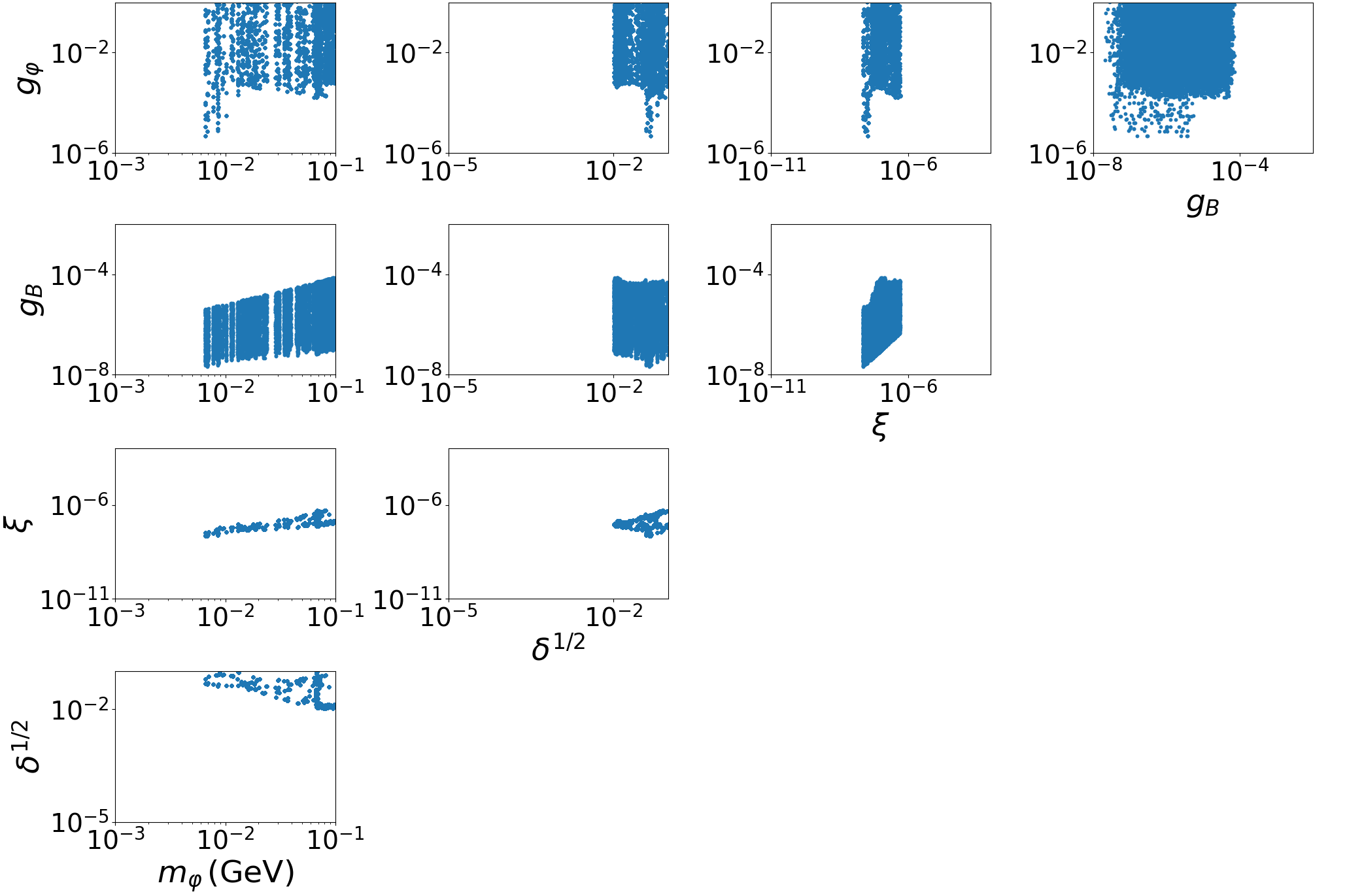}
    \caption{\small \sl The viable parameter region of the SV(B)-R scenario obtained in the MCMC sampling by imposing the conditions and constraints shown in Table\,\ref{tab: conditions & constraints}. See the main text for more details.
    }
    \label{fig: scan_SV-R-inv-B}
\end{figure}

In the MCMC sampling, the WIMP mass $m_\varphi$, the mediator particle mass $\delta$, the kinetic mixing parameter $\xi$, and the WIMP coupling to the vector mediator particle $g_\varphi$ are varied in the same ranges as those in the SV-R scenario. The U(1)$_{\rm B}$ coupling is varied in the range of $g_{\rm B} \geq \xi$ to find the parameter region with a certain branching fraction of the $\varphi\varphi^* \to \pi^0 \gamma$ process. The result of the sampling is shown in Fig.\,\ref{fig: scan_SV-R-inv-B}. First, the lower limit on the WIMP mass comes from the CMB observation in section\,\ref{subsubsec: mass CMB}, and that on $\delta$ is from the BBN observation in section\,\ref{subsubsec: BBN on annihiation}. On the other hand, the relic abundance condition in section\,\ref{subsubsec: thermal relics} and the stringent accelerator constraints on $g_{\rm B}\,(\geq \xi)$ in section\,\ref{sec: Accelerator Light} require $g_\varphi \gg g_{\rm B}, \xi$. This establishes a lower limit on $g_\varphi$ and implies that the invisible decay width of the mediator particle (via $g_\varphi$) is larger than the visible decay width (via $\xi$ and $g_{\rm B}$). Some structures of the viable parameter region can be seen in the top four panels because of the non-trivial accelerator constraints; those are also seen in the SV-R scenario with $\delta > 0$ (the left panels in Fig.\,\ref{fig: scan_SV-R-inv-DP}). Since we consider the parameter region with $g_{\rm B} \gtrsim \xi$ and the accelerator constraint on $g_{\rm B}$ gives the upper limit on $g_{\rm B}$, the upper limit on $\xi$ is obtained as $\xi < {\cal O}(10^{-8})$. Meanwhile, the relic abundance condition (at a given $m_\varphi$) and the lower limit on $\delta$ gives the lower limit on $\xi$ as $\xi \gtrsim {\cal O}(10^{-8})$. So, the $\xi$ is almost uniquely fixed at each $m_\varphi$.

\begin{figure}[t]
    \centering
    \includegraphics[keepaspectratio, scale=0.42]{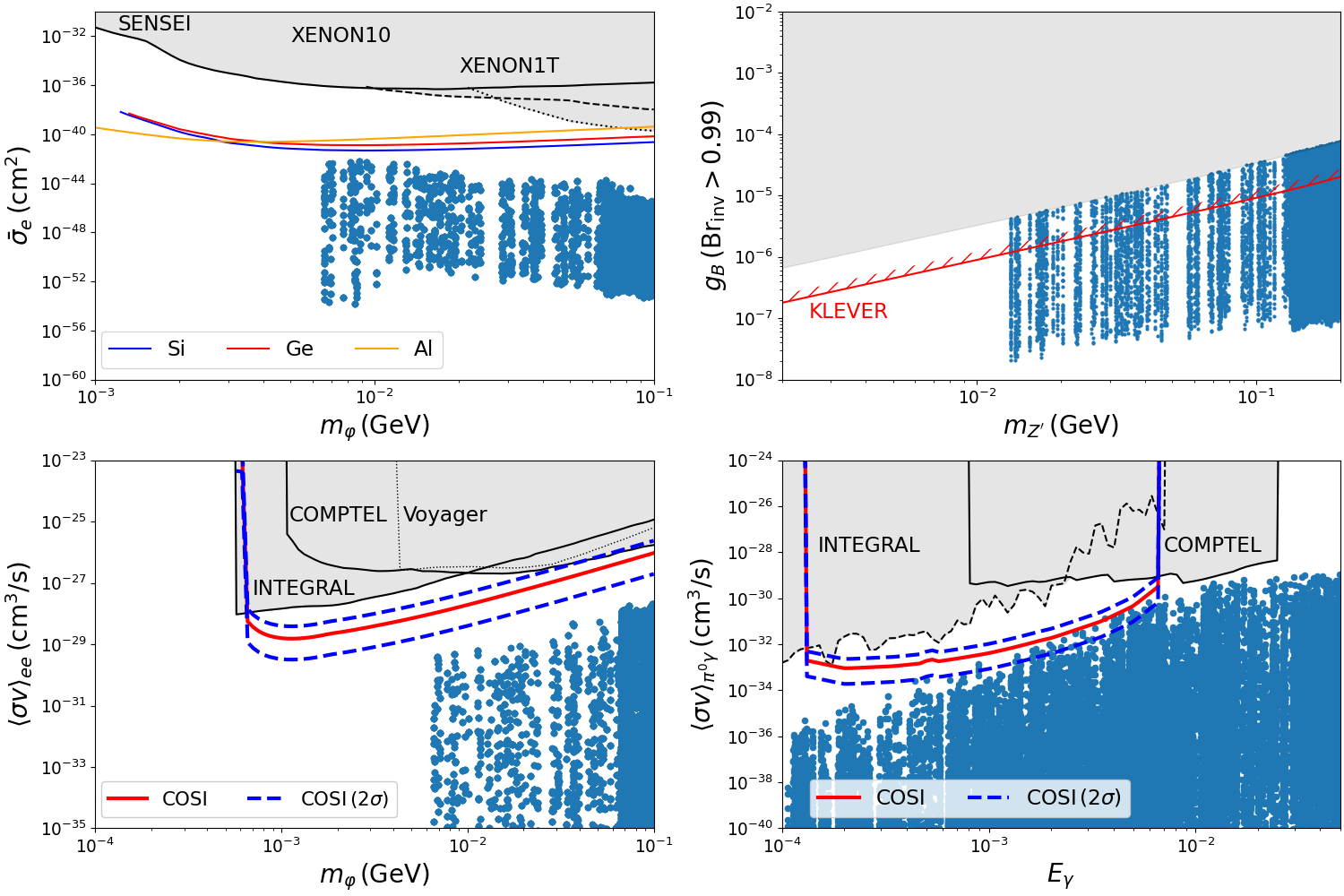}
    \caption{\small \sl The prospect of light WIMP detection in the {\bf SV(B)-R} scenario at various future dark matter and mediator particle search experiments. The prospect at the direct detection is in the top left panel, and that at the accelerator detection utilizing the invisible decay of the mediator particle is in the top right panel. The prospects at the indirect dark matter detection utilizing the continuum and line gamma-ray observations are shown in the bottom left and right panels, respectively.}
    \label{fig:  experiments_SV-R-inv-B}
\end{figure}

The viable parameter region in Fig.\,\ref{fig: scan_SV-R-inv-B} is compared with the projected sensitivity of the future experiments in Fig.\,\ref{fig:  experiments_SV-R-inv-B}. The top left panel shows the prospect for light WIMP detection (in the SV(B)-R scenario) at underground experiments (direct dark matter detection); it is difficult to detect the signal in future experiments, as the scattering cross-section between the WIMP and an electron is suppressed because of the suppressed parameter $\xi$, as addressed above. The top right panel shows the prospect at accelerator experiments (accelerator detection) utilizing the invisible decay of the mediator particle; the near-future experiment, KLEVER, will explore a part of the parameter region. The bottom left and right panels show the prospect at astrophysical observations (indirect dark matter detection utilizing continuum and line MeV gamma-ray observations), originating in the WIMP annihilation processes $\varphi \varphi^* \to e^- e^+ \gamma$ and $\varphi \varphi^* \to \pi^0 \gamma$, respectively. Detecting the WIMP utilizing the search for the continuum MeV gamma-ray seems difficult by the COSI observation. Although the strength of the continuum gamma-ray signal is similar to that of the SV-R scenario, the region with a ${\cal O}(1)$\,MeV WIMP mass, which gives a strong enough continuum gamma-ray signal in the SV-R scenario, is excluded in the SV(B)-R scenario because of the CMB constraint in section\,\ref{subsubsec: the scenrios}. On the other hand, the COSI has the potential to detect a distinctive signal, the line MeV gamma-ray from the WIMP annihilation into $\pi^0 \gamma$.

\subsubsection{The FV(B)-R scenario}

\begin{figure}[t]
    \centering
    \includegraphics[keepaspectratio, scale=0.19]{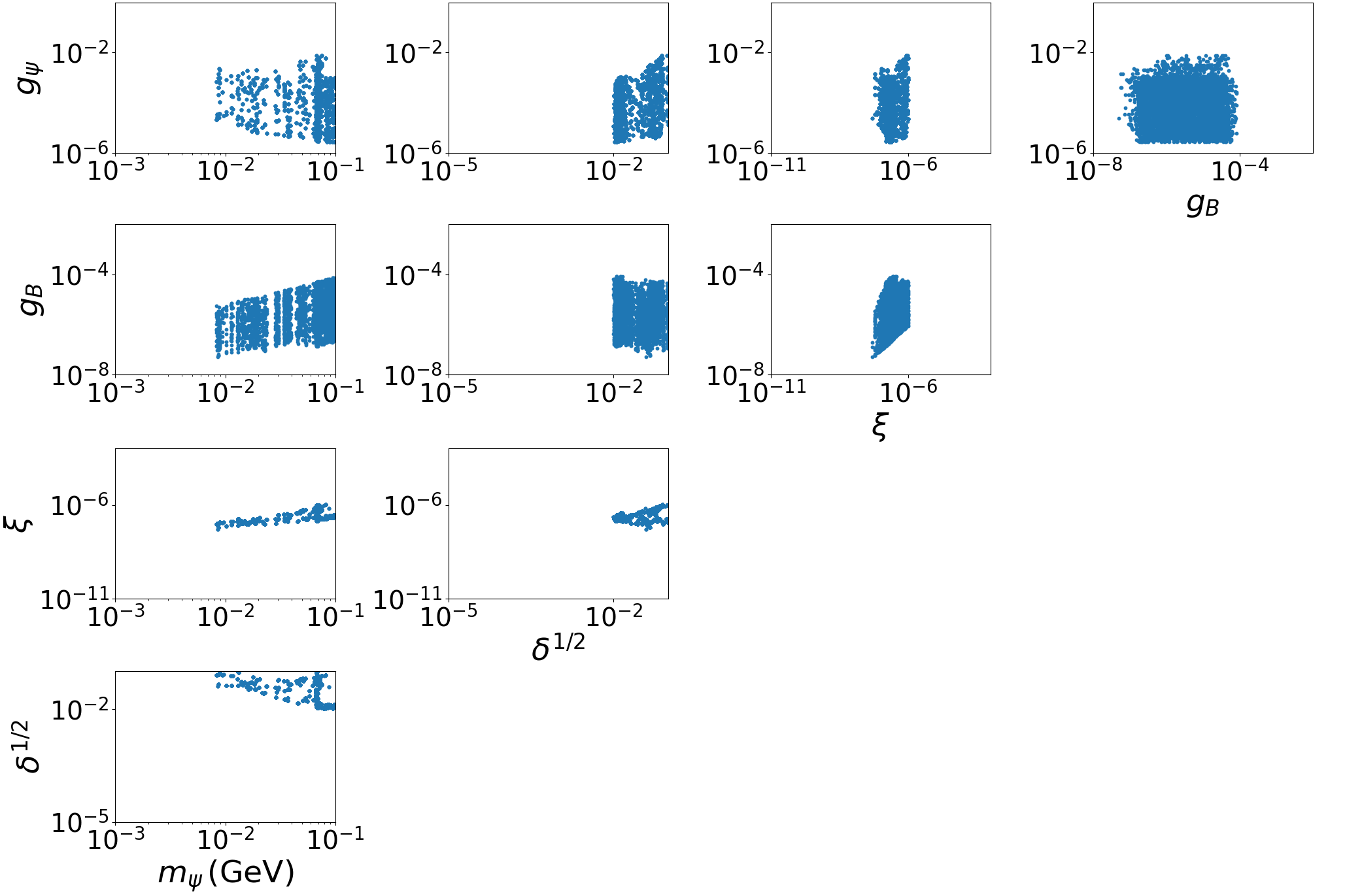}
    \caption{\small \sl The viable parameter region of the FV(B)-R scenario obtained in the MCMC sampling by imposing the conditions and constraints shown in Table\,\ref{tab: conditions & constraints}. See the main text for more details.
    }
    \label{fig: scan_FV-R-inv-B}
\end{figure}

We consider the model with the U(1)$_{\rm B}$ vector mediator particle and a fermionic WIMP on the resonance region, i.e., the FV(B)-R scenario. The independent model parameters describing the scenario are \mbox{\boldmath $(m_\psi, \delta, \xi, g_{\rm B}, g_\psi)$} and those are almost the same as those in the FV-R and SV(B)-R scenarios, so we take the same strategy as those in sections\,\ref{subsubsec: SV-R} and \ref{subsubsec: SV(B)-R}.

In the MCMC sampling, all the model parameters (WIMP mass $m_\psi$, mediator particle mass $\delta$, kinetic mixing parameter $\xi$, WIMP coupling to the vector mediator particle $g_\psi$, and U(1)$_{\rm B}$ coupling $g_{\rm B}$) are varied in the same ranges as those in the SV(B)-R scenario. The result of the sampling is shown in Fig.\,\ref{fig: scan_FV-R-inv-B}. Viable parameter regions in the panels are almost the same as those in the SV(B)-R scenario except for the upper four panels. In the panels, the coupling between the WIMP and the mediator particle $g_\psi$ is found to be smaller than about 10$^{-3}$; it is from the CMB constraint in section\,\ref{subsubsec: the scenrios}, as the fermionic WIMP annihilates in s-wave. On the other hand, the lower limit on $g_\psi$ is less severe than that on $g_\varphi$(in the SV-R scenario), as the mediator particle decays into a WIMP pair in s-wave and its invisible decay width easily dominates over the visible one (for satisfying the relic abundance condition), even if $g_\psi$ is as small as $10^{-6}$. Finally, the lower limit on the WIMP mass is severe compared to that in the SV(B)-R scenario, as the fermionic WIMP has a larger degree of freedom than the scalar WIMP. As a result, some structures seen in Fig.\,\ref{fig: scan_SV-R-inv-B} disappear in Fig.\,\ref{fig: scan_FV-R-inv-B}.

\begin{figure}[t]
    \centering
    \includegraphics[keepaspectratio, scale=0.42]{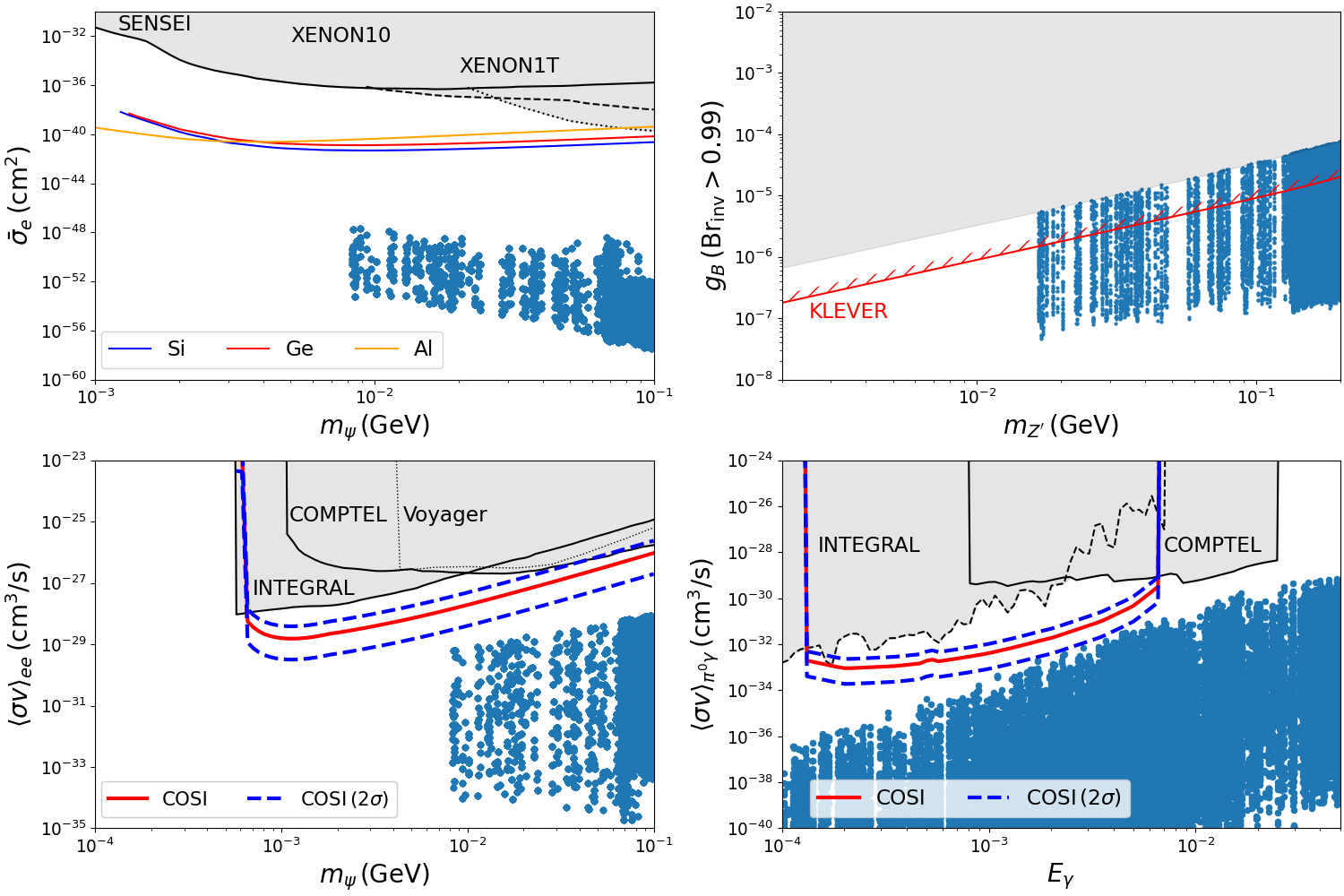}
    \caption{\small \sl The prospect of light WIMP detection in the {\bf FV(B)-R} scenario at various future dark matter and mediator particle search experiments. The prospect at the direct detection is in the top left panel, and that at the accelerator detection utilizing the invisible decay of the mediator particle is in the top right panel. The prospect at the indirect dark matter detection utilizing the continuum and line gamma-ray observations are shown in the bottom left and right panels, respectively.}
    \label{fig: experiments_FV-R-inv-B}
\end{figure}

The viable parameter region in Fig.\,\ref{fig: scan_FV-R-inv-B} is compared with the projected sensitivity of the future experiments in Fig.\,\ref{fig: experiments_FV-R-inv-B}. The result is more or less similar to that of the SV(B)-R scenario. The prospect for light WIMP detection (in the FV(B)-R scenario) at underground experiments (direct dark matter detection) shown in the top left panel indicates that it is more difficult than the SV(B)-R's to detect the signal in future experiments, as $g_\psi$ is more suppressed than $g_\varphi$ (in the SV(B)-R scenario), leading to a lower scattering cross-section between the WIMP and an electron. The prospect at accelerator experiments (accelerator detection) utilizing the invisible decay of the mediator particle shown in the top right panel is essentially the same as that in the SV-R scenario, as the viable range of the parameter $g_{\rm B}$ is the same in both scenarios. The prospect at astrophysical observations (indirect dark matter detection) looks similar to that in the SV-R scenario. It is, however, an accident; the annihilation cross-section at the present universe is $\propto g_\psi^2$, which is about $10^{-6}$ smaller than $g_\varphi^2$, as addressed above. On the other hand, the annihilation proceeds in s-wave in the FV(B)-R scenario, while in p-wave in the SV(B)-R scenario, which gives about $10^{-6}$ suppression in the latter. As a result, the signal strength becomes similar in both scenarios.

\section{Summary}
\label{sec: summary}

In this article, we systematically studied light (MeV-scale) WIMPs based on minimality and renormalizability, imposing the $Z_2$ symmetry to stabilize the WIMP. To make the discussion concrete, we considered two cases of the WIMP with a spin-0 (real/complex scalar) and a spin-1/2 (Majorana/Dirac fermion) as representatives of bosonic and fermionic DM, with two cases of the singlet mediator particle with spin-0 and 1. We considered the two models for the cases with a vector (spin-1) mediator particle; the one originates in the dark photon scenario (with a tiny U(1)$_{\rm B-L}$ gauge coupling), and the other is from the U(1)$_{\rm B}$ gauge theory. So, the six models have been discussed; the {\bf SS}, {\bf FS}, {\bf SV}, {\bf FV}, {\bf SV(B)}, and {\bf FV(B)} models.

Although light WIMPs are severely constrained by cosmological observations in general, as discussed in section\,\ref{sec: Cosmology}, the models composed of the WIMP and the mediator particle, as those mentioned above, enable us to avoid the constraints in non-trivial ways: First one is that the WIMP annihilation proceeds in p-wave thanks to the superb assignments of the WIMP and mediator particle spins. This scenario is realized in the FS, SV, and SV(B) models, called the bulk scenario in this article, as it is realized in the bulk parameter region of the models. The next one is that the WIMP annihilates dominantly into a pair of mediator particles via a forbidden channel ($m_{\rm MED} \gtrsim m_{\rm DM}$), which is realized in all the models and called the forbidden scenario. Finally, the WIMP annihilates into the SM particles via the $s$-channel resonance of the mediator particle ($m_{\rm MED} \simeq 2 m_{\rm DM}$), which is also realized in all the models and called the resonance scenario. All the scenarios predict a velocity-dependent annihilation cross-section and enable us to avoid the severe cosmological constraints, as addressed in section\,\ref{subsubsec: the scenrios}. We particularly focus on bulk scenarios in the FS and SV models, forbidden scenarios in the SS, FS, SV, and FV models, and resonance scenarios in the SS, FS, SV, FV, SV(B), and FV(B) models, which are called {\bf FS-B} \& {\bf SV-B}, {\bf SS-F} \& {\bf FS-F} \& {\bf SV-F} \& {\bf FV-F}, and {\bf SS-R} \& {\bf FS-R} \& {\bf SV-R} \& {\bf FV-R}, {\bf SV(B)-R} \& {\bf FV(B)-R} scenarios in this article.

\begin{table}[t]
    \centering
    \begin{tabular}{r|cccc}
        & Bulk & Resonance ($\delta > 0$) & Resonance ($\delta < 0$) & Forbidden \\
        \hline
        {\bf SS} & {\footnotesize N/A} & {\footnotesize No viable region} & {\footnotesize No viable region} & \textcolor{blue}{\large $\circledcirc$} \\
        {\bf FS} & \textcolor{blue}{\footnotesize Weak} & {\footnotesize No viable region} & {\footnotesize No viable region} & \textcolor{blue}{\footnotesize Weak} \\
        {\bf SV} & \textcolor{blue}{\footnotesize Weak} & \textcolor{blue}{\Large $\circ$} & \textcolor{blue}{\large $\circledcirc$} & \textcolor{blue}{\large $\circledcirc$} \\
        {\bf FV} & {\footnotesize N/A} & \textcolor{blue}{\Large $\circ$} & {\footnotesize No viable region} & \textcolor{blue}{\footnotesize Weak} \\
        {\bf SV(B)} & -- & \textcolor{blue}{\large $\checkmark$} & {\footnotesize No viable region} & -- \\
        {\bf FV(B)} & -- & \textcolor{blue}{\large $\checkmark$} & {\footnotesize No viable region} & -- \\
        \hline
    \end{tabular}
    \caption{\small \sl 
    The prospect of detecting the gamma-ray signal at the COSI from light WIMP annihilation. The symbol ``\,{\large $\circledcirc$}'' (``\,{\Large $\circ$}'') denotes that a strong enough continuum gamma-ray signal is expected widely in (a part of) the viable parameter region, which originates in the DM + DM $\to e^- e^+ \gamma$ process. The symbol ``\,{\large $\checkmark$}" denotes that a strong enough monochromatic gamma-ray signal is expected in a part of the viable parameter region, which originates in the DM + DM $\to \pi^0 + \gamma$ process. On the other hand, the words ``Weak'' and ``No viable region'' in the table mean that the gamma-ray signals are too weak to detect at the COSI and no viable parameter region can be found in our sampling with the constraints in section\,\ref{sec: conditions and constraints}. Finally, ``N/A'' and ``--'' have the same meanings as in table\,\ref{tab: light WIMP scenarios}.}
    \label{tab: sumarry}
\end{table}

We clarify the variable parameter region in each scenario using MCMC sampling, considering the cosmological constraints addressed above, as well as those from the dark matter and the mediator particle search experiments in section\,\ref{sec: Light Thermal DM: Detection}. After that, the viable parameter regions are compared with the projected sensitivity of future direct, accelerator, and indirect detection experiments. Since we are interested in the scenarios that will be explored by the near-future indirect dark matter detection, i.e., COSI, we particularly focused on the six scenarios (i.e., the {\bf SS-F}, {\bf SV-F}, {\bf SV-R}, {\bf FV-R}, {\bf SV(B)-R}, and {\bf FV(B)-R} scenarios) among those addressed above. The viable parameter regions of the scenarios are shown in Figs.\,\ref{fig: scan_SS-F}, \ref{fig:  scan_SV-F}, \ref{fig: scan_SV-R-inv-DP}, \ref{fig: scan_FV-R-inv-DP}, \ref{fig: scan_SV-R-inv-B}, and \ref{fig: scan_FV-R-inv-B}, while their prospects of light WIMP detection at the future experiments are given in Figs.\ref{fig: experiments_SS-F}, \ref{fig:  experiments_SV-F}, \ref{fig:  experiments_SV-R-inv-DP} \& \ref{fig:  experiments_SV-R-vis}, \ref{fig:  experiments_FV-R-vis}, \ref{fig:  experiments_SV-R-inv-B}, and \ref{fig: experiments_FV-R-inv-B}, respectively. We found that COSI will explore a diverse range of the above scenarios, as summarized in Table\,\ref{tab: sumarry}. Its details are as follows: First, in the forbidden scenarios of the SS and SV models (i.e., the {\bf SS-F} and {\bf SV-F} scenarios), the light WIMP annihilation produces a strong continuum gamma-ray signal (originating in the DM + DM $\to e^- e^+ \gamma$ process) widely in their parameter regions, which the COSI will efficiently explore. On the other hand, in the resonance scenarios of the SV and FV models (i.e., the {\bf SV-R} and {\bf FV-R} scenarios), the annihilation produces the same strong gamma-ray signal in their parameter regions with $m_{\rm DM} < m_{\rm MED}/2 = {\cal O}(1)$\,MeV, while the annihilation produces the same strong signal widely in the parameter region of the SV model when $m_{\rm DM} > m_{\rm MED}/2$. Finally, in the resonance scenarios of the SV(B) and FV(B) models (i.e., the {\bf SV(B)-R} and {\bf FV(B)-R} scenarios), the WIMP annihilation produces a strong monochromatic gamma-ray signal (originating in the DM + DM $\to \pi^0 \gamma$ process) when $m_{\rm DM} < m_{\rm MED}/2$, with the energy of the photon (directly from the annihilation and not from the $\pi^0$ decay) being ${\cal O}(1)$\,MeV, which will be an interesting target for COSI.

\begin{figure}[t]
    \centering
    \includegraphics[keepaspectratio, scale=0.42]{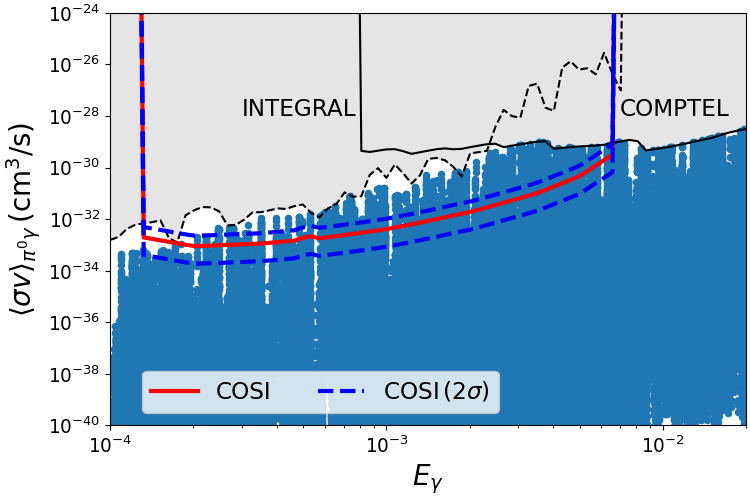}
    \caption{\small \sl The prospect of light WIMP detection in the {\bf FV(B)-R} scenario with the BBN constraint in Ref.\,\cite{Braat:2024khe} at the indirect dark matter detection using the line gamma-ray observation at the COSI.}
    \label{fig: experiments_FV-R-inv-B_wo_BBN}
\end{figure}

Before closing the section, we discuss possible improvements in the comprehensive study of the light WIMP discussed in this article. First, it concerns lower limits on the WIMP and mediator particle masses from the CMB and BBN observations when the particles simultaneously interact with electromagnetic and neutrino sectors. We referred to the references\,\cite{Matsumoto:2018acr, Ibe:2018juk, Escudero:2018mvt} in our analysis to put the limits on such light WIMP scenarios. In order to have the limits more accurately, it is mandatory to accurately evaluate how the photon and neutrino temperatures, as well as the expansion rate of the universe, evolve, which is obtained by integrating Boltzmann equations for the WIMP and mediator particle numerically. According to the estimate in Ref.\,\cite{Sabti:2021reh}, the limits are expected to be slightly more severe than those we used in this article; the masses below 1-2\,MeV would be excluded in the SV and FV models, as already mentioned in section\,\ref{subsubsec: mass BBN}. The next concern is limiting the WIMP annihilation cross-section in the resonance scenarios from the BBN observation. We put the constraint on the cross-section as explained in section\,\ref{subsubsec: BBN on annihiation}; it seems aggressive compared to that from a more realistic estimate. In fact, according to the recent paper\,\cite{Braat:2024khe}, the constraint is weaker than we adopted. To grasp how much the new constraint affects our result, we performed the same MC sampling again in the FV(B)-R scenario with this latest constraint (instead of that in section\,\ref{subsubsec: BBN on annihiation}) and reevaluate the prospect of detecting the monochromatic gamma-ray signal with COSI. The result is shown in Fig.\,\ref{fig: experiments_FV-R-inv-B_wo_BBN}; the signal can be stronger than that in the bottom right panel of Fig.\,\ref{fig:  experiments_SV-R-inv-B}, which is well above the COSI sensitivity. So, COSI may be able to detect the stronger monochromatic gamma-ray signal; we will leave its detailed quantitative discussion for future work. The last concern is the accelerator constraint on the mediator particles, particularly when they have comparable visible and invisible decay widths. In this article, we adopted the method developed in Ref.\,\cite{Ilten:2018crw} to put the constraint. Although this method is employed in many past phenomenological studies\,\cite{Brahma:2023psr, Bernreuther:2020koj, Amrith:2018yfb, Flores:2020lji}, caution also must be exercised due to inherent uncertainties; the method ignores the kinematical spread of the mediator particle's momentum and the dependence of the detection efficiency on the position that the mediator decays inside the decay volume. A more detailed simulation study will be required to refine the accelerator constraint, and we will leave this problem for future work. Here, it is also worth notifying that this refinement affects only a part of our result: It does not affect the forbidden and resonance (with $\delta < 0$) scenarios at all, as the mediator always decays visibly. It also affects little on the resonance (with $\delta > 0$) scenarios based on U(1)$_{\rm B}$ (i.e., SV(B)-R ann FV(B)-R scenarios), as the mediator decays invisibly at almost 100\,\% C.L.\footnote{
    The refinement would affect only a tiny fraction of the viable parameter space with the smallest $\delta$.}
So, the refinement affects the SV-R and FV-R scenarios with $\delta > 0$. In even such cases, most of the set of parameters in the viable parameter region predicts either the visible or invisible mediator decay with a $\sim 100$\,\% branching fraction, and the region that predicts comparable visible and invisible fractions is only a small part of the entire viable region, as it requires fine-tuning on $\xi$.

\appendix
\section{scalar interactions}
\label{app: scalar interactions}

\begin{align}
	C_{h h h} &= 3\lambda_H v_H c_\theta^3 - 3\mu_{S H} c_\theta^2 s_\theta - \mu_3 s_\theta^3
	+ 3\lambda_{S H} v_H c_\theta s_\theta^2, \nonumber \\
	C_{\varsigma h h} &= 3\lambda_H v_H c_\theta^2 s_\theta + \mu_{S H} (c_\theta^3 - 2c_\theta s_\theta^2)
	+ \mu_3 c_\theta s_\theta^2 + \lambda_{S H} v_H (s_\theta^3 - 2c_\theta^2 s_\theta), \nonumber \\
	C_{\varsigma \varsigma h} &= 3\lambda_H v_H c_\theta s_\theta^2 + \mu_{S H} (2c_\theta^2 s_\theta - s_\theta^3)
	- \mu_3 c_\theta^2 s_\theta + \lambda_{S H} v_H (c_\theta^3 - 2c_\theta s_\theta^2), \nonumber \\
	C_{\varsigma \varsigma \varsigma} &= 3\lambda_H v_H s_\theta^3 + 3\mu_{S H} c_\theta s_\theta^2
	+ \mu_3 c_\theta^3 + 3\lambda_{S H} v_H c_\theta^2 s_\theta, \nonumber \\
	C_{h h h h} &= 3\lambda_H c_\theta^4 + 6\lambda_{S H} c_\theta^2 s_\theta^2 + \lambda_S s_\theta^4, \nonumber \\
	C_{\varsigma h h h} &= 3\lambda_H c_\theta^3 s_\theta - 3\lambda_{S H} (c_\theta^3 s_\theta - c_\theta s_\theta^3)
	- \lambda_S c_\theta s_\theta^3, \nonumber \\
	C_{\varsigma \varsigma h h} &= 3\lambda_H c_\theta^2 s_\theta^2 +\lambda_{S H} (c_\theta^4 - 4c_\theta^2 s_\theta^2 + s_\theta^4)
	+ \lambda_S c_\theta^2 s_\theta^2, \nonumber \\
	C_{\varsigma \varsigma \varsigma h} &= 3\lambda_H c_\theta s_\theta^3 + 3\lambda_{S H} (c_\theta^3 s_\theta - c_\theta s_\theta^3)
	- \lambda_S c_\theta^3 s_\theta, \nonumber \\
	C_{\varsigma \varsigma \varsigma \varsigma} &= 3\lambda_H s_\theta^4 + 6\lambda_{S H} c_\theta^2 s_\theta^2 + \lambda_S c_\theta^4, \\
    \nonumber \\
    C_{h \phi \phi} &= \lambda_{H \phi} v_H c_\theta - \mu_{S \phi} s_\theta, \nonumber \\
    C_{\varsigma \phi \phi} &= \lambda_{H \phi} v_H s_\theta + \mu_{S \phi} c_\theta, \nonumber \\
    C_{h h \phi \phi} &= \lambda_{H \phi} c_\theta^2 + \lambda_{S \phi} s_\theta^2, \nonumber \\
    C_{h \varsigma \phi \phi} &= \lambda_{H \phi} c_\theta s_\theta - \lambda_{S \phi} s_\theta c_\theta, \nonumber \\
    C_{\varsigma \varsigma \phi \phi} &= \lambda_{H \phi} s_\theta^2 + \lambda_{S \phi} c_\theta^2.
    \label{eq : couplings}
\end{align}

\section{The "p-wave" scenarios}
\label{app: p-wave}

As discussed in section\,\ref{sec: conditions and constraints}, the light WIMP undergoes p-wave annihilation in the FS-B and SV-B scenarios. In the former scenario, the WIMP annihilates mainly into $e^- e^+$ or a pair of mediator particles ($\varsigma \varsigma$) decaying into $e^- e^+$. Meanwhile, the WIMP annihilation into $\gamma \gamma$ or $\varsigma \varsigma$ decaying into $\gamma\gamma$ is highly suppressed, whose cross-section is about $10^5$ times smaller than the main channel. So, based on the discussion in section\,\ref{subsec: analysis}, the annihilation cross-section producing a line signal is expected to be ${\cal O}(10^{-36})$\,cm$^3$/s, which is two orders of magnitude below the projected sensitivity of the COSI, as shown in the right panel of Fig.\,\ref{fig: indirect detection}.

In the latter model, i.e., the light WIMP with a vector mediator particle, the WIMP annihilates mainly into $e^- e^+$ and sub-dominantly into $\pi^0 \gamma$, where the latter process produces a line gamma-ray signal. The annihilation into $\pi^0 \gamma$ is highly suppressed by a loop factor, especially when the energy of the line gamma-ray is ${\cal O}(1)$\,MeV. The flux of the line gamma-ray is estimated to be $\alpha\,(E_\gamma/\sqrt{s})^3 \sim 10^{-9}$ times smaller than that of the main channel, assuming $E_\gamma \sim 1$\,MeV and $\sqrt{s} \sim m_{\pi^0}$, which is well below the projected sensitivity of the COSI.

\section{Perturbative unitarity limits}
\label{app: unitarity}

The matrix element of a scattering amplitude between the initial state ``$i$" and the final state ``$f$," denoted by ${\cal M}_{fi}$ in this subsection, is defined by the following formula
\begin{align}
    \braket{f|\hat{T}|i} =
    (2\pi)^4\,\delta^{(4)}(P_i - P_f)
    \,{\cal M}_{fi}(\sqrt{s},\cos\theta),
\end{align}
where $\sqrt{s}$ is the center-of-mass energy of the collision for the scattering, $\cos\theta$ is the scattering angle, $P_i\,(P_f)$ is the total four-momentum of the initial\,(final) state, and $\hat{T}$ is the interacting part of the $S$-matrix, i.e., $\hat{S} \equiv {\bf 1} + i {\hat T}$. Then, the scattering amplitude can be decomposed into segments with a fixed total angular momentum named $J$ as follows:
\begin{align}
    a^J_{fi}
    = \frac{1}{32\pi} \sqrt{\frac{4\abs{\mathbf{p_f}}\abs{\mathbf{p_i}}}{2^{\delta_f}\,2^{\delta_i}\,s}}
    \int^1_{-1} d(\cos\theta)\,d^J_{\mu_i\mu_f}(\theta)\,{\cal M}_{fi}(\sqrt{s},\cos\theta).
    \label{partial wave}
\end{align}
Here, $d^J_{\mu_i\mu_f}(\theta)$ is the so-called $J$-th Wigner d-function, and $\mu_{i\,(f)} \equiv \lambda_{i\,(f)\,1} - \lambda_{i\,(f)\,2}$ with $\lambda$ being the helicity, $\mathbf{p_{i\,(f)}}$ is the momentum of the scattering particle in the center-of-mass frame, and $\delta_{i\,(f)}$ is one if particles in state ``$i$\,($f$)" are identical, or zero otherwise. The preservation of the probability before and after the scattering process is guaranteed by the unitarity of the $S$-matrix, i.e., $\hat{S}^\dagger \hat{S} = 1$, and this condition gives the upper limit; ${\rm Re}( a^J_{ii} ) \leq 1/2$\,\cite{Lee:1977eg, Marciano:1989ns}.

At high energy (i.e., large $\sqrt{s}$) limit, diagrams having only dimensionless coupling constants dominantly contribute to the scattering amplitude. Scatterings between various initial ("i") and final ("f") states provide the matrix of scattering amplitudes, ${\bf a}^J = \{a^J\}_{fi}$. Imposing the unitarity limit above on the eigenvalues of ${\bf a}^J$ gives upper limits on the coupling constants\,\cite{Arhrib:2011uy, Aoki:2007ah}. The limits are approximately given by $\lesssim 4\pi$ for four-point couplings such as the self-interaction of the scalar WIMP, and $\lesssim \sqrt{4\pi}$ for three-point couplings such as Yukawa and gauge interactions. On the other hand, dimensionful coupling constants are expected to be constrained by particle scatterings at lower energy. Those are approximately constrained to be less than $\sqrt{4\pi}$ times the mass scale of the scatterings\,\cite{Schuessler:2007av, Goodsell:2018tti}.

Detailed calculation of the amplitudes will give slightly stronger constraints than the approximations addressed above. Considering all initial and final states, partial waves, and energies is difficult, even numerically. Moreover, including the precise constraints in the analysis does not significantly change the result. Hence, we adopt the approximations.

\section{Vacuum stability}
\label{app: vacuum stability}

Here, we discuss the vacuum stability of the light WIMP models with a scalar mediator particle, i.e., the SS and FS models. We first consider the FS model, where the scalar potential is composed of the SM Higgs double field $H$ and the mediator field $S$, as shown in eq.\,(\ref{eq: S}).

First, the scalar potential must be bounded from below to prevent the vacuum energy from going to a negative infinity, i.e., to make the system well-defined. This condition gives constraints on the dimensionless model parameters, $\lambda_S$, $\lambda_H$, and $\lambda_{SH}$, as follows\,\cite{Kannike:2012pe}:
\begin{align}
    \lambda_S > 0,\quad
    \lambda_H > 0,\quad
    a_{SH} \equiv \sqrt{3}\lambda_{SH} + \sqrt{\lambda_S \lambda_H} > 0.
    \label{eq: vacuum stability BfB}
\end{align}

Next, our vacuum with the vacuum expectation value of the Higgs doublet field being $v_H$ must be, at least, a local minimum, giving the following constraints on the parameters:
\begin{align}
    2 \mu^2_{S}+\lambda_{SH} v^2_{H} > 0,
    \quad
    \lambda_H (2\mu^2_{S}+\lambda_{SH} v^2_{H}) > 2 \mu^2_{SH}.
    \label{eq: local minimum cond}
\end{align}

Finally, our vacuum must be stable enough. It means that our vacuum is absolutely stable, or its lifetime is sufficiently longer than the age of the universe. The global minimum condition, i.e., the condition making our vacuum absolutely stable, can be obtained analytically\,\cite{Espinosa:2011ax}. Let us first discuss this condition. The scalar potential in question is
\begin{align}
    V(h,a)
    = \mu_{SH}\,S \left( h^2 - \frac{v_H^2}{2} \right)
    + \frac{\lambda_{SH}}{2} S^2 h^2
    + \frac{\mu_S^2}{2} S^2
    + \frac{\mu_3}{3!} S^3
    + \frac{\lambda_S}{4!} S^4
    + \frac{\lambda_H}{2} \left( h^2 - \frac{v_H^2}{2} \right)^2.
\end{align}
Here, we take the unitarity gauge, and a real scalar field $h$ is defined as $H = (h, 0)^T$ in this appendix. It can be easily seen in the above scalar potential that our vacuum is at $(h, S) = (v_H/\sqrt{2}, 0)$. The global minimum condition requests our vacuum to be deeper than other local minima. Stationary points in $V(S, h)$ are found by solving the equations,
\begin{align}
    &\frac{\partial V}{\partial h} = 2 h
    \left[
        \mu_{SH}\,S
        + \frac{\lambda_{SH}}{2} S^2
        + \lambda_H \left( h^2 - \frac{v_H^2}{2} \right)
    \right] = 0,
    \label{eq: stationary 1}\\
    &\frac{\partial V}{\partial S} =
    \mu_{SH} \left( h^2 - \frac{v_H^2}{2} \right)
    + \lambda_{SH}\, S\,h^2
    + \mu_S^2\,S
    + \frac{\mu_3}{2} S^2
    + \frac{\lambda_{S}}{6} S^3 =0.
    \label{eq: stationary 2}
\end{align}
The first equation gives two solutions of the Higgs boson field $h$ (as a function of "$S$") as
\begin{align}
    h=0,
    \quad
    h^2 = \frac{v_H^2}{2}-\frac{1}{\lambda_H}\left( \mu_{SH}\,S + \frac{\lambda_{SH}}{2} S^2 \right) \equiv D^2(S).
    \label{eq: two solutions of h}
\end{align}
Regarding the latter solution, the insertion of the solution into the equation\,(\ref{eq: stationary 2}) gives
\begin{align}
    & \frac{\partial V}{\partial S} = \frac{S}{2\lambda_H}
    (c_2\,S^2 + c_1\,S + c_0) = 0, \\
    &c_2 = \frac{\lambda_S \lambda_H}{3} - \lambda_{SH}^2,
    \qquad
    c_1 = \mu_3 \lambda_H - 3\lambda_{SH}\mu_{SH},
    \quad
    c_0 = \lambda_H (\lambda_{SH}v_H^2 + 2\mu_S^2)-2\mu_{SH}^2.
    \nonumber
\end{align}
The local minimum condition\,(\ref{eq: local minimum cond}) guarantees the positivity of $c_0$, i.e., $c_0 > 0$. When $c_1^2 < 4 c_2 c_0$, the equation $\partial V/\partial S = 0$ has only one solution $S = 0$, which corresponds to our vacuum. Otherwise, the equation has two more solutions, $S = S_\pm$. Both the two solutions correspond to maxima when $c_2 < 0$, so the potential is no longer bounded from below and contradicts the condition in eq.\,(\ref{eq: vacuum stability BfB}). The exception is when the two solutions become unphysical, i.e., when the solutions satisfy $D^2(S_\pm) < 0$. On the other hand, when $c_2 > 0$, one of the two solutions gives a minimum, while the other gives a maximum. This minimum is always shallower than our vacuum when $c_1^2 < 9 c_2 c_0/2$. To the contrary, when $c_1^2 > 9 c_2 c_0/2$, this minimum must be non-physical to make our vacuum the global minimum, so the solutions must satisfy the condition $D^2(S_+) < 0$ when $c_1 < 0$ or $D^2(S_-) < 0$ when $c_1 > 0$. We next consider another solution in eq.\,(\ref{eq: two solutions of h}), i.e., $h = 0$. Inserting it into eq.\,(\ref{eq: stationary 2}) gives
\begin{align}
    \frac{\partial V}{\partial S} =
    -\mu_{SH}\frac{v_H^2}{2}
    +\mu_S^2\,S
    +\frac{\mu_3}{2}S^2
    +\frac{\lambda_S}{6}S^3
    = 0.
\end{align}
This equation has, at most, three solutions $S_i$. Here, it is possible to prove the inequality, $V(0, S) - V(D(S), S) = D^4(S)\,\lambda_H/2 > 0$. Then, it is also possible to prove that our vacuum is deeper than the other minima obtained from the above cubic equation for $``S_i"$ when $c_1^2 < 9 c_2 c_0/2$ is satisfied. This is because $V(0, S) > V(D(S), S) > V( v_H/\sqrt2, 0)$ at any $``S"$. In another case, i.e., $c_1^2 > 9 c_2 c_0/2$, we must solve the above cubic equation and compare the potential values at the minima obtained by the cubic equation to that of our vacuum. In summary, our vacuum is absolutely stable whenever $c_1^2 < 9 c_2 c_0/2$. On the other hand, the two maxima $S_\pm$ must satisfy $D^2(S_\pm) < 0$ when $c_1^2 > 9 c_2 c_0/2$ and $c_2 < 0$, while the extra minimum $S_+$ or $S_-$ must satisfy $D^2(S_+) < 0$ or $D^2(S_-) < 0$, depending on the sign of $c_1$, when $c_1^2 > 9 c_2 c_0/2$ and $c_2 > 0$. Moreover, in the latter cases with $c_1^2 > 9 c_2 c_0/2$, the potential values at $S_i$ must be larger than that of our vacuum, i.e., $V(0, S_i) > V(v_H/\sqrt2, 0)$. So, the condition making our vacuum absolutely stable is expressed by the following statement:
\begin{align}
    &\left[ c_1^2 < \frac{9 c_2 c_0}{2} \right]
    ~\cup~
    \left[
        c_1^2 > \frac{9 c_2 c_0}{2}
        ~\cap~
        V(0, S_i) > V( v_H/\sqrt2, 0)
        ~\cap~
        \left\{
            \left( c_2 < 0 ~\cap~ D^2(S_\pm) < 0 \right)
        \right.
    \right.
    \nonumber \\
    &\qquad\left.
        \left.
            ~\cup~
            \left( c_2 > 0 ~\cap~ c_1 < 0 ~\cap~ D^2(S_+) < 0 \right)
            ~\cup~
            \left( c_2 > 0 ~\cap~ c_1 > 0 ~\cap~ D^2(S_-) < 0 \right) 
        \right\}
        \rule{0cm}{0.6cm}
    \right].
    \label{eq: global minimum cond}
\end{align}
On the other hand, our vacuum does not necessarily have to be absolutely stable. We can request that the lifetime of our vacuum be longer than the age of the universe, which is called the meta-stability condition. It gives a milder constraint than that in eq.\,(\ref{eq: global minimum cond}). The strength of the constraint is expected to be in between those of the local minimum condition\,(\ref{eq: local minimum cond}) and the global minimum condition\,(\ref{eq: global minimum cond}). Obtaining the meta-stability condition by computing the tunneling rate of our vacuum is beyond the scope of this paper. Instead, we implemented the two analyses with and without the global minimum conditions and compared them to interpret the analysis result. Our results have shown that including the global minimum condition does not significantly differ from including only the local minimum one.

We next consider the SS model, where the scalar potential is composed of the dark matter field $\phi$ in addition to the SM Higgs double field H and the mediator field S addressed above, as shown in the lagrangian\,(\ref{eq: S-phi lagrangian}). The vacuum stability condition of this model is derived in the same way as that of the FS model. First, the condition to ensure that the potential is bounded from below is the same as that given in eq.\,(\ref{eq: vacuum stability BfB}), in addition to those of
\begin{align}
    & \lambda_\phi > 0,\quad
    a_{H\phi} \equiv \sqrt{3}\lambda_{H\phi} + \sqrt{\lambda_H \lambda_\phi} > 0,\quad
    a_{\phi S} \equiv 3\lambda_{\phi S} + \sqrt{\lambda_\phi \lambda_S} > 0,
    \nonumber \\
    & \sqrt{\lambda_{H }\lambda_{S}\lambda_{\phi}} + \sqrt{3}\lambda_{SH}\sqrt{\lambda_{\phi}} + \sqrt{3}\lambda_{H\phi}\sqrt{\lambda_{S}} + 3\lambda_{\phi S}\sqrt{\lambda_{H}} + \sqrt{2 a_{SH} a_{H\phi} a_{\phi S}}  > 0.
\end{align}

Next, the local minimum condition is the same as that in eq.\,(\ref{eq: local minimum cond}), in addition to
\begin{align} 
    2\mu_\phi^2 + \lambda_{H\phi} v_H^2 >0.
\end{align}

Finally, the global minimum condition is the same as eq. (\ref{eq: global minimum cond}). We confirmed using the global minimum condition instead of the local one does not significantly alter our results.

\section*{Acknowledgments}

Yu Watanabe is supported by the JSPS KAKENHI Grant Number 23KJ0470. S. Matsumoto is supported by Grant-in-Aid for Scientific Research from MEXT, Japan; 23K20232(20H01895), 20H00153, 24H00244, 24H02244, by JSPS Core-to-Core Program; JPJSCCA20200002, and by World Premier International Research Center Initiative (WPI), MEXT, Japan (Kavli IPMU). T.M. is supported by JSPS KAKENHI grant JP22K18712 and by World Premier International Research Center Initiative (WPI), MEXT, Japan (Kavli IPMU).

\bibliographystyle{unsrt}
\bibliography{refs}

\end{document}